\documentclass[aps, prd,showpacs,amsfonts,notitlepage,amssymb,amsmath,nofootinbib,10pt,longbibliography,superscriptaddress]{revtex4-1}
\usepackage[utf8x]{inputenc}
\usepackage[english]{babel}
\usepackage[colorlinks=true,linkcolor=blue,citecolor=blue,urlcolor=blue]{hyperref}
\usepackage{graphicx}
\usepackage{amsmath}    
\usepackage{xcolor}
\usepackage{amssymb}
\usepackage{import}

\newcommand{\nn}{\nonumber \\}
\newcommand{\eq}[1]{Eq.~(\ref{#1})}
\newcommand{\eqs}[2]{Eqs.~(\ref{#1}) and~(\ref{#2})}
\newcommand{\fig}[1]{Fig.~\ref{#1}}

\newcommand{\be}{\begin{eqnarray}}
\newcommand{\ee}{\end{eqnarray}}
\newcommand{\om}{\ensuremath{\omega}}

\newcommand{\pd}{\ensuremath{\partial}}

\newcommand{\lp}{\ensuremath{\left(}}
\newcommand{\rp}{\ensuremath{\right)}}
\newcommand{\abs}[1]{\ensuremath{\left \lvert #1 \right\rvert}} 
\definecolor{darkcyan}{rgb}{0,0.7,0.7}

\begin{document}

\title{Phonon spectrum and correlations in a transonic flow of an atomic Bose gas}
\date{\today}

\author{Florent Michel} 
\email{florent.michel@th.u-psud.fr}
\affiliation{Laboratoire de Physique Th\'eorique, CNRS, Univ. Paris-Sud, Universit\'e Paris-Saclay, 91405 Orsay, France}

\author{Jean-Fran\c{c}ois Coupechoux} 
\email{Jean-Francois.Coupechoux@th.u-psud.fr}
\affiliation{\'Ecole Normale Sup\'erieure de Cachan, 94230 Cachan, France}

\author{Renaud Parentani}
\email{renaud.parentani@th.u-psud.fr}
\affiliation{Laboratoire de Physique Th\'eorique, CNRS, Univ. Paris-Sud, Universit\'e Paris-Saclay, 91405 Orsay, France}

\begin{abstract}
Motivated by a recent experiment of J.~Steinhauer, we reconsider the spectrum and the correlations of the phonons spontaneously emitted in stationary transonic flows. The latter are described by ``waterfall'' configurations which form a one-parameter family of stable flows. For parameters close to their experimental values, in spite of high gradients near the sonic horizon, the spectrum is accurately Planckian in the relevant frequency domain, where the temperature differs from the relativistic prediction by less than $10 \%$. We then study the density correlations across the horizon and the non-separable character of the final state. We show that the relativistic expressions provide accurate approximations when the initial temperature is not too high. We also show that the phases of the scattering coefficients introduce a finite shift of the location of the correlations which was so far overlooked. This shift is due to the asymmetry of the flow across the horizon, and persists in the dispersion-less regime.
Finally we show how the formation of the sonic horizon modifies both local and non-local density correlations. 
\end{abstract}

\maketitle 

\section{Introduction}

In a recent work~\cite{Steinhauer:2015saa}, J.~Steinhauer reported the observation of the spectrum and correlations of phonons emitted in a flowing condensed atomic gas~\cite{pitaevskii2003bose}. The flow was stationary to a good approximation. 
Importantly, its velocity $v(x)$ crossed the sound speed $c(x)$, so that there was a sonic horizon~\cite{Unruh:1980cg}. 
On a qualitative level, the observations agree rather well with the predictions one can draw from the analogy with black hole radiation~\cite{Barcelo:2005fc}. First, on the subsonic side, one expects to find a steady flux of phonons with a spectrum approximately thermal and with a temperature fixed by the analog surface gravity, at least when the healing length is much smaller than the scale of the horizon surface gravity~\cite{Unruh:1994je,Brout:1995wp,Corley:1996ar,Garay:1999sk,Macher:2009nz}. 
Second, in spite of dispersive effects which radically modify the propagation near the horizon~\cite{Brout:1995wp}, one expects that each of these quanta comes from a pair of entangled phonons, the partner carrying negative energy and propagating on the other side of the sonic horizon, as is the case in relativistic settings~\cite{Brout:1995rd,Massar:1996tx,Parentani:2010bn}. 
Two important features have been observed in~\cite{Steinhauer:2015saa}. On the one hand, near the sonic horizon, the background flow was observed to be close to a ``waterfall'' solution~\cite{Larre:2011mq} with a high Mach number in the supersonic region $M_+ \approx 5$. On the other hand, the initial temperature was reported to be low enough so that the initial state can be considered to be the incoming vacuum. Hence the phonons should be mainly emitted by spontaneous amplification of vacuum fluctuations, rather than stimulated by pre-existing phonons (as is the case when working at higher temperatures). In fact, from the observation of density-density correlations~\cite{Balbinot:2007de,Carusotto:2008ep}, J.~Steinhauer also reported that, for large frequencies, the intensity of the correlations fulfills an inequality which implies that the final phonon state is non-separable, as is the case when the spontaneous channel is the dominant one~\cite{Campo:2005sv,deNova:2012hm,Busch:2014bza,Boiron:2014npa,Steinhauer:2015ava,deNova:2015lva}. 

When working in vacuum and with background flows described by waterfall solutions, the spectral properties and coherence of the emitted phonons can be determined numerically by solving the Bogoliubov-de Gennes equation 
In this paper, we focus on the solution with a Mach number $M_+ = 5$ in the supersonic region. To test the sensitivity of the predictions, we also consider nearby flows with $M_+ = 5 \lp 1 \pm 0.25 \rp$. Despite their high spatial gradients (which are larger than the inverse healing length evaluated at the sonic horizon), we shall see that the spectrum accurately follows the Planck law in the relevant low frequency domain. We also observe that the scattering coefficients involving the co-propagating mode~\cite{Macher:2009tw,Macher:2009nz} are about $10$ times smaller than the coupling between the two counter-propagating modes carrying opposite energy (encoding the analog Hawking effect~\cite{Unruh:1980cg,Unruh:1994je}). Hence, to a fairly good approximation, the spectral properties can be accounted for by their relativistic expression. This stops to be true when the initial temperature of the condensate is much higher than the Hawking temperature fixed by the surface gravity. 

We then study the correlations between the phonons emitted on opposite sides of the sonic horizon. We find that the norm and the phase of the Fourier components of the term encoding these correlations are also well approximated by their relativistic expressions. 
Moreover, the weakness of the couplings to the co-propagating mode preserves the non-separable character of the final phonons up to relatively high initial temperatures. 
Focusing on the {\it phase} of the correlation term, we find that its dependence on the frequency induces a non trivial shift of the locus of the correlations with respect to the expression of~\cite{Balbinot:2007de}. 
Interestingly, this shift persists in the dispersionless limit when sending to zero the healing length. It originates from the large asymmetry of the background flow near the horizon. 

The paper is organized as follows. In Section~\ref{sp}, we first review the basic properties of waterfall solutions and the calculation of the scattering coefficients of linear density perturbations. We then analyze the spectral properties of the emitted phonons on waterfall flows with $M_+ \approx 5$. In Section~\ref{corr}, we study the strength of the density correlations of the pairs of phonons on the same flows. We conclude in Section~\ref{concl}. In Appendix~\ref{App:NPGPE}, to take into account the 3-dimensional nature of the flow, we study the waterfall solutions of the non-polynomial Schr\"odinger equation. We show that 
the phonon spectrum hardly varies with respect to that obtained using the the Gross-Pitaevskii equation (GPE).  
In Appendix~\ref{App:NLstab} we report numerical results which indicate that local perturbations are expelled from the near-horizon region. 
Appendix~\ref{App:DLS} and Appendix~\ref{App:shift} are devoted to the study of various properties of the two-point function in dispersionless settings, namely the time-dependent modifications induced by the formation of the horizon in the former and the calculation of the above mentioned shift in asymmetrical flows in the latter. Finally, Appendix~\ref{App:phase} focuses on the phase of individual scattering coefficients.  

\section{Spontaneous emission of phonons in transonic flows}
\label{sp}

\subsection{Parametrization of the background flows}

To describe the background flows, we consider a one-dimensional, dilute, weakly interacting atomic Bose-Einstein condensate (BEC) with repulsive interactions~\cite{pitaevskii2003bose}. In the mean field approximation, the condensed atoms are described by a complex field $\psi(t,x)$ which satisfies the Gross-Pitaevskii equation (GPE):
\begin{equation}\label{eq:GPE}
i \hbar \pd_t \psi = - \frac{\hbar^2}{2 m} \pd_x^2 \psi + V(x) \psi + g \psi^* \psi^2. 
\end{equation}
Here $V$ is the external potential and $g$ the effective 1-dimensional two-body coupling, see Appendix~\ref{App:NPGPE}. 
We assume that $g$ is a constant and that $V$ only contains the sharp potential drop engendering the sonic horizon. 
That is, we neglect the gradients of the longitudinal shallow harmonic potential used in the experiment~\cite{Steinhauer:2015saa}. 
In this approximation, $V$ only depends on $x$ in the frame at rest with respect to the sharp potential. Relaxing this approximation, one would obtain a time-dependent inhomogeneous system, rather similar to that numerically studied in~\cite{Tettamanti:2016ntx,Wang:2016jaj,Steinhauer:2016hfa}.

To reduce the number of parameters, it is useful to define the non-dimensional quantities $\bar{x} \equiv x / X, \; \bar{t} \equiv t/T, \; \bar{\psi} \lp \bar{x}, \bar{t} \rp \equiv \sqrt{X} \psi(x,t), \; \bar{V} \lp \bar{x} \rp \equiv \lp m X^2 / \hbar^2 \rp V(x), \; \text{and} \; \bar{g} \equiv \lp m X / \hbar^2 \rp g$, where $T = m X^2 / \hbar$. In the following we will only work with these dimensionless quantities. In this system, the healing length $\xi \equiv  \hbar / \sqrt{m g \rho}$ (where $\rho = \psi^* \psi $ is the mean atomic density) becomes $\bar{\xi} = 1/ \sqrt{\bar{g} \bar{\rho}}$. As there is no ambiguity, from now on, we shall remove the bars to avoid cumbersome notations. The GPE then becomes
\begin{equation}\label{eq:ndGPE}
i \pd_t \psi = -\frac{1}{2} \pd_x^2 \psi + V(x) \psi + g \psi^* \psi^2.
\end{equation}
We look for stationary solutions of the form
\begin{equation}\label{eq:ansatz_statsols}
\psi(x,t) = \sqrt{\rho(x)} \exp \lp i \int_0^x v(y) dy \rp,
\end{equation}
where $\rho$ and $v$ are two real-valued functions. (Note that a non-zero frequency can be absorbed by adding a constant to $V$.) Plugging this ansatz into Eq.~(\ref{eq:ndGPE}) and taking the imaginary part gives the conservation of the current $\pd_x \lp \rho v \rp = 0$. 
Setting the scale $X$ so that $\rho v = 1$, the real part of Eq.~(\ref{eq:ndGPE}) becomes
\begin{equation}\label{eq:sec_rho}
\frac{1}{2} \pd_x^2 \lp \rho^{1/2} \rp = V \rho^{1/2} + g \rho^{3/2} + \frac{1}{2 \rho^{3/2}}.
\end{equation}
As a simple model of the sharp variation of the potential used in~\cite{Steinhauer:2015saa}, we consider a step-like potential of the form 
\begin{equation} \label{eq:Vstep}
V(x) = \left\lbrace
\begin{array}{cc}
V_- & x<0 \\
V_+ & x>0
\end{array}
\right.,
\end{equation}
where $\lp V_+, V_- \rp \in \mathbb{R}^2$. In what follows, the subscript $\pm$ denotes the sign of $x$. In general, integrating Eq.~(\ref{eq:sec_rho}) over $x$ in a region of homogeneous potential gives the square of $\pd_x \rho$ as a polynomial of degree 3. In our case we have two polynomials: one defined on the positive half-line and one on the negative half-line. We now focus on ``waterfall'' solutions~\cite{Larre:2011mq}.
Since these solutions are asymptotically uniform on both sides, the two integration constants must be chosen so that each polynomial has a double root: 
\begin{equation}
\lp \pd_x \rho \rp^2 = 4 g \lp \rho(x) - \rho_{1,\pm} \rp^2 \lp \rho(x) - \rho_{2,\pm} \rp, 
\label{droot}
\end{equation}
where the constants $\rho_{1,\pm}$, $\rho_{2,\pm}$ obey $2 \rho_{1,\pm} + \rho_{2,\pm} = -2 V_\pm / g$ and $\rho_{1,\pm}^2 \rho_{2,\pm} = 1 / g$. 
$\rho_{1,\pm}$ is the asymptotic value of $\rho$ at $x \to \pm \infty$, while $\rho_{2,-}$ is the density at the bottom of the (unique) stationary soliton solution in the left region. A waterfall solution is obtained by matching a uniform configuration $\rho = \rho_{1,+}$ for $x>0$ with a half-soliton in the region $x<0$. This solution thus requires $\rho_{2,-} = \rho_{1,+}$.
As a result, the solution is fully determined by the asymptotic densities $\rho_{1,-}$ and $\rho_{1,+}$. 

At this point, it is useful to notice that the GPE is invariant under the rescaling 
\begin{equation}
\label{lresc}
\psi \to \lambda \psi, \quad x \to \lambda^2 x, \quad t \to \lambda^4 t, \quad V \to \lambda^{-4} V, \quad g \to \lambda^{-6} g ,
\end{equation}
which preserves the condition $J = 1$. 
In the following, unless explicitly stated, the numerical values we will give (explicitly or in plots) involve only quantities invariant under this rescaling. This allows us to work with a one-dimensional set of waterfall solutions, which can be parametrized by the ratio $\rho_{1,-}/\rho_{1,+}$. 

Introducing the Mach number $M(x) \equiv v(x)/c(x)= 1/(g \rho^3(x))^{1/2}$, one gets $M_+ \equiv \mathop{\rm lim}_{x \to \infty} M(x) = \rho_{1,-}/\rho_{1,+}$ and $M_- \equiv \mathop{\rm lim}_{x \to -\infty} M(x)  = M_+^{-1/2}$. 
Since the waterfall solution is supersonic for $x \to +\infty$, this imposes $\rho_{1,-}/\rho_{1,+} > 1$, i.e., $V_- < - 1.5 g \rho_{1,+}$. The other quantities can be expressed in terms of $\rho_{1,-}$ and $\rho_{1,+}$, namely, $g = \rho_{1,+}^{-1} \rho_{1,-}^{-2}$, $\rho_{2,+} =\rho_{1,+} M_+^{2}$, $V_+ = -g \lp 2 \rho_{1,+}+\rho_{2,+} \rp / 2$, $V_- = -g \lp 2 \rho_{1,-} + \rho_{2,-} \rp / 2$. 

\begin{figure}  
\begin{center}
\includegraphics[width=0.4 \linewidth]{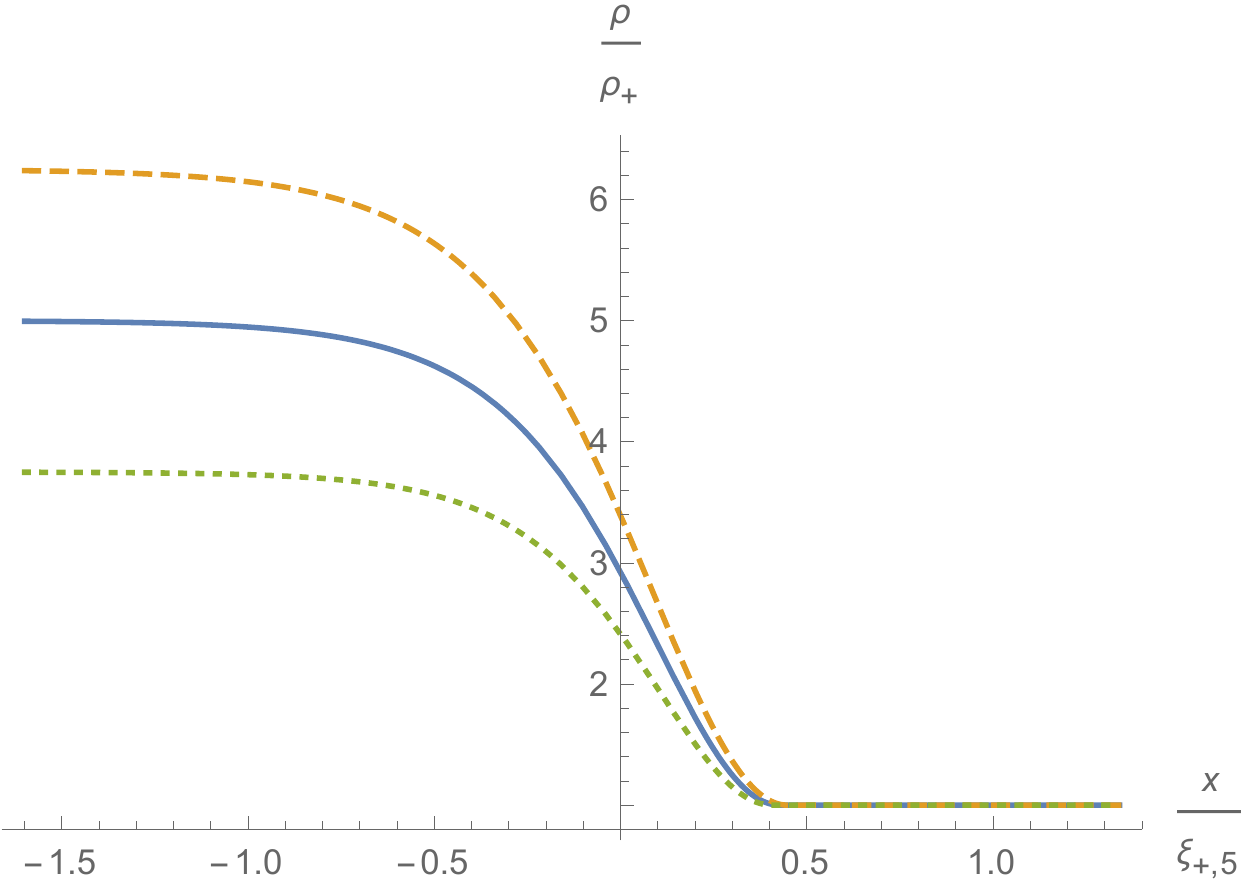} 
\includegraphics[width=0.4 \linewidth]{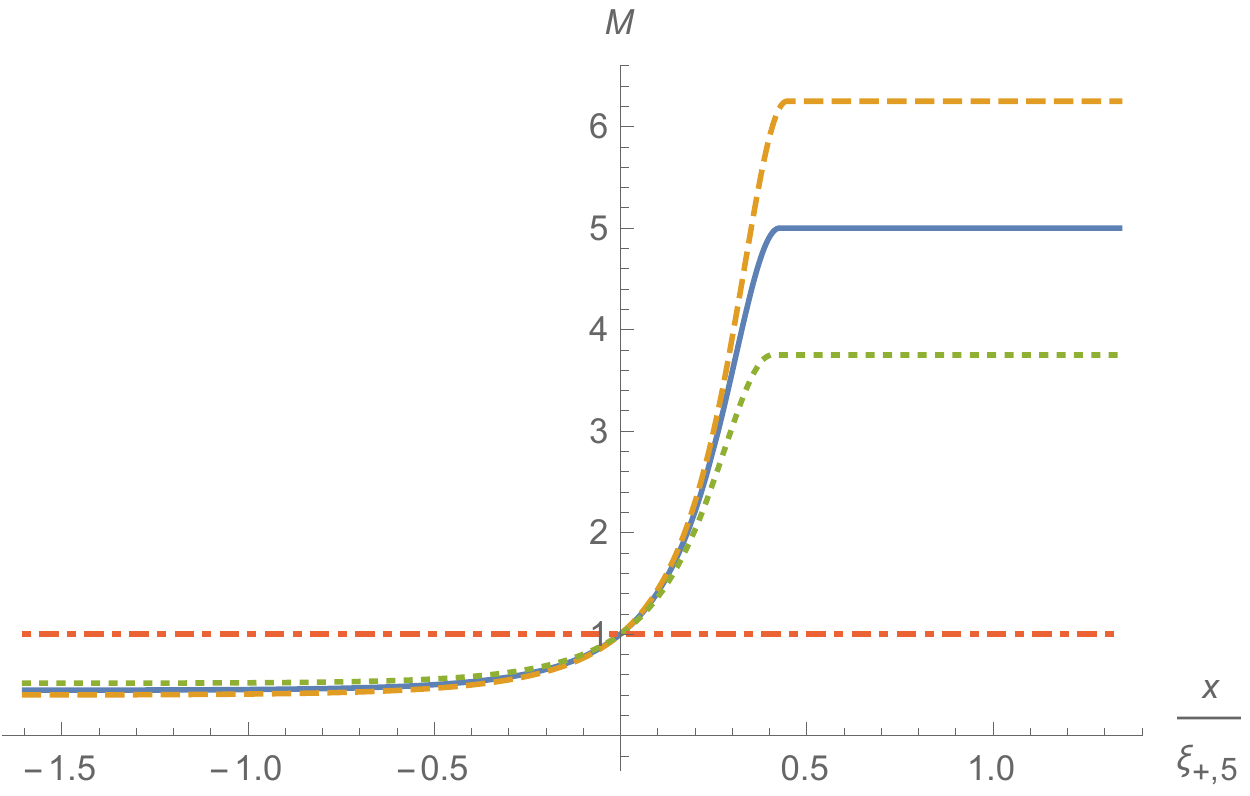} 
\includegraphics[width=0.4 \linewidth]{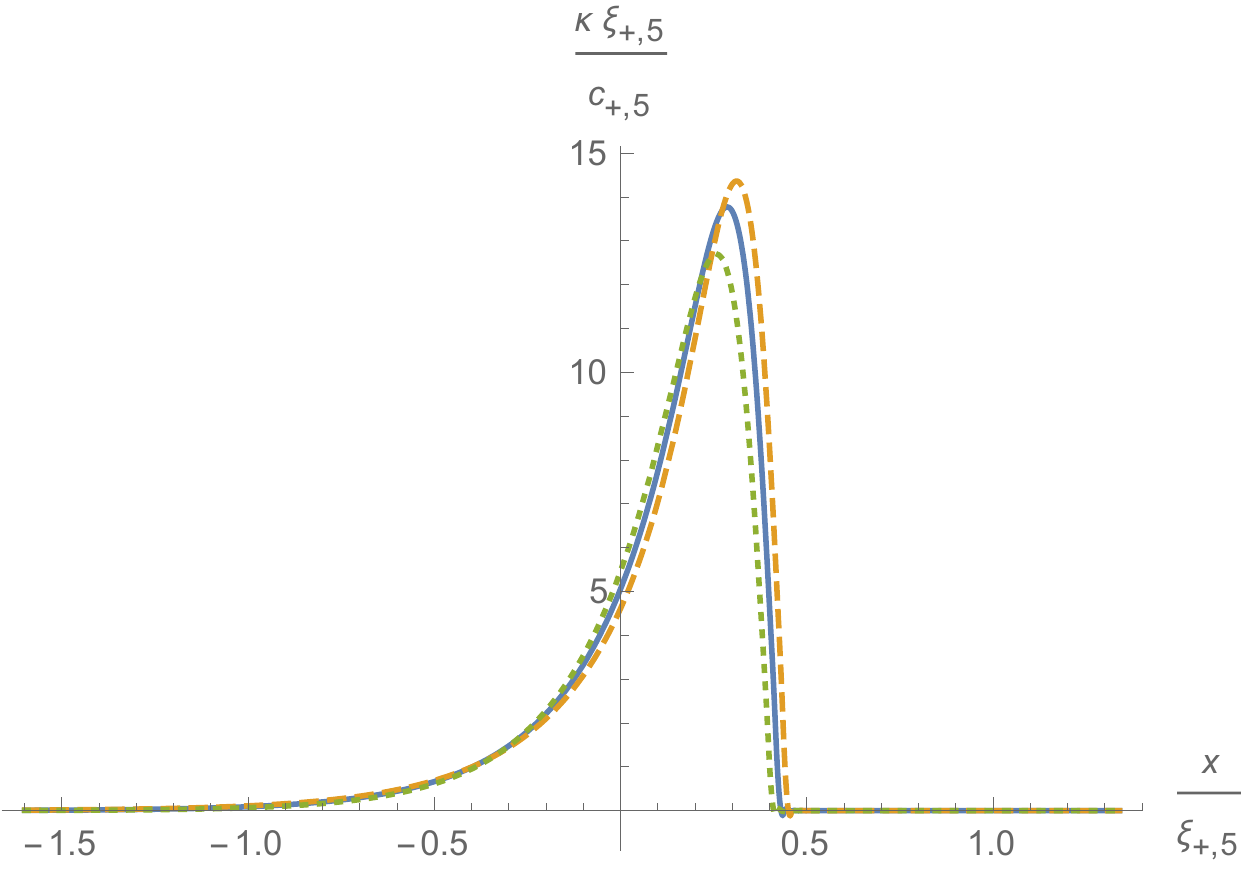} 
\includegraphics[width=0.4 \linewidth]{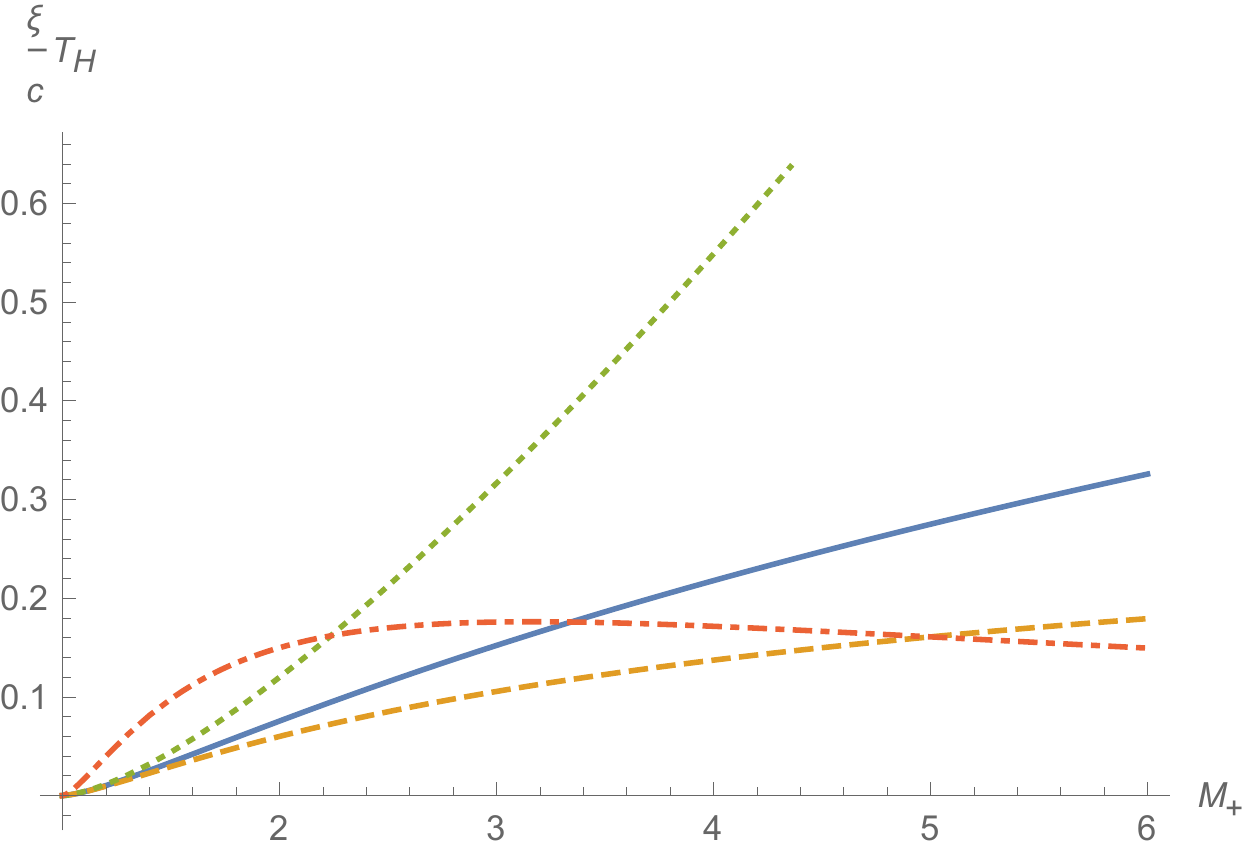} 
\end{center}
\caption{Plots of the rescaled atomic density $\rho/\rho_+$ (top, left) and the Mach number (top, right) for three waterfall solutions similar to that realized in~\cite{Steinhauer:2015saa}. The flow is from left to right and the subsonic region is on the left side. The asymptotic values of the Mach numbers for the green (dotted), blue (continuous), and orange (dashed) curves are, respectively, $M_- \approx 0.52, 0.45, 0.4$ and $M_+ = 3.75, 5, 6.25$. The unit of the horizontal axis is $\xi_{+,5}$, the healing length in the supersonic region for the flow with the central value $M_+ = 5$ (blue curves). The plots are shifted so that the sonic horizon $M=1$ (marked by an horizontal red dashed line the upper right panel) is at $x= 0$. The bottom left plot shows the profile of the adimensionalized gradient $\kappa(x) \xi_{+,5} / c_{+,5}$, where $c_{+,5}$ is the asymptotic downstream sound velocity for the flow with $M_+ = 5$. One clearly sees that $\kappa(x)$ identically vanishes on the right of the potential barrier located near $x/\xi_{+,5} = 0.4$. When evaluated at $x = 0$, $\kappa$ gives the surface gravity $\kappa_H$ of \eq{eq:surgra}. As functions of $M_+$, the bottom right plot shows the Hawking temperature $T_H = \kappa_H/2\pi$ adimensionalized by $\xi_H/c_H$ (blue, continuous), $\xi_+/c_+$ (green, dotted), $\xi_-/c_-$ (orange, dashed), and $\xi_{-,5}/c_{-,5}$ (red, dot-dashed), for $\rho_+ = 1$.
The first three values increase with  $M_+$, unlike the last one which is not monotonic.} 
\label{fig:bckgd}
\end{figure}

In brief, when working with a uniform $g$ and a step-like $V$, up to an overall scale fixed by $\lambda$, there is a one-dimensional series of inequivalent solutions parametrized by $M_+ >1$.~\footnote{In physical terms, for a given type of atoms and when assuming that $g$ does not vary with $x$, two independent parameters also characterize these waterfall solutions. These can be taken to be the density $\rho_{1,-}$ (which is determined by the shallow longitudinal harmonic potential and the total number of atoms) and the depth of the potential $V_+ - V_-$. Then, requiring that the solution be stationary and asymptotically homogeneous on both sides fixes the value of the current $J$, and thus the relative velocity of the sharp potential drop with respect to the harmonic potential 
used in~\cite{Steinhauer:2015saa}.}
Explicitly, these solutions read
\begin{equation}\label{eq:waterfall}
\rho(x)/\rho_{1,+} = \left\lbrace
\begin{array}{cc}
M_+ + \lp 1 - M_+  \rp \lp \cosh \lp \sigma x \rp \rp^{-2} & x \leq 0 \\
1 & x \geq 0
\end{array}
\right.,
\end{equation}
where $\sigma = \sqrt{M_+ - 1} / \xi_+$. In the upper plots of \fig{fig:bckgd} we show the (non-dimensional) density profile and Mach number for three nearby flows. The central blue curve corresponds to $M_+ = 5$, 
close to its value in the experiment~\cite{Steinhauer:2015saa}. The two others are obtained with relative changes of $25 \%$, i.e., $M_+ = 6.25$ (orange) and $3.75$ (green). We shall use these flows to illustrate the typical behavior of the scattering coefficients and their sensitivity to $M_+$. It should be also noticed that these flows are stable, and act as attractors in that localized perturbations propagate outwards from the sonic horizon leaving the solution intact, see Appendix~\ref{App:NLstab}. 
Their stability and the smallness of the non-polynomial parameter (discussed in Appendix~\ref{App:NPGPE}) probably explain why the flow observed in Ref.~\cite{Steinhauer:2015saa} has a profile in rather good agreement with that of \eq{eq:waterfall}, see Fig.~1b in~\cite{Steinhauer:2015saa}.~\footnote{\label{ftn1} The upstream and downstream values of the condensate velocity $v$ and sound speed $c$ reported in~\cite{Steinhauer:2015saa} seem to be incompatible with the conservation of the atom flux. Indeed, the product $c^2 v$ is constant for any stationary solution of the one-dimensional Gross-Pitaevskii equation, while it varies by $\sim 20 \%$ when using the reported values. When taking into account the three-dimensional character of the flow, $c^2 v$ can vary, see Appendix~\ref{App:NPGPE}. 
However, in the experimental conditions, the modification associated with this refined description is at most of the order of $10\%$, which is too small to explain the discrepancy. As a result the value of $M_+$ obtained by using the downstream values of $v$ and $c$ differs from that obtained with the upstream values and the hypothesis that the flow is described by a waterfall solution.} 

In the lower left plot of \fig{fig:bckgd} we represent the gradient $\kappa(x) = \partial_x (v - c)$ for the following reason. If the analogy with gravity is accurate~\cite{Unruh:1980cg,Unruh:1994je,Macher:2009nz}, the spectrum of phonons emitted from the sonic horizon should closely follow the Planck law with an effective temperature given (in units where the Boltzmann and Planck constants are equal to 1) by $T_H =\kappa_H/2\pi$, where $\kappa_H \equiv \kappa(x_H)$ is the analog surface gravity and $x_H$ gives the location of the sonic horizon where $M(x) = 1$. In the waterfall flows, it is given by~\cite{Larre:2011mq}:  
\begin{equation}\label{eq:surgra}
\kappa_H =  3 \frac{c_+}{\xi_+} M_+^{1/3} \lp M_+^{1/3} - 1 \rp^{3/2} \lp M_+^{1/3} + 1 \rp^{1/2}. 
\end{equation} 
It can be seen in the figure that $\kappa_H$ is only $\approx 35 \%$ of the maximal value of $\kappa(x)$. This is in sharp contrast with the  symmetrical flows considered in~\cite{Macher:2009nz}. It implies that the deviations from the Planck spectrum will be larger than in symmetrical flows with the same $\kappa_H$~\cite{Finazzi:2011jd}. 

It should also be noticed that $\kappa_H \approx 5.1 \, c_{+,5}/\xi_{+,5}$, where $c_{+,5}$ and $\xi_{+,5}$ are the asymptotic downstream sound velocity and healing length for the flow with $M_+ = 5$. By comparison, the dispersive frequency evaluated at the horizon for the same flow is $c(x_H)/ \xi(x_H) \approx 2.9 \, c_{+,5}/\xi_{+,5}$. Hence $\kappa_H \xi_H / c_H \approx 1.7$. Since this ratio is larger than unity, one could a priori expect that the relativistic expressions will {\it not} provide an accurate description of the emission spectrum. However, we shall see that this is not the case. The validity of the relativistic expressions 
comes from the fact that the flows we consider are deeply supersonic since $M_+ \sim 5$, see below and~\cite{Finazzi:2012iu}. 

We finally notice that the values of $\kappa_H \xi_{+,5} /c_{+,5}$ for the three represented flows characterized by $M_+ = 3.75$, $5$, and $6.25$ are $5.5$, $5.1$, and $4.6$, respectively. Contrary to what could be expected, $\kappa_H$ is larger for the flows with smaller $M_+$. To further study the variations of $T_H$ with $M_+$ in the unit of various dispersive scales, on the lower right plot of \fig{fig:bckgd}, we represent $T_H$ multiplied by $\xi_H /c_H$, $\xi_+ /c_+$, $\xi_-  /c_-$, and $\xi_{-,5} /c_{-,5}$. 
The first three products go to zero like $\lp M_+ - 1 \rp^{3/2} / \lp \sqrt{6} \pi \rp$ when $M_+ \to 1$. 
When $M_+ \to \infty$, they behave differently: $\xi_H T_H /c_H \sim 3 M_+^{1/3} / (2 \pi)$, $\xi_- T_H /c_- \sim 3 / (2 \pi)$, and $\xi_+ T_H /c_+ \sim 3 M_+ / (2 \pi)$. 
Interestingly, when divided by a fixed frequency, e.g., $c_{-,5}/\xi_{-,5}$, $T_H$ is a {\it non-monotonic} function of $M_+$, the maximum being reached for $M_+ = 3.17$. It behaves as $T_H \xi_{-,5}/c_{-,5} \mathop{\sim} \sqrt{2/3} (M_+ - 1)^{3/2}$ for $M_+ \to 1$ and $T_H \xi_{-,5} /c_{-,5} \mathop{\sim} 3 M_+^{-1}$ for $M_+ \to \infty$.

\subsection{Spontaneous emission of phonons, generalities}

To describe the propagation of linear density fluctuations in the above flows, we use quantum mechanical settings~\cite{pitaevskii2003bose}. 
We follow~\cite{Macher:2009nz} where more details can be found. 
It is convenient to write the atomic field operator as 
\begin{equation}
\hat \psi(x,t) = \psi_0(x,t) \lp 1 + \hat \phi(x,t) \rp,
\label{phi}
\end{equation}
where $\psi_0$ is a known stationary solution of \eq{eq:ndGPE} with mean density $\rho(x)$ and velocity $v(x)$. To first order in $\hat \phi$, one obtains the Bogoliubov-de Gennes (BdG) equation, which here reads:
\begin{equation}\label{eq:BdG}
i \lp \pd_t + v(x) \pd_x \rp \hat \phi = - \dfrac{1}{2 \rho(x)} \pd_x \left[ \rho(x) \pd_x 
\hat \phi\right] + g \rho(x) \lp \hat \phi + \hat \phi^\dagger \rp.
\end{equation}
Since the background flow is stationary, we look for stationary solutions of the form
\be
\hat \phi_\om(t,x) = e^{- i \om t} \phi_\om(x) \, \hat a_\om +  \lp e^{-i \om t} \varphi_\om(x) \, \hat a_\om \rp^\dagger.
\label{phivar}
\ee
The operators  $\hat a_\om$ and $ \hat a_\om^\dagger$ destroy and create a phonon of frequency $\om$, and obey the usual bosonic commutation relations. This particular form of the decomposition of the field operator follows from the antilinear term in \eq{eq:BdG}. It can be easily shown that the stationary c-number mode doublet  $\left( \phi_\om(x), \varphi_\om(x) \right)$ obeys 
\begin{eqnarray}\label{eq:BdG_W} 
\lp \lp \om + i v \pd_x \rp - \dfrac{1}{2 \rho} \pd_x \rho \pd_x - c^2 \rp \phi_\om & =&  c^2 \varphi_\om, \nonumber \\
- \lp \lp \om + i v \pd_x \rp + \dfrac{1}{2 \rho} \pd_x \rho \pd_x + c^2 \rp \varphi_\om & =&  c^2 \phi_\om. 
\end{eqnarray} 
Introducing the notation $W_1= (\phi_1,\varphi_1)$ and $W_2= (\phi_2,\varphi_2)$, the inner product reads
\begin{equation} \label{eq:scalp}
\lp W_1 \vert W_2 \rp\equiv \int_{-\infty}^{+\infty}dx\,  \rho(x) \left(  \phi_1^* \phi_2 -\varphi_1^* \varphi_2  \right).
\end{equation}
One verifies that it is conserved in time for any pair of solutions of \eq{eq:BdG}.
We call $(W_1 \vert W_1)$ the norm of the solution represented by the doublet $W_1$. 
\begin{figure}
\begin{center}
\includegraphics[width=0.49 \linewidth]{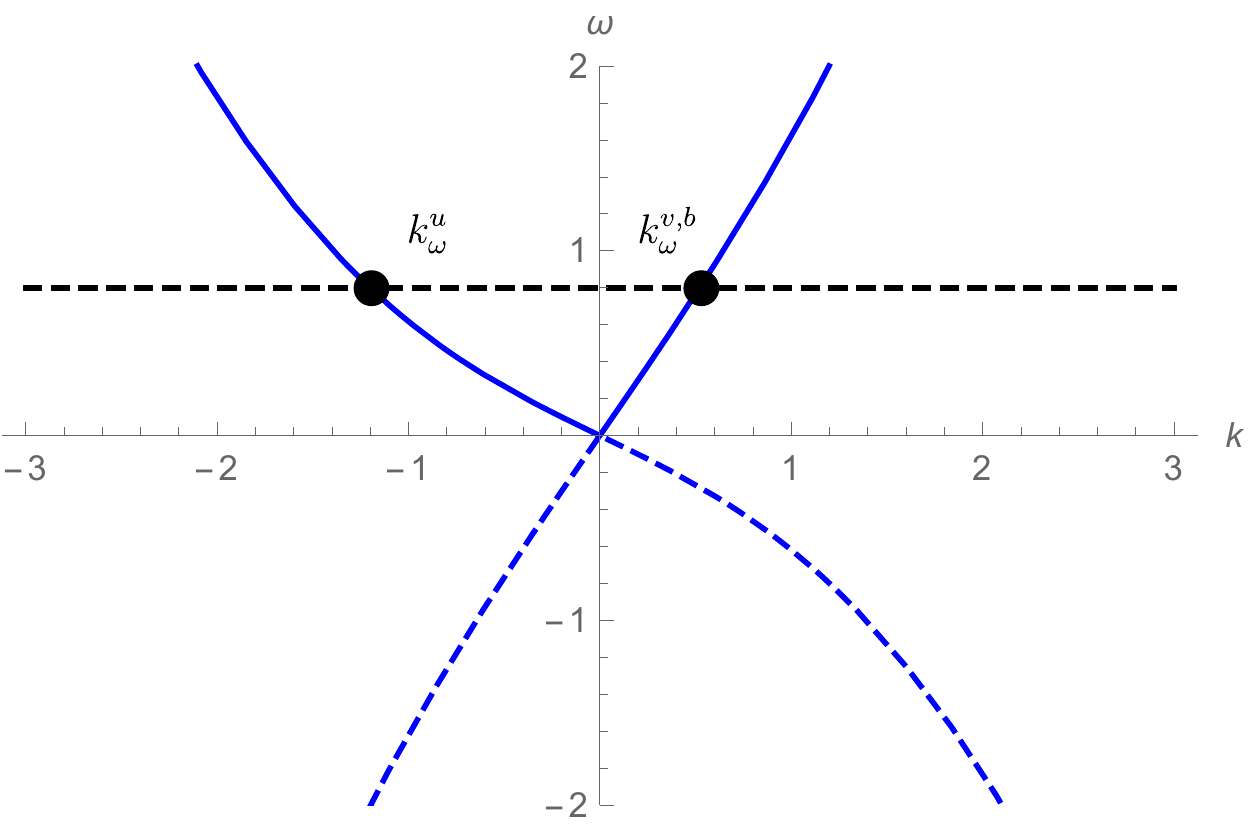}
\includegraphics[width=0.49 \linewidth]{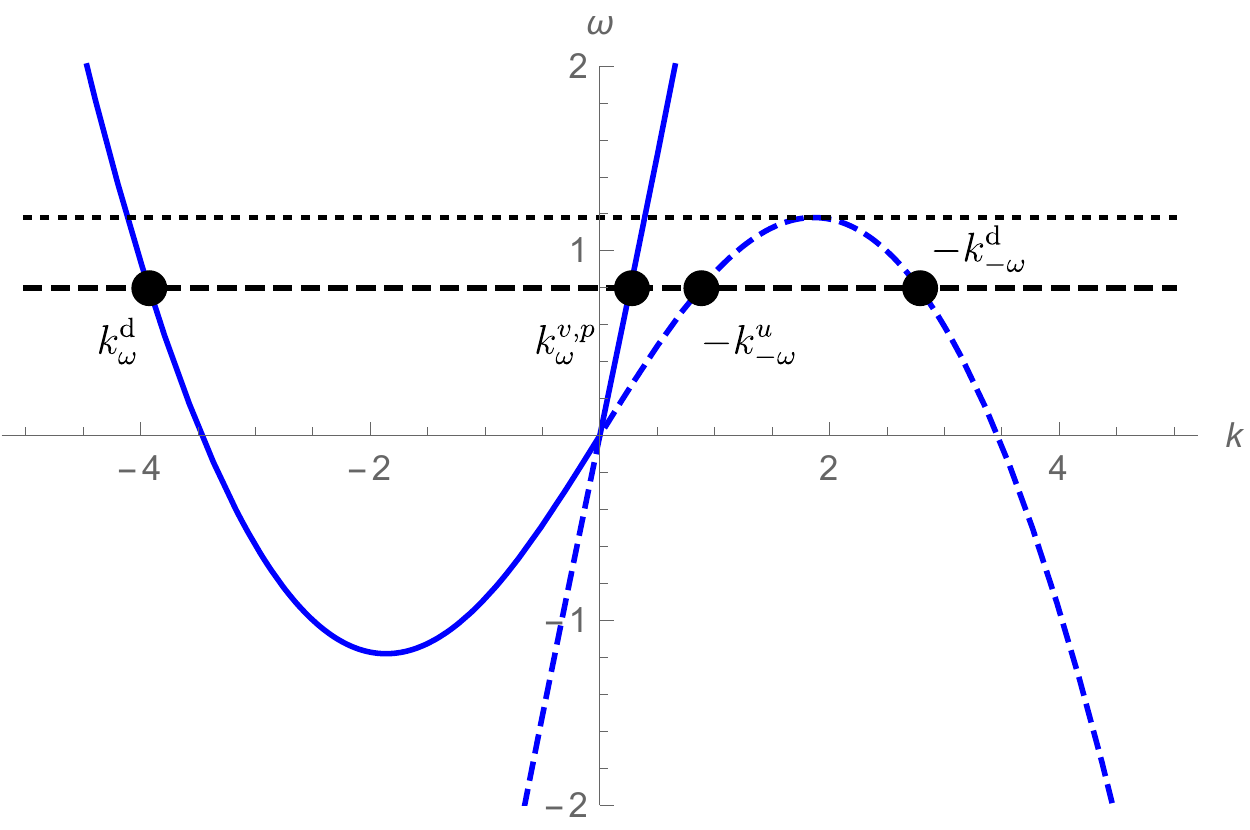}
\end{center}
\caption{Plot of the dispersion relation in homogeneous subsonic (left) and supersonic (right) flows. The horizontal, dashed, black line corresponds to a fixed frequency. The continuous curves correspond to positive values of $\Omega = \om - vk$ and the dashed ones to negative values of $\Omega$. One notices that the two extra roots on the right of the right plot have $\Omega < 0$. As explained in the text, they describe phonons carrying a negative energy in the frame where the potential $V$ is stationary. These two extra roots become complex when $\om$ reaches the critical frequency $\om_{\rm max}$ of \eq{ommax}, which is here indicated by a dotted line. 
}\label{fig:DR}
\end{figure}

In the asymptotic regions where $v$ and $\rho$ are uniform, any solution of \eq{eq:BdG_W} can be written as a superposition of plane wave doublets $W_{\om,k}(t,x) = (U_{k},V_{k}) \exp \lp -i \om t +i k x \rp$, where $\om$ and $k$ are related by the dispersion relation
\begin{equation}\label{eq:DR}
\Omega^2 = (\om - v k)^2 = g \rho k^2 + \frac{k^4}{4}.
\end{equation}
Here $\Omega \equiv \om - v k$ is the angular frequency in the rest frame of the condensed atoms.  
This dispersion relation is represented graphically in \fig{fig:DR}. In the following we consider only 
the case $\om > 0$. 
Then, for $\Omega > 0 $, $U_{k}$ and $V_{k}$ are related by $V_{k} = D(k,\rho) U_{k}$, where 
\begin{equation}\label{eq:Dk}
D(k,\rho) = \frac{\sqrt{g \rho k^2 + {k^4}/{4}} }{g \rho} - \lp \frac{k^2}{2 g \rho} + 1\rp . 
\end{equation}
When $U_{k}$ and $V_{k}$ satisfy the usual relation $|U_{k}|^2 - |V_{k}|^2 = 1$, the doublets $W_{\om,k}$ obey
\begin{equation}\label{norm}
\lp W_{\om,k} \vert W_{\om',k'} \rp = 2 \pi \rho \,  \delta(k - k').
\end{equation}
For $\Omega < 0$, the solutions of \eq{eq:BdG_W} are doublets $\bar W_{\om,k}$ obtained by exchanging the two components of those with $\Omega > 0$ and taking their complex conjugate. When working with $i \partial_t = \om > 0$, these doublets are thus given by $\bar W_{-\om,-k}(t,x) \equiv (V_{k}^*,U_{k}^*) \exp \lp - i \om t  + i k x \rp$. 
Importantly, they have a negative norm: $\lp \bar W_{\om,k} \vert \bar W_{\om',k'} \rp = - \lp W_{\om,k} \vert W_{\om',k'} \rp$. The phonons described by $W_{-\om,-k}$ carry a negative energy equal to $- \om$ (in units where $\hbar = 1)$. 

We now study separately subsonic and supersonic flows to identify the number of independent solutions. In a subsonic flow, i.e., $0 < v < c$, there are two real roots in $k$ for $\om >0$: 
\begin{itemize} 
\item $k_\om^u$ is counter-propagating (its group velocity is negative in the rest frame of the condensate) and left-moving in the frame of $V(x)$ of \eq{eq:GPE}; 
\item $k_\om^{v, b}$ is co-propagating and right-moving.
\end{itemize}
The corresponding modes are described by positive-norm doublets $W_{\om,k}$. There are also two complex roots with equal and opposite imaginary parts. A superscript $b$ has been added to the co-propagating root in order to distinguish it from the root $k_\om^{v, p}$ found in a supersonic flow. 
 
In a supersonic flow parametrized by $M_+ > 1$, there is a critical frequency given by~\cite{Macher:2009nz} 
\begin{equation} 
\om_{\rm max} = \frac{c_+}{\xi_+}  
\sqrt{M_+ + \sqrt{\rule{0mm}{2.ex} M_+^2+8}} 
\lp \frac{2 (M_+^2 - 1) 
}{3 M_+ + \sqrt{M_+^2 + 8}} \rp^{3/2}\, .
\label{ommax}
\end{equation}
When $\om$ crosses $\om_{\rm max}$ by increasing values, the two largest roots merge and become complex. For $0 < \om < \om_{\rm max}$, the four roots $k_\om$ are real. From left to right in the right panel of \fig{fig:DR}, they are
\begin{itemize}
\item $k_\om^d$ is counter-propagating and left-moving; 
\item $k_\om^{v, p}$ is co-propagating and right-moving;
\item $- k_{-\om}^u$ is counter-propagating and right-moving;
\item $-k_{-\om}^d$ is counter-propagating and left-moving.
\end{itemize}
The superscript ``$d$'' on the first and last roots indicates that they are dispersive, i.e., that they do not vanish in the limit $\om \to 0$. The minus signs in front of the last two roots indicate that the corresponding modes are negative-norm doublets $\bar W_{-\om,-k}(t,x)$ describing negative-energy phonons. These two roots merge when $\om$ reaches $\om_{\rm max}$. 

\begin{figure}
\begin{center}
\includegraphics[width=0.5\linewidth]{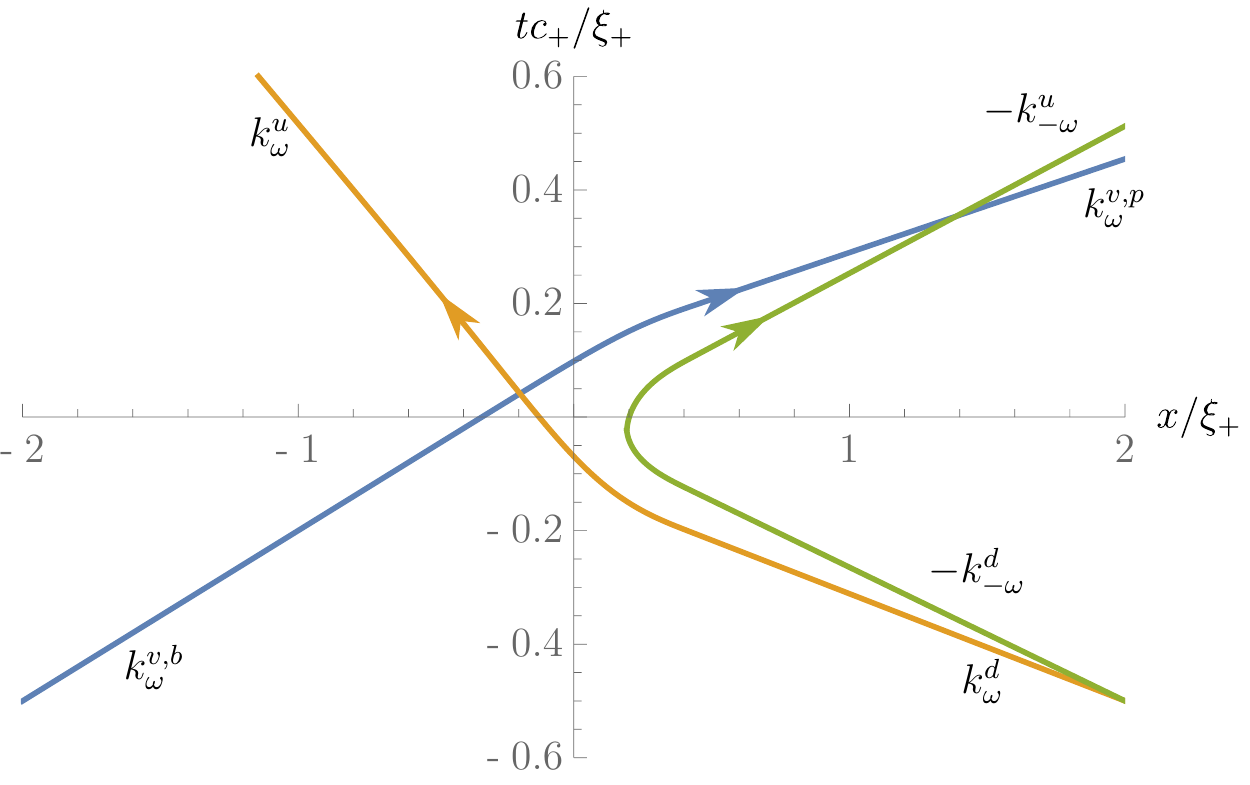}
\end{center}
\caption{Space-time diagram of the characteristics associated with the three types of stationary modes propagating in the transonic waterfall flow (from left to right) of \fig{fig:bckgd} with $M_+ = 5$. 
The sonic horizon is located at $x = 0$, and the characteristics are computed for $\om = 2.5\,  c_+ / \xi_+$. 
As explained in~\cite{Brout:1995wp,Jacobson:2007jx}, they obey \eq{eq:DR} treated as a Hamilton-Jacobi equation. 
The initial (final) asymptotic values of their wave vectors are given by the corresponding roots, indicated in the Figure at early (late) time. The arrows give the orientation of the group velocity in the rest frame of the potential $V(x)$.
The dispersive roots $k_\om^d, -k_{-\om}^d$ and the co-propagating one $k_\om^{v, b}$ characterize the three incoming modes.  
}\label{fig:charact}
\end{figure} 
When considering transonic stationary flows which interpolate from a subsonic to a 
supersonic region, these asymptotic modes will be mixed by the scattering on the region where $\rho$ and $v$ depend on $x$. 
Then three globally defined and linearly independent doublets are found for $\om < \om_{\rm max}$. 
Three of the above roots, namely $k_\om^d, -k_{-\om}^d$, and $k_\om^{v, b}$, characterize the 3 incoming modes, each of which containing asymptotically only one wave with a group velocity oriented towards the horizon, see \fig{fig:charact}. 
The three other roots characterize the 3 outgoing modes, which each contains only one asymptotic wave with a group velocity oriented away from the horizon. Following~\cite{Macher:2009nz}, we write the $3 \times 3$ matrix relating these two mode bases as
\begin{equation}\label{eq:scatt_coeff}
\begin{pmatrix}
\phi_\om^{d, {\rm in}} \\
\lp \varphi_{-\om}^{d, {\rm in}} \rp^*\\ 
\phi_\om^{v, {\rm in}}
\end{pmatrix} = 
\begin{pmatrix}
\alpha_\om & \beta_{-\om} & \tilde{A}_\om \\
\beta_\om^* & \alpha_{-\om}^* & \tilde{B}_\om^* \\
A_\om & B_\om & \alpha_\om^v
\end{pmatrix}
\begin{pmatrix}
\phi_\om^u \\
\lp \varphi_{-\om}^u \rp^*\\ 
\phi_\om^v
\end{pmatrix}
\end{equation}
where the superscripts on the modes have the same meaning as those of the wave vectors. To avoid any ambiguity, we labeled the $in$ modes by the superscript ``in''. For the $out$ modes instead, the superscript ``out'' is implicit. In each basis, the 3 globally-defined doublets $W^a_\om(x) = (\phi^a_\om(x), \varphi^a_\om(x))$ are orthogonal to each others and have a positive unit norm
\begin{equation} 
\lp W^a_{\om} \vert W^b_{\om'} \rp = \delta^{ab}\, \delta(\om - \om') .
\label{norm_om}
\end{equation}
This normalisation differs from that of \eq{norm} because we here exploit the stationarity of the flow (since the homogeneity is broken near the sonic horizon). Because of the negative energy phonons described by doublets of the form $\bar W_{-\om} = (\varphi_{-\om}^*,\phi_{-\om}^*)$ in \eq{eq:scatt_coeff}, the $3 \times 3$ matrix is an element of $U(1,2)$. As a result, for instance, the coefficients of the first line obey
\begin{equation} 
\label{unit}
|\alpha_\om|^2 - |\beta_{-\om}|^2 + |\tilde{A}_\om|^2 = 1.
\end{equation}
For more details about these relations, we refer to~\cite{Busch:2014bza}. The two sets of modes are orthonormal and complete. Using for instance the $out$ set, the Fourier component of the field operator with $\om > 0$ thus reads
\begin{equation}\label{hatphi}
\hat{\phi}_\om (x) = \hat{a}_\om^u \phi_\om^u (x) + \hat{a}_\om^v \phi_\om^v (x) + \hat{a}_{-\om}^{u \dagger} \lp \varphi_{-\om}^u(x) \rp^*.
\end{equation}
The three operators $\hat{a}_\om^u$, $\hat{a}_\om^v$, and $\hat{a}_{-\om}^u$ destroy respectively an outgoing phonon with wave vector $k_\om^u$, $k_\om^{v, p}$, and $k_{-\om}^u$.  When starting from the initial vacuum state, the mean numbers of outgoing phonons spontaneously emitted by the scattering on the flow are
\begin{equation}
n_\om^u = \abs{\beta_{\om}}^2, \, n_\om^v = |\tilde B_{\om}|^2.
\end{equation}
For negative-energy phonons, by energy conservation, we have $n_{-\om}^u =n_\om^u +  n_\om^v $.

\subsection{Spectral properties in waterfall background flows}

\begin{figure}
\begin{center}
\includegraphics[width=0.49 \linewidth]{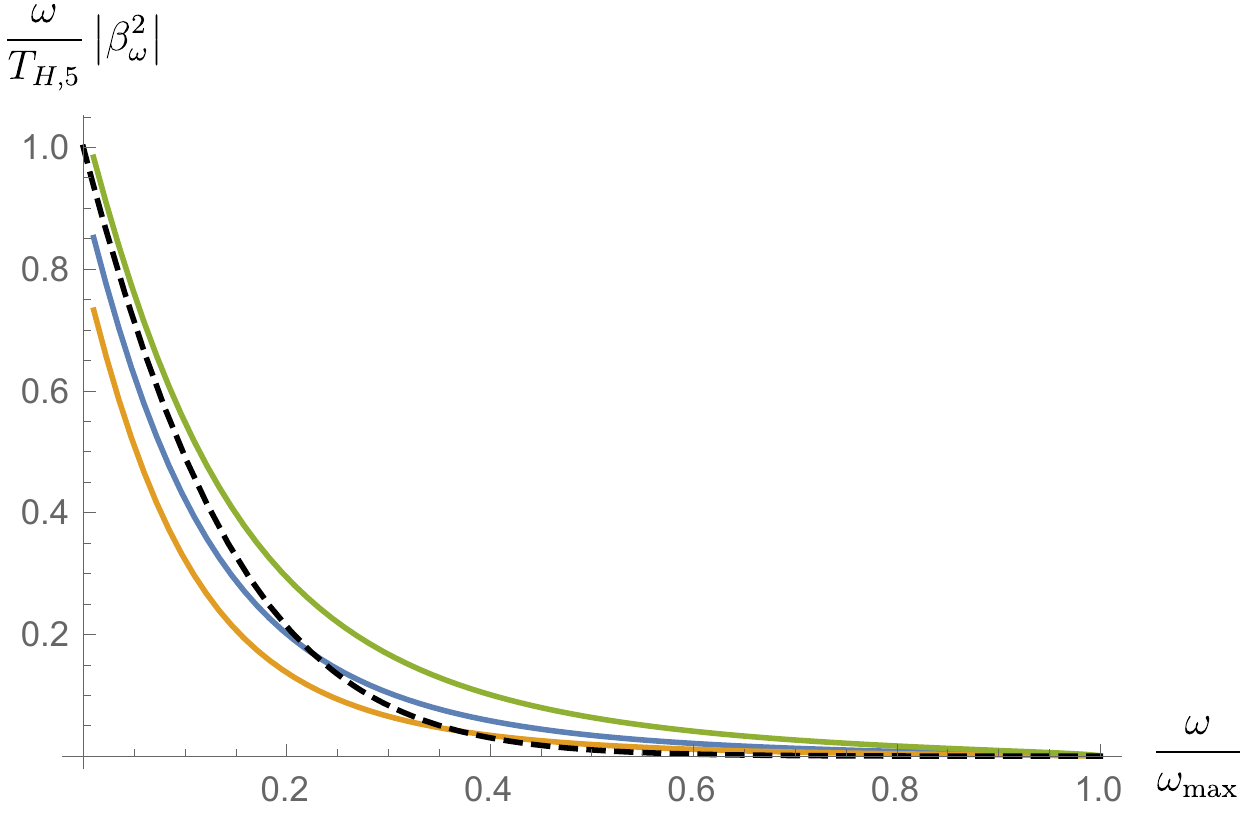}
\includegraphics[width=0.49 \linewidth]{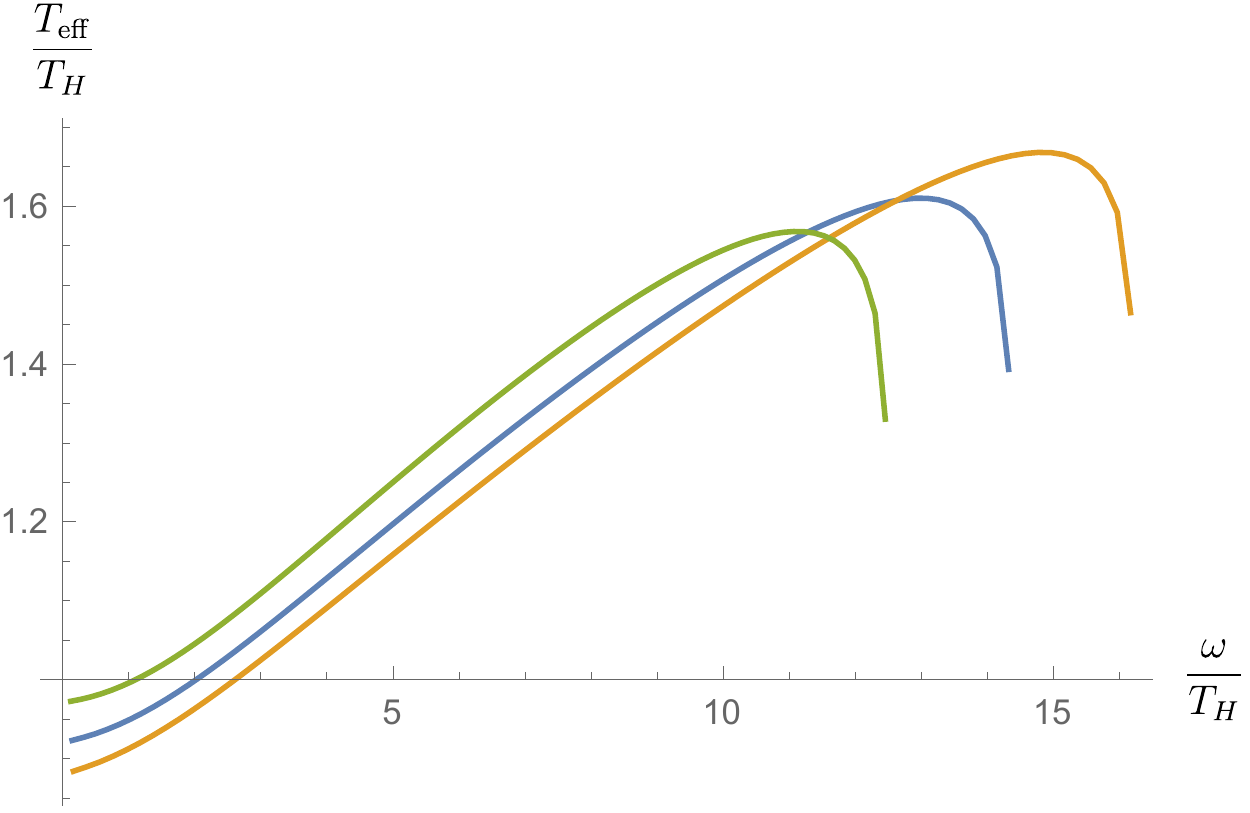}
\end{center}
\caption{On the left panel, we show the adimensionalized energy spectrum $\epsilon_\om/T_{H,\, 5}$ of the outgoing phonons spontaneously emitted by the scattering on the three flows of \fig{fig:bckgd}, where $T_{H,\, 5}  =\kappa_{H,\, 5}/2\pi$ is the Hawking temperature for the flow with the central value of $M_+ = 5$. The black dashed line shows the adimensionalized Planck energy spectrum evaluated for this flow. On the right panel, we show $T_{\rm eff}(\om)$ in units of the corresponding value of $T_H(M_+)$ for the same three flows. One clearly sees that $T_{\rm eff}(\om)$ becomes significantly larger than $T_H$, but this occurs in a domain where $\epsilon_\om$ is very small. $T_{\rm eff}(\om)$ abruptly drops to zero when $\om$ reaches $\om_{\rm max}$.} 
\label{fig:Teff}
\end{figure}
To obtain the scattering coefficients in the three waterfall solutions of \fig{fig:bckgd}, we numerically integrated \eq{eq:BdG_W} following a procedure similar to that of~\cite{Macher:2009nz}, here implemented in {\it Mathematica}~\cite{Mathematica}. 
We first consider the energy spectrum $\epsilon_\om = \om \, \abs{\beta_{\om}}^2$ of positive energy $u$-phonons. On the left plot of \fig{fig:Teff}, for the three flows of \fig{fig:bckgd}, as a function of $\om/\om_{\rm max}$ (where $\om_{\rm max}$ is the corresponding value of the critical frequency of \eq{ommax}), we represent $\epsilon_\om /T_{H,\, 5}$ where $T_{H,\, 5}$ is the Hawking temperature of the central flow with $M_+ = 5$. We see that the three energy spectra are quite similar. We also see that they closely follow the (adimensional) Planck spectrum $\epsilon^{T_{H,\, 5}}_\om /T_{H,\, 5}= (\om /T_{H,\, 5}) /(\exp \lp \om / T_{H,\, 5} \rp - 1)$ evaluated for the central flow with $M_+ = 5$. In fact, for this flow the maximum value of the difference $\abs{(\epsilon_\om - \epsilon^{T_{H,\, 5}}_\om)/T_{H,\, 5}}$ is less than $9\%$. We also see that $\epsilon_\om$ becomes larger than $\epsilon^{T_{H,\, 5}}_\om$ for $\om \gtrsim 0.2 \om_{\rm max} \approx 3 T_H$, something which indicates that $T_{\rm eff}(\om)$, the effective temperature of \eq{eq:Teff}, should grow with $\om$.

To study more closely the Planckianity of the spectrum, on the right plot of \fig{fig:Teff}, we represent $T_{\rm eff}(\om)/T_H$ for the same flows, where the effective temperature $T_{\rm eff}$ is defined by
\begin{equation}  \label{eq:Teff}
\abs{\beta_{\om}}^2 = \frac{1}{\exp \lp \om / T_{\rm eff}(\om) \rp - 1}.
\end{equation}
In the limit $\om \to 0$, for each of the three flows, $T_{\rm eff}$ goes to a value close to the corresponding $T_H$, with a difference of the order of $10 \%$. Moreover, the slope  $d T_{\rm eff} / d \om$ evaluated near $\om = T_H$ is smaller than $0.05$.
Yet, relative deviations become large when increasing $\om$ (reaching a maximum of $\sim 70\%$). 
But these occur only for large values of $\om / T_H$ where the energy spectrum is very small. For instance, when $T_{\rm eff}$ differs from $T_H$ by $20\%$, $\epsilon_\om/T_H$ is less than $8 \%$. Although $T_H \xi_H/c_H \approx 1.7$ as discussed above, the values of $T_H/\om_{\rm max}$ for the three flows we consider are $0.062$, $0.069$, and $0.080$. It is the smallness of this ratio which guarantees that the deviations from the Planck spectrum with a temperature $\kappa_H/2\pi$ are, in effect, so small~\cite{Macher:2009nz,Finazzi:2012iu}. 

\begin{figure}
\begin{center}
\includegraphics[width=0.49 \linewidth]{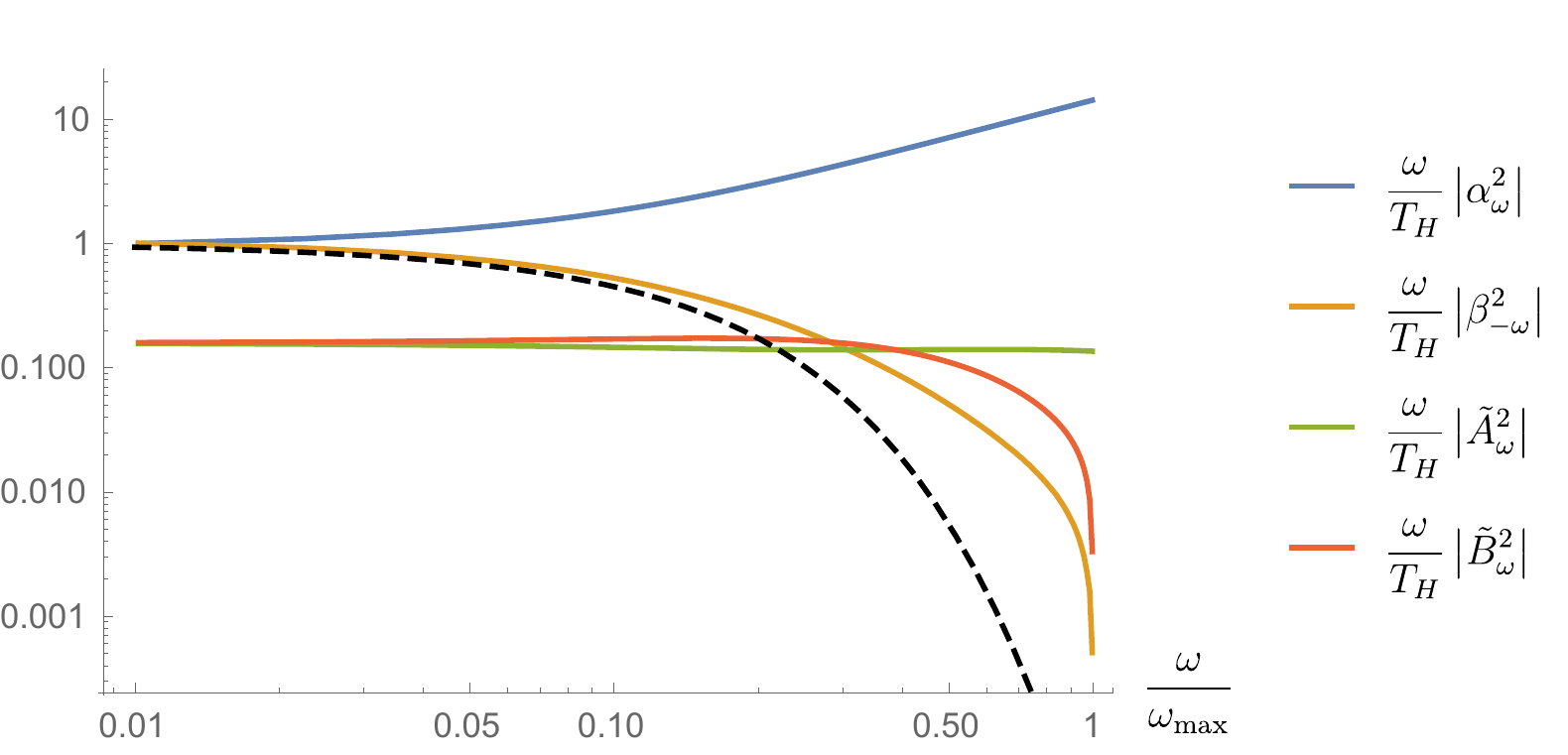}
\includegraphics[width=0.49 \linewidth]{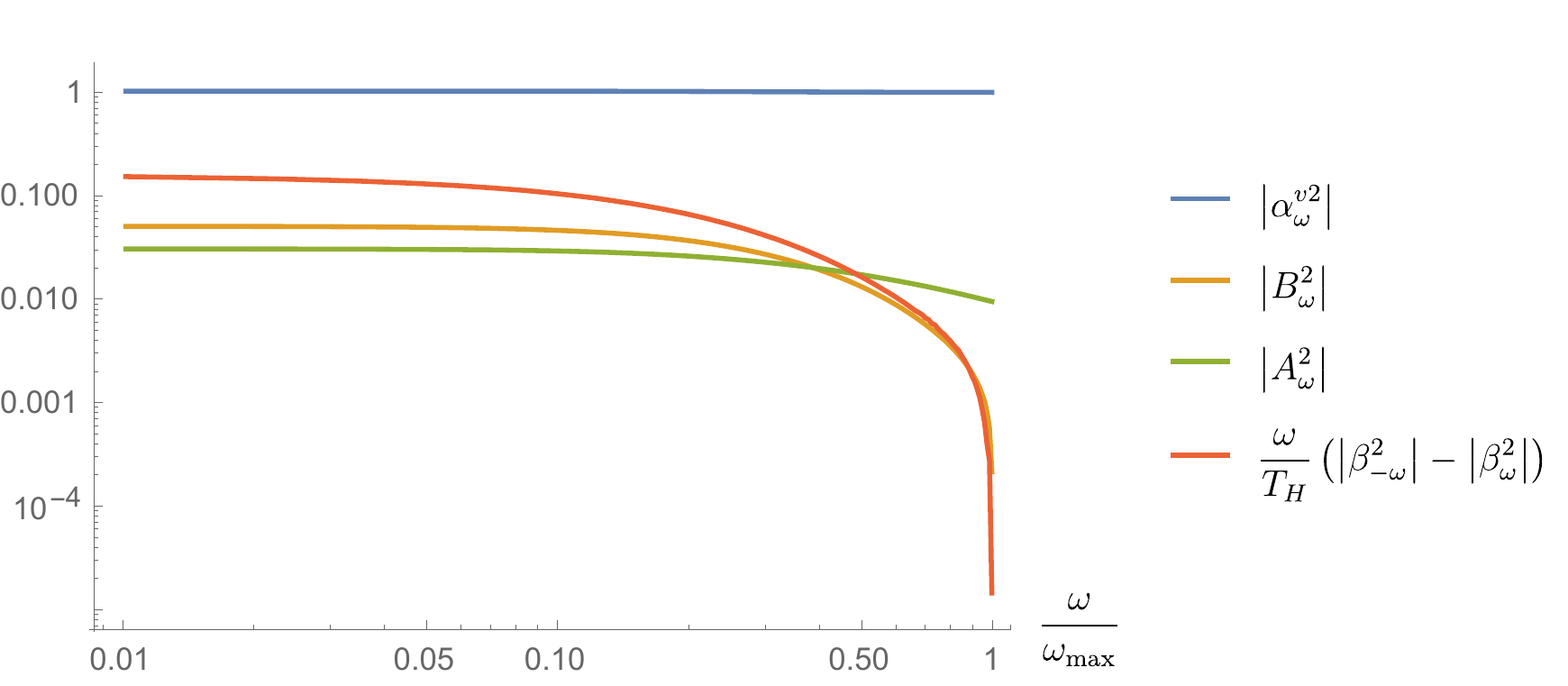}
\end{center}
\caption{Plots of the squared absolute values of the coefficients of \eq{eq:scatt_coeff} for the ``waterfall'' solution with $M_+ = 5$. The left panel shows the coefficients of the first line as well as $\abs{\tilde{B}_\om^2}$ and the Planck distribution with temperature $T_{H,\, 5}$ (dashed line), all multiplied by $\om / T_{H,\, 5}$. 
The right panel shows the coefficients involving the incoming co-propagating $v$-mode and the difference $(\abs{\beta_{-\om}^2} - \abs{\beta_\om^2}) \om / T_{H,\, 5}$.  
The smallness of these quantities reveals the weakness of the coupling between the $v$-mode and the two $u$-modes.
}\label{fig:scatcoeffs}
\end{figure} 
To pursue the analysis of the scattering, it is instructive to study the other coefficients of \eq{eq:scatt_coeff}. Here we only consider the flow with $M_+ = 5$. It is then appropriate to separate the coefficients whose norm squared diverges like $1/\om$ for $\om \to 0$, from those which remain regular in this limit. (One can verify that the first ones involve one of the two counter-propagating dispersive incoming waves.) On the left plot of \fig{fig:scatcoeffs}, we show the absolute values of the squared scattering coefficients of the first line of \eq{eq:scatt_coeff} and that of $\abs{\tilde{B}_\om^2}$. 
Besides the Planckianity already discussed, we learn here that for all $\om$, $\abs{\tilde{A}_\om^2}$ and $\abs{\tilde{B}_\om^2}$ both remain approximately $6$ times smaller than $\abs{\alpha_\om^2}$. The co-propagating $v$-mode is thus relatively weakly coupled to the two $u$-modes. This is confirmed by the right plot of \fig{fig:scatcoeffs}, where we show the absolute values of the squared coefficients of the third line involving the $v$-mode. We see that $\abs{A_\om^2}$ and $\abs{B_\om^2}$ are smaller than $0.06$ for all values of $\om$. The weakness of the coupling of the $v$-mode also explains why $\abs{\beta_{-\om}}$ remains close to $\abs{\beta_{\om}}$, as can be seen by the red curve in the right panel. Indeed, the difference $\abs{\beta_{-\om}^2}-\abs{\beta_{\om}^2}$ can be shown to be equal to $\abs{\tilde{B}_\om^2}$ for $\om \ll T_H$~\citep{Macher:2009nz}. 

\begin{figure}
\includegraphics[width=0.49 \linewidth]{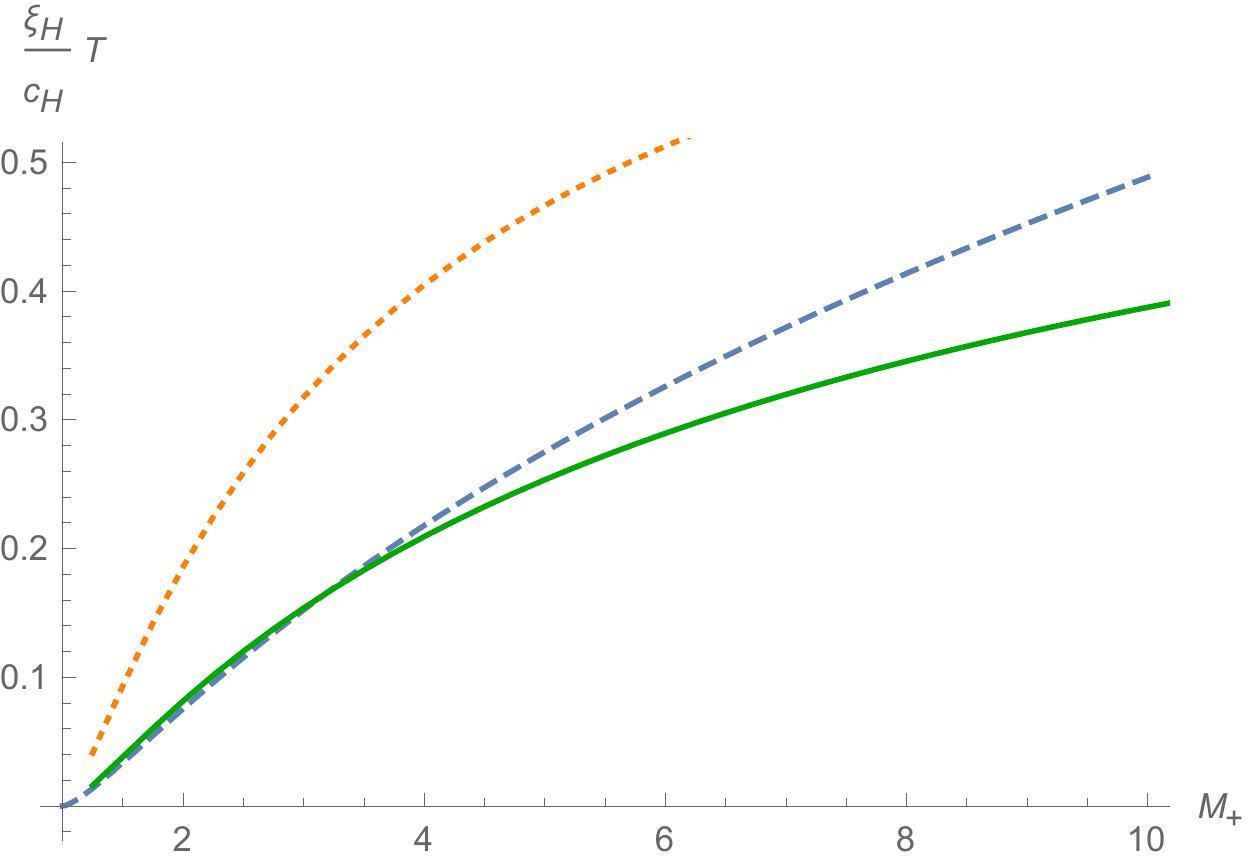} 
\includegraphics[width=0.49 \linewidth]{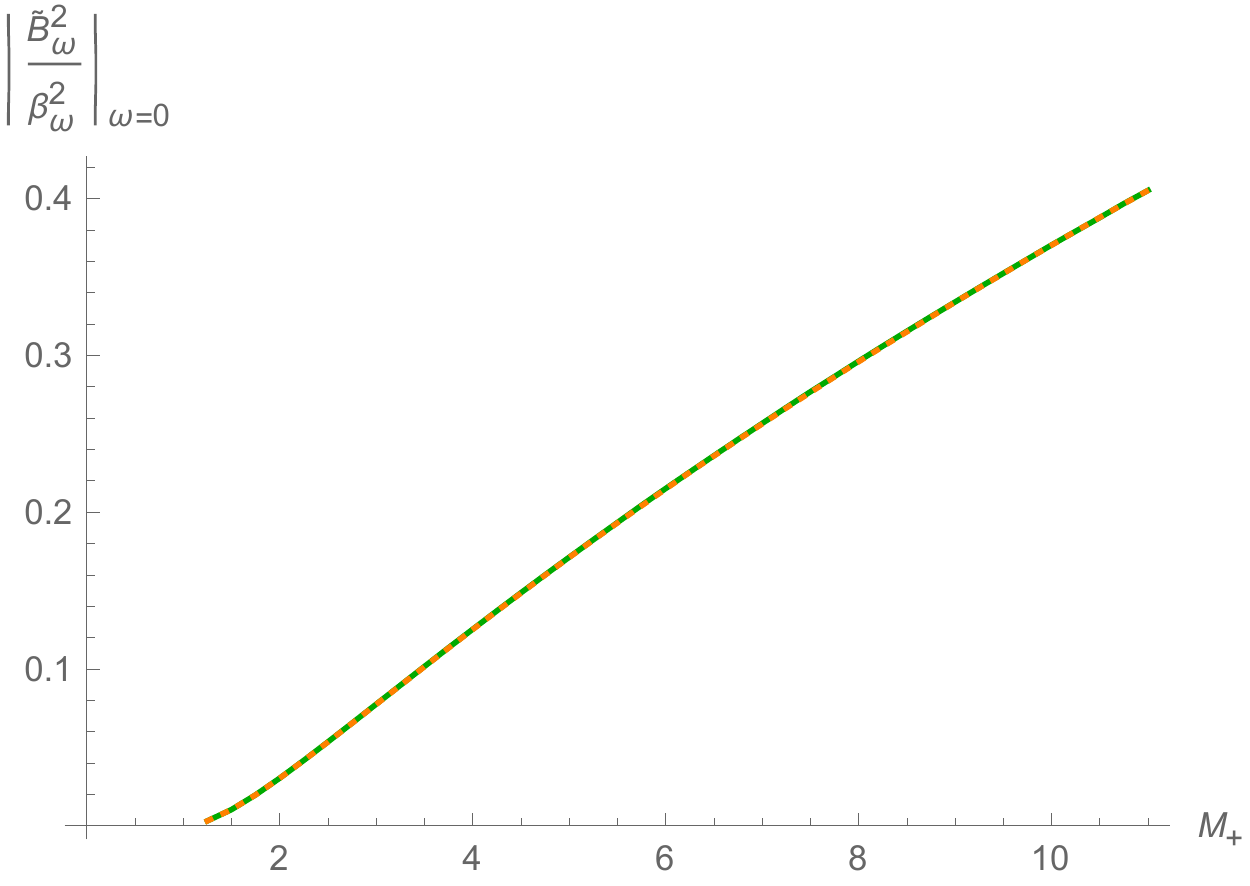} 
\caption{Left panel: As functions of the supersonic Mach number $M_+$, we show the limit $\om \to 0$ of $T_{\rm eff}$ of \eq{eq:Teff} (green, continuous), the Hawking temperature $T_H$ (blue, dashed), and  $T_{\rm step}$ obtained with the density profile of \eq{tanh} (orange, dotted), all adimensionalized by multiplication by the dispersive time-scale $\xi_H/c_H$. 
Right panel: Ratio of $\abs{\tilde{B}_{\om = 0}^2/ \beta_{\om = 0}^2 }$ for the waterfall (green, continuous) and for the ``step-like'' (orange, dotted) profiles. The very close agreement indicates that, for $\om \to 0$, the ratio $\abs{\tilde{B}_{\om}^2/ \beta_{\om}^2 }$ only depends on the asymptotic values of $v$ and $c$. 
} \label{fig:TandB}
\end{figure} 
To complete the analysis, we study in \fig{fig:TandB} two key properties characterizing the spectrum for the entire series of waterfall solutions. On the left plot, as functions of $M_+$, we represent the low frequency effective temperature $T_{\rm eff}$ and $T_H$, both adimensionalized by $c_H/\xi_H$. For all values of $M_+$, we see that $T_{\rm eff}/T_H$ remains in the interval $[0.75,1.25]$, thereby indicating that the low-frequency effective temperature is always well approximated by $T_H = \kappa_H/2\pi$. 

To estimate the largest value of the effective temperature one can obtain for a monotonic flow given the asymptotic values of $v$ and $c$ on both sides, we also represent $T_{\rm step}$, the low frequency effective temperature for the flow characterized by the density
\begin{equation}
\frac{\rho_{\rm step}(x)}{\rho_+} = \frac{1+M_+}{2} + \frac{1-M_+}{2} \tanh \lp \frac{x}{\sigma} \rp .
\label{tanh}
\end{equation} 
In our simulations, we took $\sigma = \xi_- / 8$ (decreasing $\sigma$ does not significantly modify the results.)~\footnote{Notice that the expressions in Appendix B of~\citep{Finazzi:2012iu} for the spectrum in the sharp profile limit cannot be used here, as Eq.~(B13) of that reference requires that $\log \lp v(x)/v_H \rp$ be symmetric with respect to $x_H$ in the step-like limit, while the waterfall solutions become very asymmetric for $M_+ \gg 1$. However, $T_{\rm step}$ should be computable using the same techniques, modifying Eq.~(B13) to account for the flow asymmetry.} 
For all values of $M_+$, we observe that $T_{\rm step}$ is larger than $T_{\rm eff}$ by a factor close to $2$. 
The temperature observed in~\cite{Steinhauer:2015saa} is $(T_{\rm eff} \xi_- / c_-)^{\rm obs} \approx 0.36$. 
This is larger than $T_{\rm step} \xi_- / c_- \approx 0.29$ obtained for the flow with $M = 5$ and $T_{\rm step} \xi_- / c_- \approx 0.25$ obtained for $M = 4$, which is close to the value reported in the published version of~\cite{Steinhauer:2015saa}.  
We currently have no explanation for this excess. 
(It could be related to the uncertainties in the precise values of the flow properties, see footnote~\ref{ftn1}, which could affect the estimation of $\xi_- / c_-$. It could also be partially due to the difficulty of measuring the temperature with accuracy by considering the density fluctuations in a rather small domain in the subsonic flow.) 
 
On the right plot of \fig{fig:TandB}, as functions of $M_+$, we represent the zero-frequency limit of the ratio $\abs{\tilde{B}_{\om}^2}/\abs{{\beta}_{\om}^2}$ for the waterfall solution and the step-like profile of \eq{tanh}.   
This quantity characterizes the relative importance of the $u-v$ pair creation channel with respect to the standard one involving the two $u$-modes. Since $\abs{\tilde{B}_{\om}^2} $ and $\abs{{\beta}_{\om}^2}$ both diverge as $1/\om$ for $\om \to 0$, their ratio is a constant at low frequency. We observe that it is roughly linear in $M_+ - 1$ and becomes important for large values of $M_+$. 
We also observe that the curves are almost identical which means that this ratio only depends on the asymptotic values of $v$ and $c$.  

In brief, for the flows here considered and when working in the initial vacuum state, $n_\om^u$, the mean occupation number of outgoing positive frequency phonons, closely follows the relativistic expression in the relevant frequency domain $\om / T_H \lesssim 1$, both in the Planckian character of the spectrum and in the value of the effective temperature. The spectrum of negative energy phonons, $n_{-\om}^u = n_\om^u + n_\om^v$, which includes the spontaneous production pairs of $u - v$ of phonons, is larger than $n_\om^u$ by $\approx 15 \%$ at low frequency when $M_+ = 5$. 

\section{The two-point correlation function}
\label{corr} 
\subsection{Generalities}

Following Refs.~\cite{Balbinot:2007de,Carusotto:2008ep}, J.~Steinhauher measured the density-density correlation function at a given time after the formation of the sonic horizon. In the body of the text we only consider the stationary regime, whereas in Appendix~\ref{App:t-dep} we study the time-dependent case in a simplified dispersionless model. 
To be close to the expression used in~\cite{Steinhauer:2015saa}, we study the adimensional two-point function 
\begin{equation}
G_2(x,x') = \sqrt{\frac{\xi_+ \xi_-}{{\rho}_+ {\rho}_-}} \langle \delta {\rho}(x) \delta {\rho}(x')\rangle . 
\label{G2J}
\end{equation}
The prefactor has two effects. First, it obviously adimensionalizes the density fluctuations. More importantly, when working with a given phonon state, e.g., the vacuum, $G_2$ is invariant under the rescaling of \eq{lresc}, 
unlike $\langle \delta \rho(x) \delta \rho(x')\rangle / (\rho(x) \rho(x'))$.~\footnote{\label{lamb} 
The invariance of $G_2(x,x')$ under \eq{lresc} can be verified by using \eq{structf} which shows that $\langle \delta {\rho}(x) \delta {\rho}(x')\rangle$ scales like $\rho/\xi$.} 
Because of the stationary of the system, the correlation function can be written as a single integral over $\om$: $G_2(x,x') = \int_{-\infty}^\infty d\om \, G_\om(x,x')$, see Secs. IV.D and IV.F of~\cite{Macher:2009nz}. We briefly review the main points of that analysis. 

When considering density fluctuations, since \eq{phi} gives $\delta \hat \rho(t,x)/\rho(x) = \hat \phi(t,x) + \hat \phi(t,x)^\dagger$, it is appropriate to introduce the modes $\chi_\om^a = \phi_\om^a + \varphi_\om^a $, as $\chi_\om^a$ is the only combination which enters $G_\om(x,x')$. We now assume that the temperature is sufficiently low that the initial state is well approximated by the vacuum, as seems to be the case in the experiment of~\cite{Steinhauer:2015saa}. In this case, $G_\om(x,x')$ can be written in terms of the negative frequency modes only. For $\om > 0$, one gets
\begin{equation} 
G^{\rm vac}_\om(x,x') = 
\sqrt{\frac{\xi_+ \xi_-}{{\rho}_+ {\rho}_-}} {\rho}(x) {\rho}(x') \lp \chi_{-\om}^{{\rm in},u} (x) \rp^* \chi_{-\om}^{{\rm in},u} (x') \, . 
\label{rchi}
\end{equation}
This expression can be straightforwardly generalized to account for initial states which are incoherent, i.e., fully described by the mean occupation numbers $\bar{n}_\omega^{{\rm in},u}$, $\bar{n}_\omega^{{\rm in},v}$, and $\bar{n}_{-\om}^{{\rm in},u}$ of the three types of incoming phonons, see~\cite{Macher:2009nz}. 

When $x < 0$ and $x' > 0$ are taken sufficiently far away from the horizon in the sub- and supersonic homogeneous regions, the in-modes $\chi_{-\om}^{\rm in, u}$ are superpositions of 4 asymptotic modes $\chi_\om^{a, \rm as}$: the outgoing mode $\chi_\om^{u, \rm as}$ on the sub sonic side, and three modes in the supersonic side (the incoming one, and two outgoing ones), see \fig{fig:charact}. 
When the initial state is incoherent, only the three outgoing modes interfere constructively when integrating over $\om$ to obtain $G_2(x,x')$.~\footnote{It should be noticed that this is not the case when working in momentum space with $G_2(k,k')$. 
Indeed, in that case, even when the initial state is vacuum, the fluctuations of the incoming mode with frequency $\om$ constructively interfere with those of both outgoing modes with the same frequency. 
These $in-out$ interferences have been observed in a water tank experiment~\cite{Euve:2015vml} in the stimulated regime.} 
Explicitly they are given by 
\be
\chi_\om^{u, \rm as}(x< 0) &=& S^u_-(\om) \, e^{i k^{u}_\om x }, \\ \nonumber
\chi_{-\om}^{u, \rm as}(x'> 0) &=& S^u_+(-\om) \, e^{i k^{u}_{-\om} x'}, 
\\ \nonumber
\chi_\om^{v, \rm as}(x'> 0) &=& S^v_+(\om) \, e^{i k^{v, p}_\om x'}, 
\label{3chi}
\ee
where $S^a_\pm(\om)$ is the structure factor evaluated on the right (+) or left (-) asymptotic side. 
With our normalization conventions, it is given by 
\begin{equation}\label{structf}
S^a_\pm(\om) = \frac{U_{k^{a, \pm}_\om} + V_{k^{a, \pm}_\om}}{\sqrt{2 \pi \rho_\pm}} \abs{\frac{d k^{a, \pm}_\om}{d\om}}^{1/2},  
\end{equation}
where $U_k$ and $V_k$ are the usual coefficients obeying $|U_k|^2 - |V_k|^2 = 1$ and $V_{k_\om^{a,\pm}} = D(k_\om^{a,\pm},\rho_\pm) U_{k_\om^{a,\pm}}$, see \eq{eq:Dk}. Importantly, the three outgoing modes have a vanishing wave number in the limit $\om \to 0$. 
The correlation pattern is thus a low-wavenumber one which could be well described in dispersionless settings. 
This is unlike what is found when considering white hole flows~\cite{Mayoral:2010ck,Coutant:2012mf}.   
 
Keeping only the above outgoing modes, for $\om > 0$, one obtains: 
\begin{equation} 
\frac{1}{\sqrt{\xi_+ \xi_- {\rho}_+ {\rho}_-}} G^{\rm vac}_\om(x,x') =   
e^{i k^{u}_\om x } \times \lp \mathbb{A}_\om \,  e^{- i k^{v, \, p}_\om x'} 
+  \mathbb{B}_\om \,  e^{i k^{u}_{-\om} x'} \rp .
\label{AB} 
\end{equation}
(Including the prefactor of the left-hand side in the coefficients $\mathbb{A}_\om ,  \mathbb{B}_\om$ would multiply their norm by $\sqrt{\xi_+ \xi_- {\rho}_+ {\rho}_-}$, which is close to $25$ for the flow with $M_+ = 5$ and for $\rho_+$ equal to the value reported in~\cite{Steinhauer:2015saa}.) Using Eqs.~(\ref{rchi}, \ref{3chi}) and the second line of \eq{eq:scatt_coeff}, one finds 
\begin{eqnarray} \label{eq:AandB}
\mathbb{A}_\om &= & S^u_-(\om)  S^v_+(\om) \, \beta_\om^* \tilde{B}_\om, 
\nn
\mathbb{B}_\om &= & S^u_-(\om) S^u_+(-\om) \, \beta_\om^* \alpha_{-\om}.
\end{eqnarray}
The first accounts for the correlations between $v$-phonons and positive energy $u$-phonons, while the second accounts for correlations between $u$-phonons of opposite energies, see~\cite{Busch:2014bza} for more details. 

\subsection{Strength of correlations and their dispersionless pattern} 

\begin{figure}
\begin{center}
\includegraphics[width=0.49\linewidth]{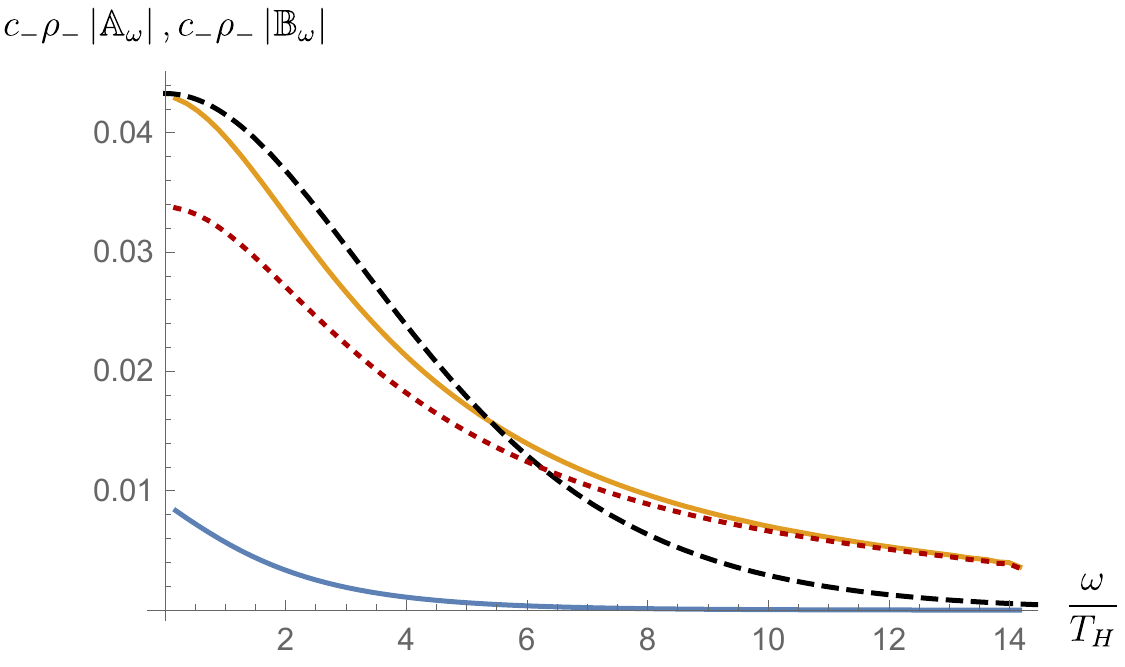}
\caption{We represent the absolute value of the spontaneous correlation terms $\mathbb{A}_\om$ (blue) and $\mathbb{B}_\om$ (orange) defined in \eq{eq:AandB}, calculated for the waterfall solution with $M_+ = 5$, and adimensionalized by $c_- \rho_-$, so as to get a result invariance under a rescaling of $\lambda$, see footnote~\ref{lamb}. 
The dotted (red) line shows $\vert \mathbb{B}_\om\vert$ when the initial state of the co-propagating $v$-modes is a thermal state with an initial temperature $T_{\rm in} = 10 T_H$. 
The black, dashed curve shows the relativistic limit for $\mathbb{B}_\om$ given in \eq{eq:Bomrelat}. 
}\label{fig:AandB}
\end{center} 
\end{figure} 
As can be seen in \fig{fig:AandB}, for the flow with $M_+ = 5$, $\abs{\mathbb{A}_\om}$ is smaller than $\abs{\mathbb{B}_\om}$ by a factor $\approx 5$. It should be noticed that the ratio $\abs{\mathbb{A}_\om}/\abs{\mathbb{B}_\om}$ significantly varies with $M_+$, 
but remains smaller than $1$ for the flows we are considering. For instance, for the three flows of \fig{fig:bckgd}, its limit $\om \to 0$ is close to $0.13$ for $M_+ = 3.75$, $0.19$ for $M_+ = 5$, and $0.24$ for $M_+ = 6.25$. (This dependence is corroborated by the curves shown on the right plot of \fig{fig:TandB}.) Neglecting the $vu$ correlations weighted by $\abs{\mathbb{A}_\om}$ is thus a fairly good approximation. 

In preparation for the subsequent analysis, in \fig{fig:AandB} we have also represented by a dotted line the strength of the $uu$ correlations when the initial state of the co-propagating incoming $v$-modes is a thermal state with temperature $T_{\rm in} = 10 T_H$ in the fluid frame. (To get this result, we used the complete expression of the $uu$-correlation which includes the stimulated processes, see Eq.(50) in~\cite{Macher:2009nz}.) We see that increasing significantly the initial temperature of $v$-modes only slightly decreases $\mathbb{B}_\om$ for low frequencies. We also see that the strength of correlations remains largest at low frequency. 

It should be also noticed that $\abs{\mathbb{B}_\om}$ closely follows the corresponding relativistic expression, 
\begin{equation}\label{eq:Bomrelat} 
\mathbb{B}_\om^{\rm relat.} = \frac{e^{\om/(2 T_H)}}{e^{\om / T_H} - 1} e^{i \arg{\lp\alpha_\om^{\rm relat.}/\beta_{-\om}^{\rm relat.}\rp}} \frac{\om \sqrt{\xi_+ \xi_-/\rho_+ \rho_- }}{4 \pi \abs{v_+ - c_+} \abs{v_- - c_-}}, 
\end{equation}
which is indicated by a dashed line. 
The first factor comes from the fact that $|\beta_\om^{\rm relat.}|^2$ (exactly) follows the Planck law with temperature $T_H = \kappa_H/2\pi$ in the present settings where there is no coupling between the $v$-mode and the two $u$-modes. 
The phase plays no role here and shall be studied below. 
The normalization comes from taking the dispersionless limit ($\xi_\pm \to 0$) of the structure factors of \eq{structf}. 
To obtain this expression, we used the low wavenumber behaviors $U_k + V_k \approx (k/2c)^{1/2}$ and $k \approx \om (c- v)$, both valid when $\xi k \to 0$.

Having shown that the vacuum relativistic expressions give reliable approximations at fixed $\om$, it is instructive to integrate them over $\om$ to get the dispersionless limit of the equal-time correlation function of \eq{G2J}.
When $x$ and $x'$ are sufficiently far away from the horizon such that $\rho$, $v$ and $c$ have reached their asymptotic values $\rho_\pm$, $v_\pm$ and $c_\pm$, one obtains unambiguous expressions. 

\begin{figure} 
\includegraphics[width=0.49 \linewidth]{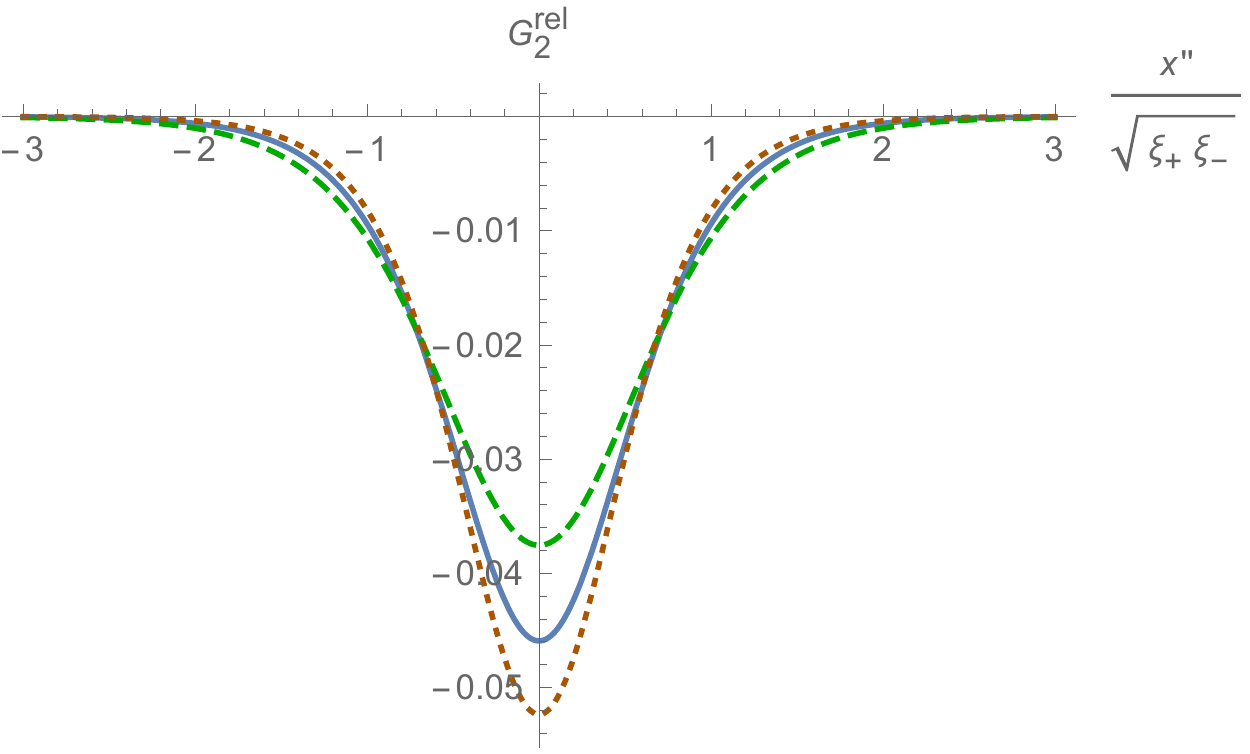}
\includegraphics[width=0.4 \linewidth]{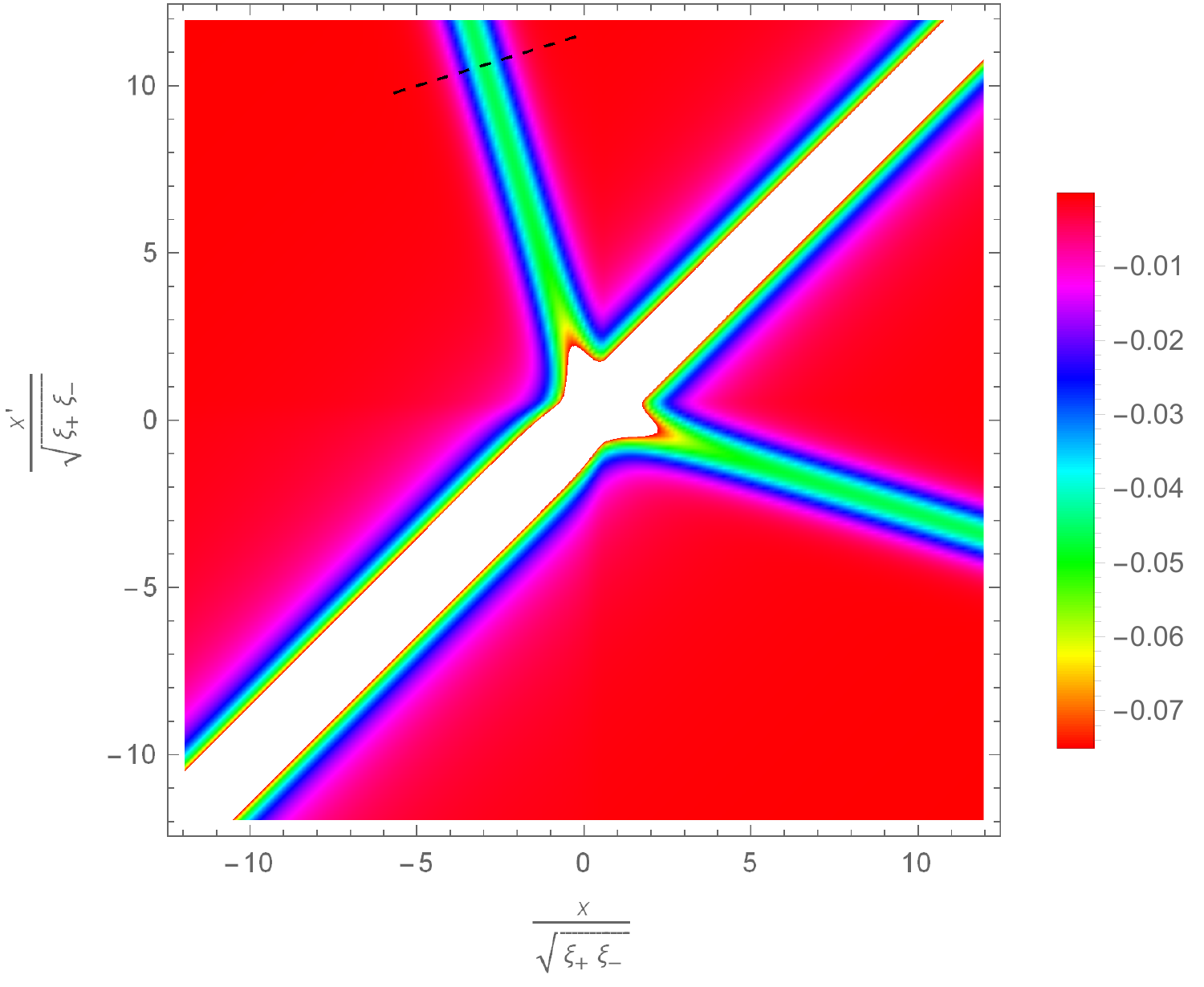}
\caption{Left: We show the profile of $G_2^{\rm rel}(x,x')$ of \eq{eq:relG2} evaluated along a line orthogonal to the locus of its minima and located far from the horizon. $x''$ is a coordinate along this line, defined by $x'' = 0$ when $G_2$ reaches its minimum and $\abs{dx''} = \sqrt{dx^2 + dx'^2}$. 
The three curves show the correlation profile for the waterfall solutions with $M_+ = 3.75$ (green, dashed), $5$ (blue, continuous), and $6.25$ (orange, dotted) considered in \fig{fig:bckgd}. 
Right: 
We show $G_2^{\rm rel}(x,x')$ of \eq{eq:Gxx2} as a function of $x$ and $x'$ for the waterfall solution with $M_+ = 5$. 
The dashed segment indicates the domain used to represent the correlations on the left panel. The broad oblique white band centered along $x = x'$ corresponds to values of $G_2$ outside the range represented in colors. 
}
\label{fig:G2rel}
\end{figure} 
For points on opposite sides of the horizon $x x' < 0$ with $x< 0$, one gets, see~\cite{Balbinot:2007de}:
\begin{equation}\label{eq:relG2} 
G_2^{\rm rel}(x,x') \approx \frac{ - \pi T_H^2 \xi_+ \xi_-}{4 \abs{v_+ - c_+} \abs{v_- - c_-} \cosh^2 \lp \pi T_H \lp u_{L}(x) - u_{R}(x') \rp \rp }\ ,
\end{equation}
where $u_{R/L}(x)$ give the values (at a given common time) of the  outgoing null lightlike coordinate on each side of the horizon, see \eq{CRL}.
To make contact with Ref.~\cite{Steinhauer:2015saa}, we study the behavior of this function in the three waterfall solutions considered in \fig{fig:bckgd}. 
In the left panel of \fig{fig:G2rel}, we show the corresponding profiles of $G_2^{\rm rel}(x,x')$ evaluated along a segment orthogonal to the locus of the minima. 
(The segment is represented by a dashed line on the right panel.)  
Using $\sqrt{\xi_+ \xi_-}$ as a unit of distance along this segment, we see that the width of the hollow hardly varies for these values of $M_+$. 
We also notice that it is about twice as large as the value reported in~\cite{Steinhauer:2015saa}, while the depth of the hollow is only half that reported in the experiment. These results are qualitatively consistent with the fact that the effective temperature of the flow with $M_+ = 4$ is significantly smaller than the measured one, see the discussion below \eq{tanh}. 

It is also interesting to study the angle $\theta$ between the horizontal and the line of maxima of the correlations in the domain $x < 0, x' > 0$.  It is given by   
\be
\label{thet}
\theta &=& \arctan \lp (c_+ - v_+)/(c_- - v_-) \rp + \pi \\ \nonumber
 &= & \pi - \arctan \lp 1 + M_+^{1/2} \rp. 
\ee
For the three values of $M_+$ we used ($3.75, 5, 6.25$), this gives $1.90$rad, $1.87$rad, and $1.85$rad. The value reported in~\cite{Steinhauer:2015saa} ($\theta^{\rm obs}\approx 2.2$rad) is slightly larger than the maximal value of \eq{thet} accessible with waterfall solutions, which is $\approx 2.03$rad.
 
When working on the same side of the horizon $xx' > 0$ and far away from the horizon, the auto-correlations of $u$-modes are also modified by the Hawking temperature~\cite{Parentani:2010bn}, whereas those involving the co-propagating $v$-modes are essentially unchanged, as can be understood from Eqs.~(\ref{eq:du}, \ref{eq:Gxx1}, \ref{eq:Gxxv2}). Explicitly, in the subsonic side $x < 0$ one gets 
\be 
\label{eq:relG3} 
G_2^{\rm rel}(x,x') &= &G_{2,u}^{\rm rel}(x,x') + G_{2,v}^{\rm rel}(x,x') \nonumber \\ 
&\approx&  
- \frac{\xi_-^2}{4 \pi} 
\sqrt{\frac{\xi_+ {\rho}_-}{\xi_- {\rho}_+}} \left[ 
\frac{(\pi T_H)^2}{\lp c_- - v_- \rp^2 \sinh^2 \lp \pi T_H \lp u_{L}(x) - u_{L}(x') \rp \rp } 
+ \frac{1}{\abs{x - x'}^2} 
 \right].
\ee 
A similar expression applies on the supersonic side. 
It should be noticed that, in the coincidence point limit, the divergence of the first term (which describes a thermal flux of outgoing $u$-phonons at temperature $T_H$, see \eq{eq:Gxx3}) is the same as that of the second term which describes $v$-phonons in their ground state.
In both cases, one has $G_{2,u/v}^{\rm rel}(x \to x')/\sqrt{\lp \xi_+ {\rho}_- \rp / \lp \xi_- {\rho}_+ \rp} \sim -  \xi_-^2/\lp 4 \pi (x-x')^2 \rp$. 
 
To complete this study, we represent in the right panel of \fig{fig:G2rel} a generalized version of $G_{2}^{\rm rel}$, 
given by \eq{eq:Gxx2}, defined in the whole $(x,x')$ plane. For this figure we only considered the waterfall solution with $M_+ = 5$. 
(Unlike for the above asymptotic expressions, it should be noticed that there is some ambiguity in obtaining this expression in the near horizon region where $v$ and $c$ significantly vary, see footnote~\ref{foot:lda}.) From this figure one clearly sees how the asymptotic $uu$-correlations of \eq{eq:relG2} emerge from the diverging auto-correlations near the sonic horizon $|x| \approx |x'| = O(\sqrt{\xi_+ \xi_-})$. 

\subsection{Non-separability} 

\begin{figure}
\begin{center}
\includegraphics[width=0.49 \linewidth]{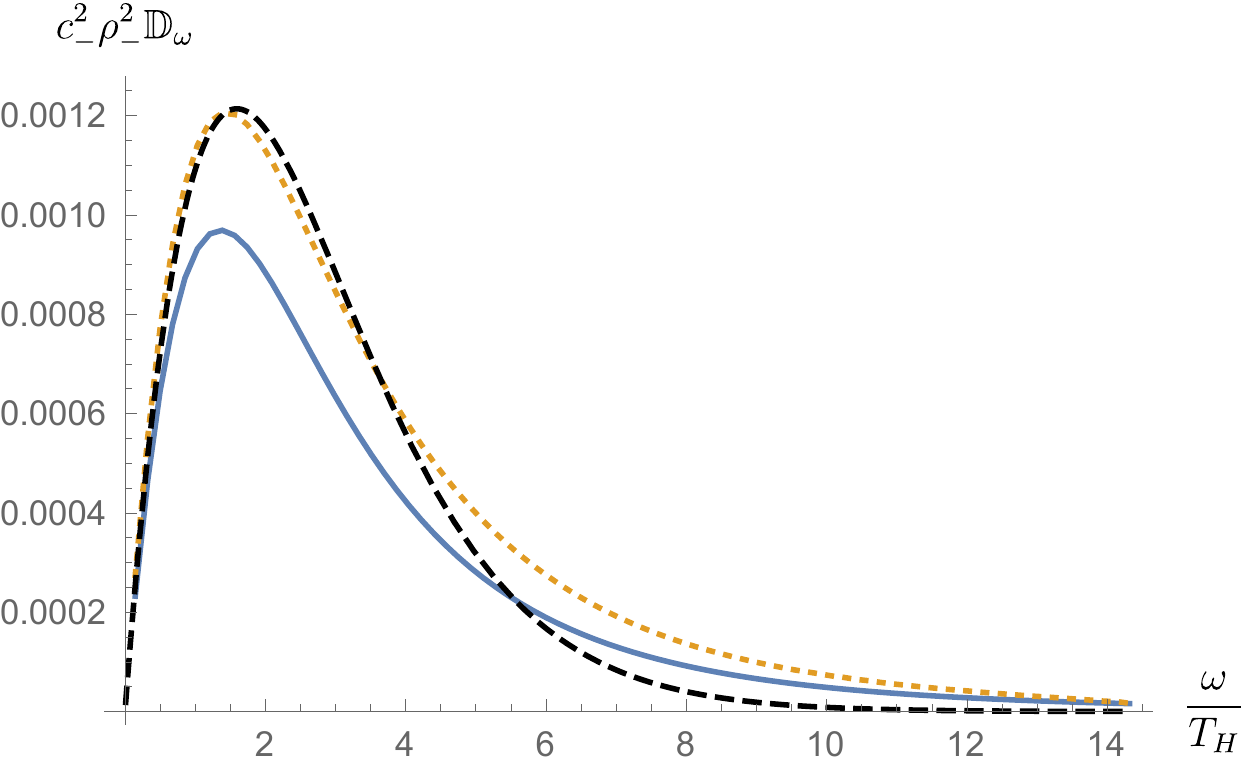} 
\includegraphics[width=0.49 \linewidth]{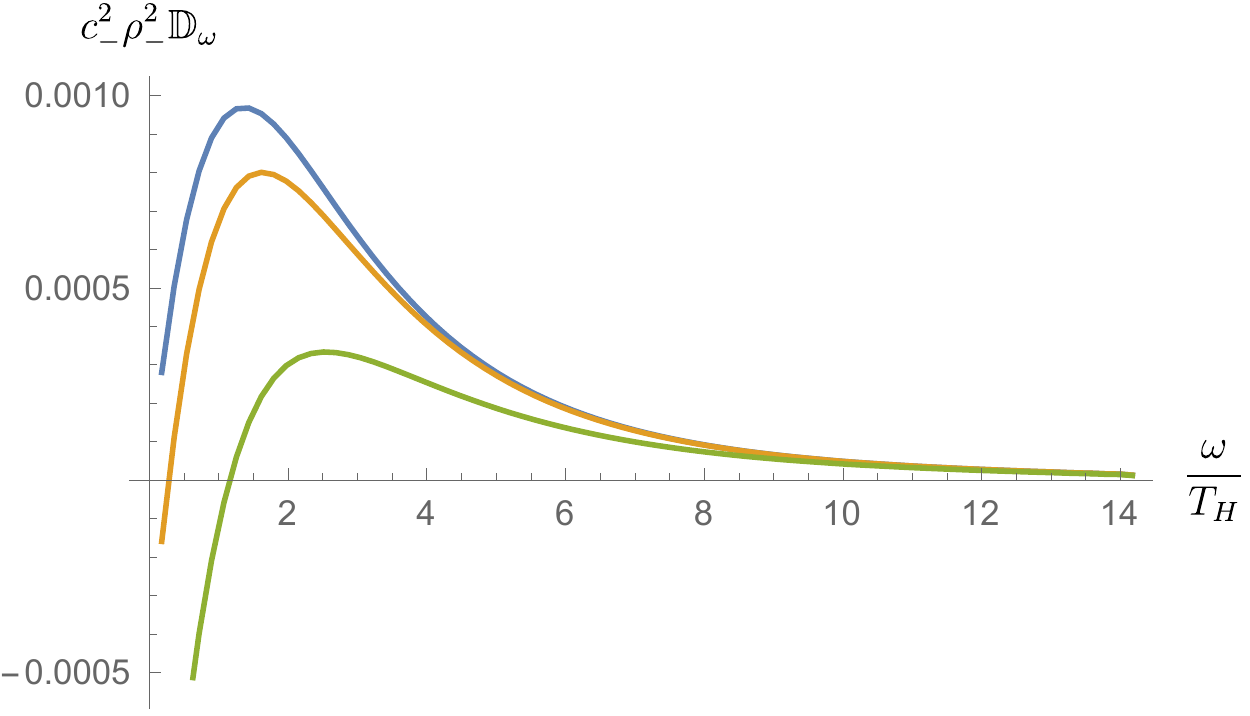}
\end{center}
\caption{Plot of the quantity $\mathbb{D}_\om$ of \eq{eq:Dom} computed for the waterfall solution with $M_+ = 5$, and adimensionalized by $c_-^2 \rho_-^2$. On the left panel, the blue continuous line gives the numerical value predicted by the Bogoliubov-de Gennes equation, while the black, dashed line shows its relativistic limit given in \eq{eq:Domrel}. 
The dotted orange line shows the maximum value of $\mathbb{D}_\om$ of \eq{eq:Dommax}. 
On the right panel, we show how $\mathbb{D}_\om$ varies when assuming that the co-propagating incoming $v$-modes have an initial temperature $T_{\rm in}$. Curves are shown for $T_{\rm in}/T_H = 10^{-6}$ (blue), $3$ (orange), and $10$ (green).
} 
\label{fig:Dom}
\end{figure} 
As discussed in~\cite{deNova:2012hm,Busch:2014bza,Boiron:2014npa,Steinhauer:2015ava}, the relative strength of the correlation, 
governed by $\abs{\mathbb{B}_\om}$ of \eq{AB}, with respect to the final mean occupation numbers $n_\om^u n_{-\om}^u$ can be used as a reliable criterion for asserting that the state is non-separable, which implies that the spontaneous amplification of vacuum fluctuations contributes more that the stimulated processes induced by the initial population of phonons. More precisely, whenever the difference $\mathbb{D}_\om$, defined by
\begin{equation}\label{eq:Dom}
\mathbb{D}_\om \equiv \abs{\mathbb{B}_\om^2} - n_\om^u n_{-\om}^u \lp S_+^u(\om) S_-^u(-\om) \rp^2,  
\end{equation}
is positive, the final state of the $u$-phonons of frequency $\pm \om$ is non-separable.~\footnote{It should be noticed that the non-separability of a quantum state does not seem to imply the non-classicality as defined in~\cite{Finke:2016wcn}.} 
When the initial state is vacuum, $\mathbb{D}_\om$ can be shown to be positive definite. Yet it is of value to study its behavior as a function of $\om$. 
On the left panel of \fig{fig:Dom}, it is represented by a continuous line. We see that it reaches its maximum for $\om \approx T_H$. 
On the same panel, the dashed line gives its dispersionless limit. 
Using \eq{eq:Bomrelat} and $n_\om^{u, \rm relat.} =  n_{-\om}^{u, \rm relat.} = (e^{\om/T_H} -1)^{- 1}$, one easily finds that it follows 
\begin{equation} 
\mathbb{D}_\om^{\rm relat.} =  \lp \frac{\om  \sqrt{\xi_+ \xi_-/\rho_+ \rho_-}
}{4 \pi \abs{v_+ - c_+} \abs{v_- - c_-}} \rp^2 \frac{1}{e^{\om / T_H}-1}.\label{eq:Domrel}
\end{equation}
We clearly see that the two curves closely agree, as can be understood from the near Planckianity of the spectrum and the weakness of the coupling to the co-propagating $v$-mode (which affects the difference $n_\om^u - n_{-\om}^u$, see the red curve of the left panel 
of \fig{fig:scatcoeffs}.). 

It should be also pointed out that $\mathbb{D}_\om$ is bounded from above, see~\cite{Campo:2005sy,Adamek:2013vw,Busch:2014bza}. 
When working in the initial vacuum, a Cauchy-Schwarz inequality implies that $\mathbb{D}_\om $ is smaller than (see Eq.~(B4) in~\cite{Adamek:2013vw}) 
\begin{equation}\label{eq:Dommax}
\mathbb{D}_\om^{\rm max} \equiv |\beta_{-\om}|^2 \lp S_+^u(\om) S_-^u(-\om) \rp^2 .
\end{equation}
This maximal value is represented by a dotted line on the left panel of \fig{fig:Dom}. Working with relativistic settings in the initial vacuum, the situation is simpler because $\mathbb{D}_\om^{\rm relat.}$ of \eq{eq:Domrel}, the dashed curve, already gives the maximal value. This is due to the fact that the coupling to the co-propagating $v$-mode identically vanishes in these settings.

To test the dependence of non-separability with respect to the initial state of the phonons, we assume that the initial state is thermal in the frame of the condensed fluid in the subsonic region. 
To compute the initial mean occupation numbers of the counter-propagating dispersive $u$-modes in the supersonic region requires in principle the knowledge of the whole time dependence and the fall-off of the condensate density for $x \to +\infty$. 
Irrespective of these details, one finds that low-frequency modes with $\om \approx \kappa$ in the black-hole frame correspond to large frequencies $\Omega^u \approx |k  v_+|$ in the rest frame of the fluid. 
For dispersive modes, $|k^{u, d}_\om v_+|$ is of the order of $\om_{\rm max}$, much larger than $T_H \approx \kappa_H/2\pi$. 
As a result, their initial population will be suppressed. In a first approximation, one can thus neglect their contribution and consider only the initial occupation number of incoming $v$-modes 
\begin{equation}
n_\om^{v, \rm in} = \lp e^{\Omega^v / T_{\rm in}} - 1 \rp^{-1},
\end{equation}
where $\Omega^v = \om -  k_\om^{v, b} v_- $.  
The right panel of \fig{fig:Dom} shows $\mathbb{D}_\om$ for $T_{\rm in}/T_H = 10^{-6}$, $3$, and $10$. 
As was found in~\cite{deNova:2012hm,Busch:2014bza,Boiron:2014npa}, the temperature has the tendency to reduce the non-separability of the state, with low-frequency modes becoming separable before high-frequency ones. 
Overall, the non-separability is strongly reduced only when $T_{\rm in}$ becomes of the order of $10 T_H$. 
This is another consequence of the relative weakness of the couplings involving the co-propagating $v$-mode.

\subsection{Phase of \texorpdfstring{$uu$}{uu} correlations} 

In the previous subsection we studied the strength of the correlations, which is governed by the absolute value of $\mathbb{B}_\om$. Here we consider the phase $\arg \mathbb{B}_\om$, which is equal to $\arg (\alpha_{-\om} \beta_\om^*)$. It does not depend on the arbitrary phase of the (globally defined) incoming modes, but it does depend on the phases of the asymptotic outgoing modes of wave numbers $k_{\pm \om}^u$. Each of them is asymptotically given by $k_{\pm\om}^{u} (x-x_H) + C_{\pm\om}$, where $C_{\pm\om}$ are two real constants. In this paper, we work with $C_{\pm\om} = 0$, see \eq{3chi}. (In the body of the paper we fix the origin of $x$ so that $x_H = 0$.) 

To see the consequence of the $\om$ dependence of $\arg \mathbb{B}_\om$, we consider the trajectories in the $x-x'$ plane where the equal-time correlations among $u$-phonons reach their maximal intensity, see \fig{fig:G2rel} for their relativistic counterpart. 
To get the locus of constructive interferences at a given time, one should impose that the phase of the $\mathbb{B}_\om$ term of \eq{AB} is stationary~\cite{Macher:2009nz}, i.e., 
\begin{equation}
\label{cin}
(\partial_\om k^{u}_\om ) x + (\partial_\om k^{u}_{-\om}) x' = -  \partial_\om \arg \lp \alpha_{-\om} \beta_\om^* \rp. 
\end{equation}
We thus see that $\partial_\om \arg \lp \alpha_\om \beta_{-\om}^* \rp$ introduces a non-trivial shift. To our knowledge it has not been studied before in the present context, although its existence was mentioned in~\cite{Macher:2009nz}.~\footnote{A phase similar to $\arg \lp \alpha_\om \beta_{-\om}^* \rp$ governs the loci of the nodes of the stationary zero-frequency modulation emitted in transonic flows which are analogous to white holes, see~\cite{Coutant:2012mf,Busch:2014hla}. 
 When working in homogeneous time-dependent settings, such as in inflationary cosmology~\cite{Campo:2005sy} and in condensed matter~\cite{Carusotto:2009re,Busch:2013sma}, a similar phase, also given by the argument of the product $\alpha \beta^*$ of two Bogoliubov coefficients, fixes the location of the nodes of the equal-time correlations.} 
 When $\partial_\om \arg \lp \alpha_{-\om} \beta_\om^* \rp \neq 0$, the asymptotic straight line $x(x')$ solution of \eq{cin} will not cross exactly the sonic horizon $x_H = 0$ when $x'= 0$. Rather $x$ will be equal to $x'$ at the point $x_M$ given by
\begin{equation}
x_M(\om) = - \frac{\partial_\om \arg \lp \alpha_{-\om} \beta_\om^* \rp}{\partial_\om k_\om^{u} + \partial_\om k_{-\om}^{u}}, 
\label{xM}
\end{equation}
where the denominator is evaluated in the asymptotic region. 
\begin{figure} 
\begin{center}
\includegraphics[width=0.49 \linewidth]{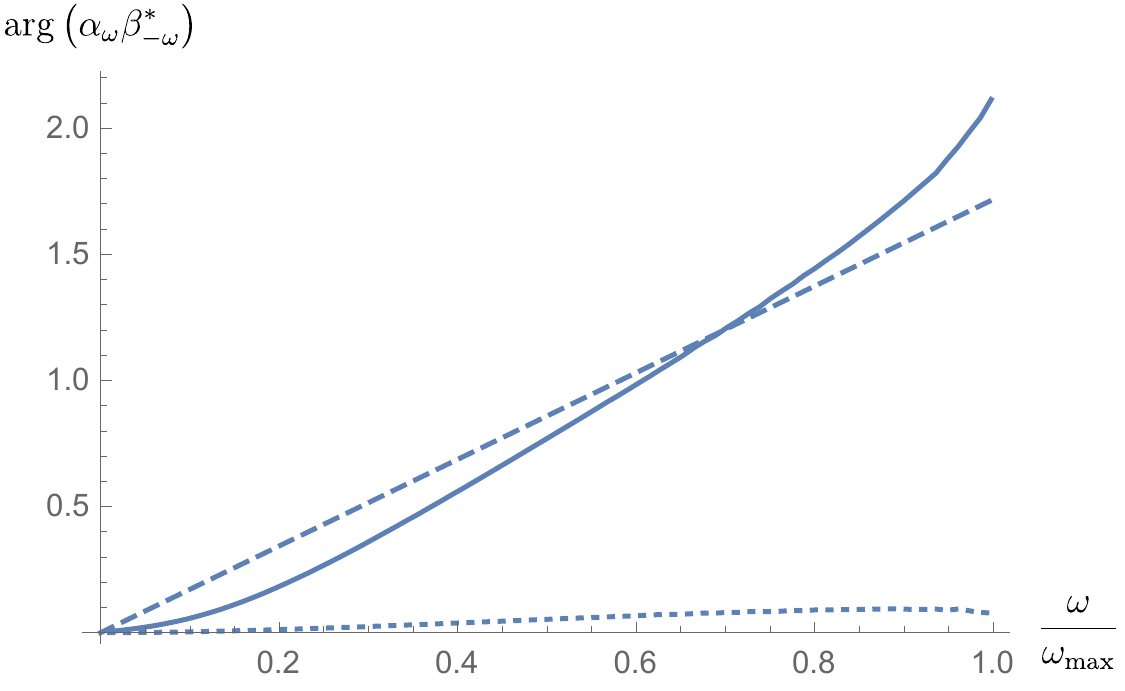} 
\includegraphics[width=0.49 \linewidth]{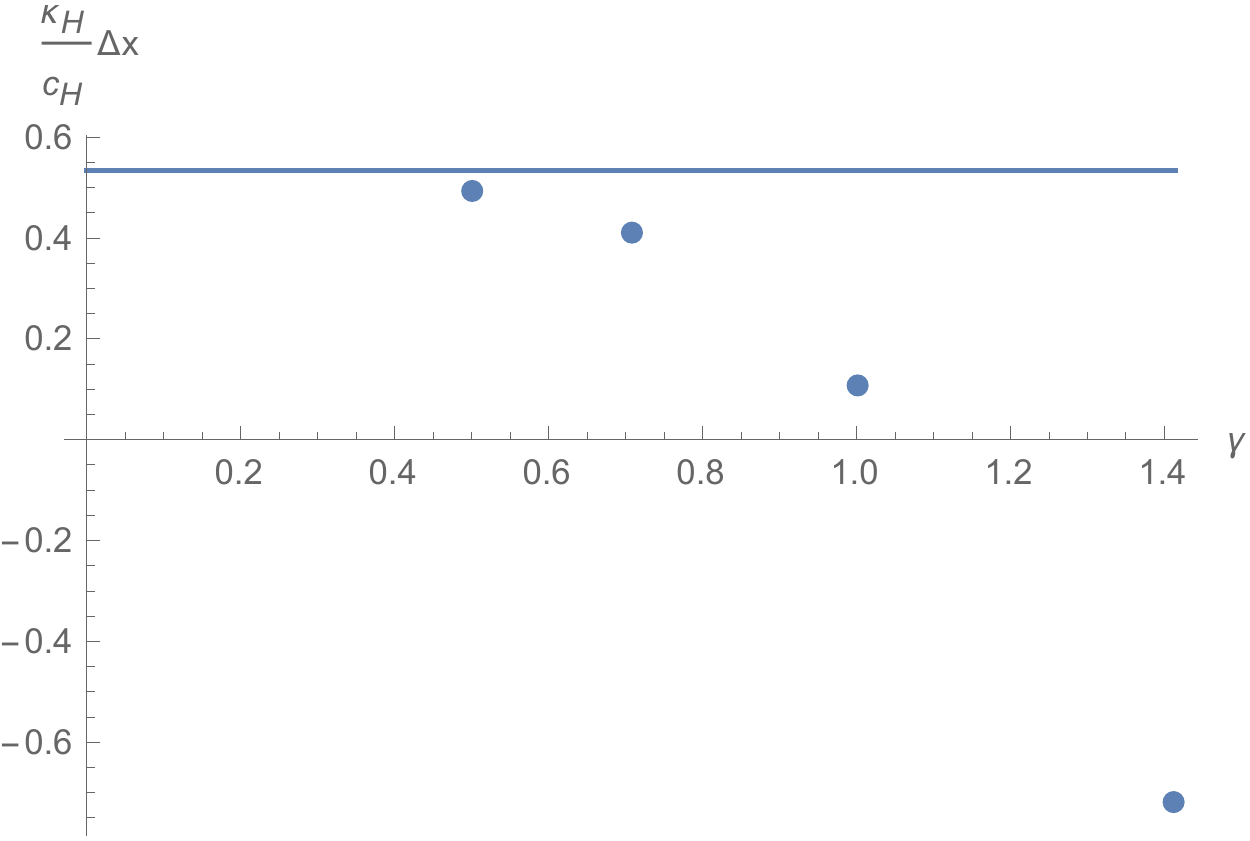} 
\end{center}
\caption{Left: As a function of $\om/\om_{\rm max}$, we show the phase of $\alpha_\om \beta_{-\om}^*$ 
for the flow of \fig{fig:bckgd} with $M_+ = 5$. The dashed line shows the relativistic phase which is obtained from \eq{eq:diffB}. The dotted line shows $\arg \lp \alpha_{-\om} \beta_\om^* \rp - \arg \lp \alpha_\om \beta_{-\om}^* \rp$. 
Its smallness is another test of the validity of the relativistic approximation, which predicts that this difference should vanish. 
Right: As a function of $\gamma$ of \eq{eq:resc_BdG}, we show the shift $\Delta x = x_M(0) - x_H$, see \eq{xM}, evaluated in the small-frequency limit and adimensionalized by the surface gravity length $c_H/\kappa_H$. 
The four dots are obtained from the numerical integration of \eq{eq:resc_BdG}, whereas the horizontal line shows the relativistic result of \eq{eq:relshift}.
} 
\label{fig:cartphaseandtorel}
\end{figure}

In the left panel of \fig{fig:cartphaseandtorel}, the continuous line shows $\arg \lp \alpha_\om \beta_{-\om}^* \rp$ as a function of $\om / \om_{\rm max}$ for the flow of \fig{fig:bckgd} with $M_+ = 5$. As can be seen, its slope is nearly constant except for $\om \approx \om_{\rm max}$ and $\om \approx 0$. 
In the intermediate frequency domain, the shift $x_M(\om)$ is thus nearly independent of $\om$. The dashed line shows $\arg \lp \alpha_{-\om} \beta_\om^* \rp$ for the relativistic field propagating in the same background flow. Its value is computed in Appendix~\ref{App:shift}. 
Quite surprisingly, its slope does not vanish and agrees rather well with the slope of \fig{fig:cartphaseandtorel} in the intermediate frequency domain. It is therefore interesting to study the relativistic limit to see the residual role played by short distance dispersion. 
To this end, we numerically computed the low-frequency slope when decreasing the healing length. Specifically, we solved the rescaled BdG equations
\begin{eqnarray} 
\label{eq:resc_BdG}
\lp \lp \om + i v \pd_x \rp - \dfrac{\gamma^2}{2 \rho} \pd_x \rho \pd_x - \frac{c^2}{\gamma^2} 
\rp \phi_\om & =& \frac{c^2}{\gamma^2} \varphi_\om, \nonumber \\
- \lp \lp \om + i v \pd_x \rp + \dfrac{\gamma^2}{2 \rho} \pd_x \rho \pd_x + \frac{c^2}{\gamma^2} 
\rp \varphi_\om & =& \frac{c^2}{\gamma^2} \phi_\om,
\end{eqnarray}
for several values of $\gamma$. This rescaling neither modifies the background flow nor the conserved scalar product, but multiplies the dispersive length scale by $\gamma$. The relativistic limit thus corresponds to $\gamma \to 0$, while \eq{eq:BdG_W} is recovered when $\gamma = 1$. Numerical results are shown in \fig{fig:cartphaseandtorel}, right panel. 
Although we were not able to obtain trustworthy values of the shift $\Delta x$ for $\gamma < 1/2$, the figure indicates that $x_M - x_H$ converges to the value obtained from \eq{eq:relshift} when decreasing $\gamma$ towards 0.
We hope that this shift  can be measured in forthcoming experiments. We also hope that $\arg \left( \alpha_\om \beta_{-\om}^* \right)$ itself will be measured in water tank experiments where one can work at fixed $\om$, see~\cite{Euve:2015vml}.

\section{Conclusions} 
\label{concl}

In this paper we studied the spectral properties and the coherence of the phonon pairs emitted in transonic flows which are similar to those experimentally realized in~\cite{Steinhauer:2015saa}. In Section~\ref{sp}, we first analyzed the stationary, asymptotically homogeneous transonic background flows which are solutions of the one-dimensional GPE in a step-like potential. These are described by a one-parameter family of stable waterfall solutions. When $M_+$, the Mach number in the asymptotic supersonic region, is significantly larger than 1, they have high spatial gradients near the sonic horizon where $M= 1$. 
Indeed, for $M_+ \gtrsim 3$, the surface gravity $\kappa_H$ is larger than the dispersive scale measured on the horizon. These flows are highly asymmetrical with respect to the horizon: for large $M_+$, the gradients of flow parameters such as the sound speed $c$ or the velocity $v$ increase significantly on the supersonic side. As a byproduct of the strong asymmetry, when $M_+ \gtrsim  3$, the surface gravity (measured on the horizon) decreases when increasing $M_+$. These results are summarized in  \fig{fig:bckgd}. 
The stability of these solutions is studied in Appendix~\ref{App:NLstab} where it is shown that perturbations are expelled from the   
near horizon region. In Appendix~\ref{App:NPGPE} we studied a generalized version of waterfall solutions which better takes into account the three-dimensional character of the background flow. 

We then studied the spectrum of the phonons spontaneously emitted when the initial state is vacuum. In spite of the fact that the surface gravity $\kappa_H$ is larger than the dispersive scale, for $M_+ \lesssim 6$, we found that the emission spectrum is well approximated by its relativistic prediction, namely a Planck spectrum governed by the temperature $T_H= \kappa_H/2 \pi$. This can be understood from the fact that the critical frequency $\om_{\rm max}$ above which the emission spectrum vanishes is about 15 times larger than $T_H$ for $M_+ = 5$. We also found that the spontaneous production of phonons involving a co-propagating mode (which is not related to the standard Hawking effect) is subdominant. We finally studied the behavior of $T_{\rm step}$, the low-frequency temperature computed in step-like flows which possess the same asymptotic properties as the waterfall solutions. We expect that $T_{\rm step}$ gives a reliable estimate of the maximal temperature in flows with a monotonic $\rho(x)$ and we found that it is about twice the temperature in waterfall flows. When compared to the experimental data of~\cite{Steinhauer:2015saa}, the observed temperature is about $20\%$ higher than $T_{\rm step}$ when using $M_+ = 5$ and $40 \%$ higher when using $M_+ = 4$, which is closer to the reported value of $M_+$. 

In Section~\ref{corr}, we studied the strength and phases of the correlations between phonons and their partners emitted on the other side of the sonic horizon. In agreement with the above study of spectral properties, we found that the frequency dependence of the strength of the correlations is well approximated by its relativistic expression. We showed that the correlation strength is hardly affected when attributing an initial temperature to these phonons, as can be understood from the smallness of the coupling terms involving co-propagating mode. We then studied the pattern of equal-time correlations in the dispersionless limit, and we showed that their profile has a narrow width, of the order of two healing lengths evaluated at the horizon. This is a consequence of the above noticed fact that the surface gravity $\kappa_H$ is larger than the dispersive scale measured on the horizon. We also showed that that this width hardly changes when varying $M_+$ from $3.75$ to $6.25$. We finally noticed that it is about twice the value reported in~\cite{Steinhauer:2015saa}. At present, together with the intensity of the correlations, which is about half the reported value, this is the largest discrepancy between the observed properties and the predictions we draw by studying phonon propagation over waterfall flows. We hope that the present analysis can help sorting out these questions and be used in forthcoming experimental works.
 
We also studied the phase of the product of scattering coefficients which enters in the long distance correlation pattern. We found that there is a non trivial, almost linear, dependence in $\om$ which induces a finite shift of the location of the equal-time correlations in the $(x,x')$ plane. When considered in the waterfall solution with $M_+ = 5$, we showed that it is a significant fraction of the typical horizon width $c_H/\kappa_H$. We also showed that this shift persists when taking the dispersionless limit, and studied its behavior in various background flows in Appendix~\ref{App:shift}. We hope it could be measured in the near future. In Appendix~\ref{App:phase} we briefly studied the phase of individual scattering coefficients as functions of the frequency. It would be interesting to measure them in analog gravity experiments where one can work at fixed frequency, as is the case when studying water waves in flumes. 

Finally, in Appendix~\ref{App:t-dep}, we studied the time-dependent modifications of the density correlations which are induced by the formation of a sonic horizon. We worked in a simple relativistic model to characterize in analytical terms both the growth of the non-local correlations and the modifications of the auto-correlations that were so far overlooked.  

\acknowledgements
We thank Jeff Steinhauer for many interesting discussions and feedback. We also thank Alessandro Fabbri, Stefano Liberati and Scott Robertson for useful comments. 
This work was supported by the French National Research Agency by the Grants No. ANR-11-IDEX-0003-02 and ANR-15-CE30-0017-04 associated respectively with the projects QEAGE and HARALAB. 

\appendix
 
\section{Non-polynomial Scr\"odinger equation}
\label{App:NPGPE}

In this appendix we study the waterfall solutions and the phonon spectrum using the non-polynomial Schr\"odinger equation (NPSE)~\cite{2002PhRvA..65d3614S} coming from integration of the three-dimensional GPE over the two orthogonal directions in a harmonic trap. 
The results are then compared with those of the main text based on \eq{eq:GPE}. Before doing the explicit calculation, it is useful to keep in mind the expected order of magnitude of the deviations. The one-dimensional GPE corresponds to the leading order in an expansion of the NPSE in the non-dimensional parameter $a_s \rho$, where $a_s$ is the scattering length of the atoms and $\rho$ their one-dimensional number density. In the experiment of~\cite{Steinhauer:2015saa}, the condensate is made of $^{87}{\rm Rb}$ atoms with $a_s \approx 5 \times 10^{-9} \, {\rm m}$. On the other hand, the maximum value of $\rho$ reported in~\cite{Steinhauer:2015saa} in the region used for analyzing the data is close to $2 \times 10^7 \, {\rm m^{-1}}$. The maximum value of $a_s \rho$ is thus close to $0.1$, indicating that the one-dimensional GPE should be a relatively good approximation, although some corrections from the next orders in $a_s \rho$ could be visible. In the following subsections, we first explain how the knowledge of the two asymptotic densities and of the Mach number on one side fully determines the waterfall solution. We then compute the effective temperature of \eq{eq:Teff} and relative phase of the coefficients $\alpha_\om$ and $\beta_\om$, and we compare them with results derived in the main text. 

\subsection{Waterfall solutions}

The NPSE may be written as~\cite{2002PhRvA..65d3614S,Tettamanti:2016ntx}
\begin{equation}\label{eq:dimNPSE}
i \hbar \pd_t f = \lp -\frac{\hbar^2}{2m} \pd_x^2 + V(x) + \frac{g_{\rm 3D} N}{2 \pi a_\perp^2} \frac{\abs{f^2}}{\eta} \rp f + \frac{\hbar \om_{\perp}}{2} \lp {\eta + \frac{1}{\eta}} \rp f,
\end{equation}
where $\eta \equiv \sqrt{1+2 a_s N \abs{f^2}}$, and $g_{\rm 3D}$ is the three-dimensional coupling. (The constant $g$ used in \eq{eq:GPE} is given by $g = {g_{\rm 3D}/{2 \pi a_\perp^2}}$.)  In this expression, $f$ is the longitudinal part of the condensate wave function, $N$ the number of atoms, $a_s$ the scattering length, $\om_\perp$ the transverse frequency of the trap (assumed to take the same value in the two transverse directions), and $a_\perp \equiv \sqrt{\hbar / (m \om_\perp)}$. 
Notice that the lengths $a_s$ and $a_\perp$ give two additional scales with respect to the one-dimensional GPE. As a result, when working with dimensionless quantities, waterfall solutions are described by three independent parameters instead of one in the case studied in the main text. 

It is useful to define the dimensionless quantities $\psi \equiv \sqrt{2 N a_s} f$, $\bar{t} \equiv \om_\perp t$, $\bar{x} \equiv x / a_\perp$, $\bar{V} \equiv V / (\hbar \om_\perp)$, and $\bar{g}_{\rm 3D} \equiv g_{\rm 3D} / \lp 4 \pi \hbar \om_\perp a_\perp^2 a_s \rp$. Since we will work only with these variables, we shall remove the bars in the following. The NPSE then becomes
\begin{equation}\label{eq:NPSE}
i \pd_t \psi = \lp - \frac{1}{2} \pd_x^2 + V(x) + \frac{1}{2} \lp \eta + \frac{1}{\eta} \rp + g_{\rm 3D} \frac{\abs{\psi^2}}{\eta} \rp \psi,
\end{equation}
where $\eta = \sqrt{1+\abs{\psi^2}}$. As was done in the main text, we assume $g_{\rm 3D}>0$, and 
we look for stationary solutions of the form
\begin{equation}
\psi(x,t) = e^{-i \om t} \sqrt{\rho(x)} e^{i \int^x v(y) dy},
\end{equation}
where $\rho$ and $v$ are two real functions, and $\om \in \mathbb{R}$. Taking the imaginary part of \eq{eq:NPSE} gives $J \equiv \rho v = Cst$. 
The real part of \eq{eq:NPSE} then gives
\begin{equation}\label{eq:NPSE_r}
-\frac{1}{2\sqrt{\rho}} \pd_x^2\sqrt{\rho} + V_{\rm eff}(\rho,x) - \om = 0,
\end{equation}
where the effective potential $V_{\rm eff}$ is
\begin{equation}
V_{\rm eff} (\rho,x) = \frac{J^2}{2 \rho^2} + V(x) + \frac{1}{2} \lp \eta + \frac{1}{\eta} \rp + g_{\rm 3D} \frac{\rho}{\eta}.
\end{equation}
When considering a region of homogeneous potential $V$, the possible homogeneous solutions are given by $V_{\rm eff} (\rho) - \om = 0$. To determine the number and properties of these solutions, we compute
\begin{equation}\label{eq:derVeff}
\frac{\partial V_{\rm eff}}{\partial \rho} = \lp 1+\rho \rp^{-3/2}  \lp \frac{\rho}{4} + g_{\rm 3D} \lp 1 + \frac{\rho}{2} \rp \rp - \frac{J^2}{\rho^3}.
\end{equation}
After multiplication by the (strictly positive for $\rho > 0$) factor $\rho^3$, the right-hand side of \eq{eq:derVeff} is a monotonically increasing function of $\rho$, which is negative for $\rho \to 0^+$ and changes sign at a value $\rho_c > 0$ of $\rho$. So, $V_{\rm eff}(\rho)$ is a monotonically decreasing function of $\rho$ for $0 < \rho \leq \rho_c$ and an increasing function for $\rho > \rho_c$. Moreover, $V_{\rm eff} \to \infty$ in the two limits $\rho \to 0^+$ and $\rho \to \infty$. The existence of homogeneous (or solitonic) solutions thus requires $\om \geq V_{\rm eff} (\rho_c)$. For $\om > V_{\rm eff} (\rho_c)$ there are two homogeneous solutions: a supersonic one with density $\rho_p < \rho_c$ and a subsonic one with density $\rho_b > \rho_c$ (their super- and subsonic characters are proven in subsection~\ref{ssec:NPGPE_lin}). 

To characterize the soliton solutions, one can integrate once \eq{eq:NPSE_r} after multiplication by $\sqrt{\rho} \pd_x \sqrt{\rho}$. This gives
\begin{equation}\label{eq:intNPGPE}
-\frac{1}{4} \lp \pd_x \sqrt{\rho} \rp^2 - \frac{J^2}{4 \rho} +\frac{V - \omega}{2}\rho+\frac{\sqrt{1+\rho}}{2} +\frac{1}{6} \lp 1+\rho \rp^{3/2} + g_{\rm 3D} \rho \sqrt{1+\rho} - \frac{2}{3}g_{\rm 3D} \lp 1+\rho \rp^{3/2} + C = 0,
\end{equation}
where $C$ is an integration constant. The soliton solution is then obtained by choosing $C$ such that \eq{eq:intNPGPE} be satisfied for $\rho = \rho_b$ and $\pd_x \rho = 0$. The bottom of the soliton is given by the largest root $\rho_s$ of the left-hand side of \eq{eq:intNPGPE} for $\pd_x \rho = 0$ in the interval $\left]0,\rho_b\right[$. 
Its existence is guaranteed by the facts that $V_{\rm eff}'(\rho_b) > 0$ and that the left-hand side of \eq{eq:intNPGPE} goes to $-\infty$ for $\rho \to 0^+$. 

We can now look for waterfall solutions in a step-like potential given by \eq{eq:Vstep}. That is, we look for a half-soliton for $x<0$ matched with a homogeneous supersonic solution for $x>0$.  
In our non-dimensional system of units, this leaves three free parameters: $g_{\rm 3D}$, $J$, and $V_- - \om$. 
The value of $V_+$ is then fixed by imposing that $\rho_{p,+} = \rho_{s,-}$. (As in the main text, a subscript ``$-$'' (respectively ``$+$'') denotes a quantity evaluated in the left (resp. right) region.) 
\begin{figure}
\includegraphics[width=0.49 \linewidth]{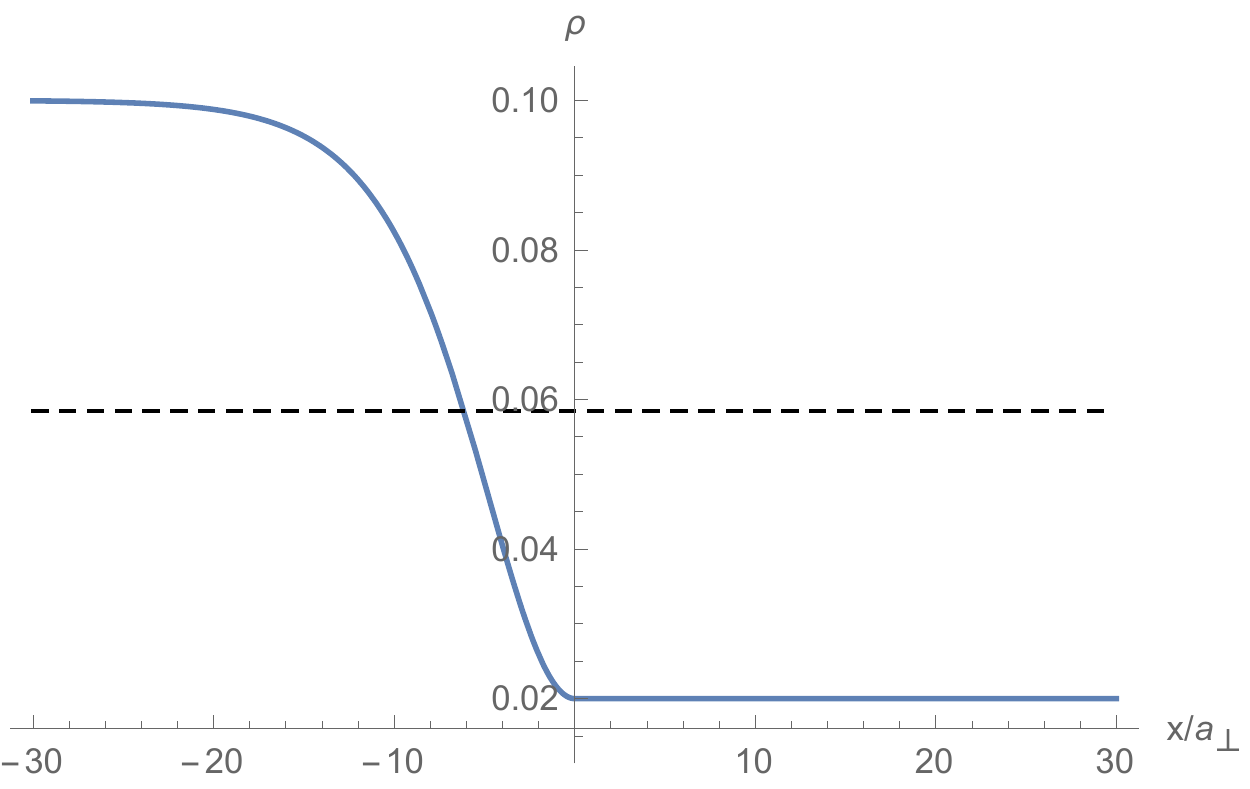} 
\includegraphics[width=0.49 \linewidth]{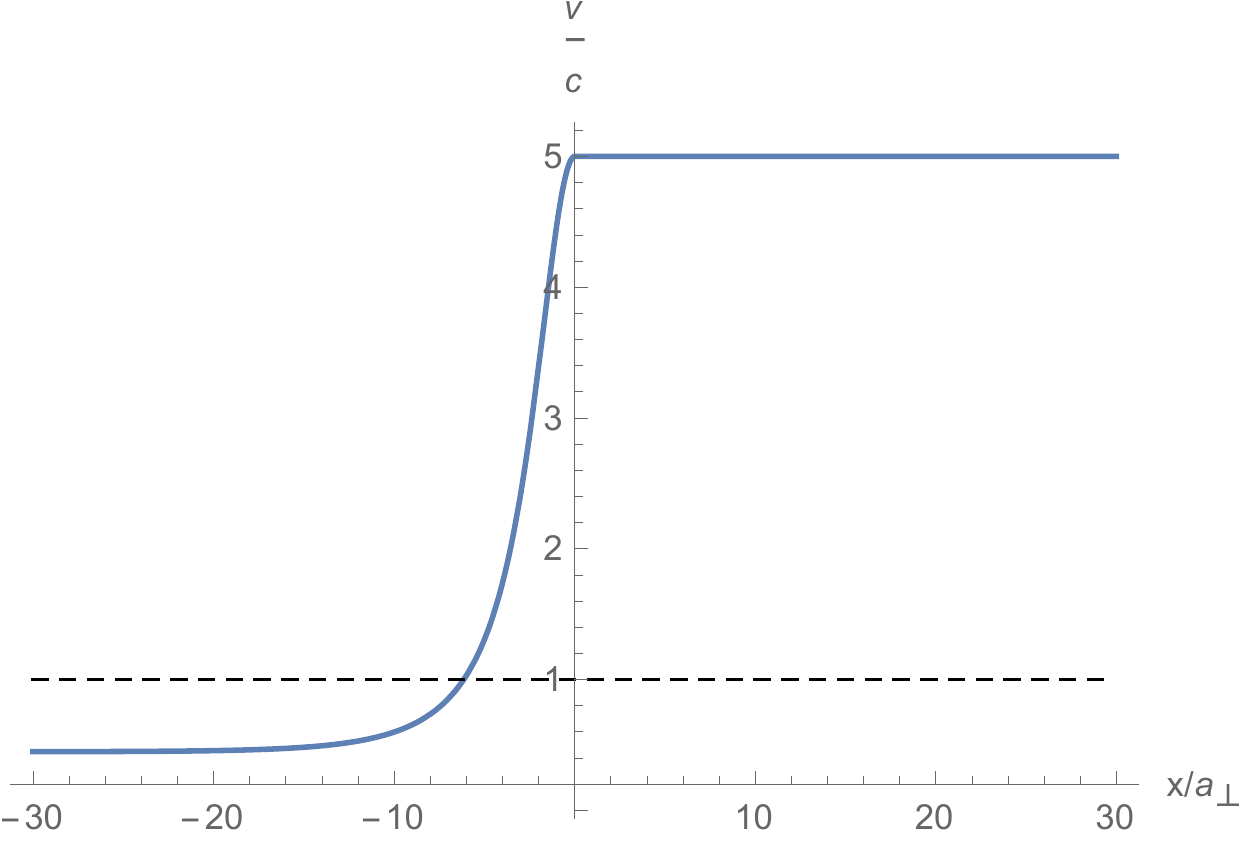}
\caption{Plots of the local density adimensionalized by $a_s$ (left) and Mach number (right) for the waterfall solution of the NPS \eq{eq:NPSE} in a step-like potential. The asymptotic densities $\rho_\pm$ and the value of $M_+$ are chosen to model the flow of~\cite{Steinhauer:2015saa}, see text. The horizontal dashed line shows the density at the horizon (left) and $M=1$ (right). The unit of the 
coordinate $x$ is $a_\perp$. 
}\label{fig:NPGPEsol}
\end{figure}

To determine the three-dimensional waterfall solution which matches what has been observed in~\cite{Steinhauer:2015saa}, it is appropriate to use the values of the asymptotic density on each side and the asymptotic Mach number on one side. These fix the values of the three free parameters, and thus the whole solution. From the inset of Fig.~1.b in~\cite{Steinhauer:2015saa}, we find (in our non-dimensional units) $\rho_{b,-} \approx 0.1$, $\rho_{s,-} = \rho_{p,+} \approx 0.02$. To see the modifications brought in by the NPSE with respect to the results of the main text, we work with $M_+ = 5$.  
The corresponding flow is shown in \fig{fig:NPGPEsol}. 
To estimate the difference with the flow obtained using the one-dimensional GPE, we consider the quantity 
\begin{equation} 
\chi \equiv x_H \lp \frac{dM}{d x} \rp_{x=x_H},
\end{equation}
where $x_H$ is the position of the sonic horizon relative to that of the potential step. $\chi$ is proportional to the surface gravity and has no dimension; hence it is insensitive to the adimensionalization procedure.
(It is also independent of the scale $\lambda$ of \eq{lresc} and thus can be used to directly compare the flows obtained with the two equations.) For the solution shown in \fig{fig:NPGPEsol}, we obtain $\chi \approx -1.26$. By comparison, for the flow of \fig{fig:bckgd} with $M_+ = 5$ we obtain $\chi \approx - 1.12$. The relative difference is of the order of $12\%$, which is close to the maximum value of $\eta^2 - 1 \approx 0.1$ (reached in the subsonic region). As a last remark, we note that $c^2 v$ is nearly uniform, varying by less than $1 \%$ between the two asymptotic regions. We found larger variations when changing the asymptotic parameters by $\approx 10 \%$, but never more than $15 \%$. This suggests that the large difference between the asymptotic values of $c^2 v$ observed in~\cite{Steinhauer:2015saa} is not only due to the three-dimensional nature of the flow. (However, such differences could be reached with larger asymptotic densities.) 

\subsection{Equation on linear perturbations}
\label{ssec:NPGPE_lin}

We now look for perturbed solutions of the form 
\begin{equation}
\psi(x,t) = \psi_0(x,t) \lp 1 + \phi(x,t) \rp,
\end{equation}
where $\psi_0$ is a known stationary solution with angular frequency $\om_0$, local density $\rho_0 \equiv \abs{\psi_0^2}$, and velocity $v_0 \equiv \Im \lp \lp\pd_x \psi_0 \rp / \psi_0 \rp$. To first order in $\phi$, \eq{eq:NPSE} becomes
\begin{equation}\label{eq:linNPSE}
i \lp \pd_t + v_0 \pd_x \rp \phi = -\frac{1}{2 \rho_0} \pd_x \lp \rho_0 \pd_x \phi \rp + \frac{d}{d \rho_0} \lp V_{\rm eff}(\rho_0) - \frac{J^2}{2 \rho_0^2} \rp \rho_0 \lp \phi + \phi^* \rp.
\end{equation}
When the background flow is homogeneous, one can find a basis of solutions of the form
\begin{equation}
\phi(x,t) = U_k e^{-i \om t + i k z} + V^*_k e^{i \om^* t - i k^* z}, 
\end{equation}
where $\lp U_k,V_k,\om,k \rp \in \mathbb{C}^4$. 
The angular frequency $\om$ and wave vector $k$ are related by the dispersion relation
\begin{equation}\label{eq:drNPSE}
\lp \om - v_0 k \rp^2 = \frac{d}{d \rho_0} \lp V_{\rm eff}(\rho_0) - \frac{J^2}{2 \rho_0^2} \rp \rho_0 k^2 + \frac{k^4}{4}.
\end{equation}
The sound velocity $c_0$ is thus related to the background flow velocity $v_0$ through
\begin{equation}\label{eq:candvNPSE}
c_0^2 = V_{\rm eff}'(\rho_0) \rho_0 + v_0^2 
= g_{\rm 3D} \rho_0 + \lp \frac{1}{4} - g_{\rm 3D} \rp \rho_0^2 + O \lp \rho_0^3 \rp. 
\end{equation}
As expected, the first deviations with respect to the one-dimensional expression are governed by 
the density $\rho$ in units of $a_s$. One verifies that $c_0^2- v_0^2 = V_{\rm eff}' \rho_0 $ is positive for $\rho_0 = \rho_b$ and negative for $\rho_0 = \rho_p$, showing that the former is subsonic while the latter is supersonic. 
Apart from the new expression of the sound velocity, \eq{eq:linNPSE} is identical to the BdG equation \eq{eq:BdG}. In particular, it has the same conserved inner product and mode structure. 

\begin{figure}
\includegraphics[width=0.49\linewidth]{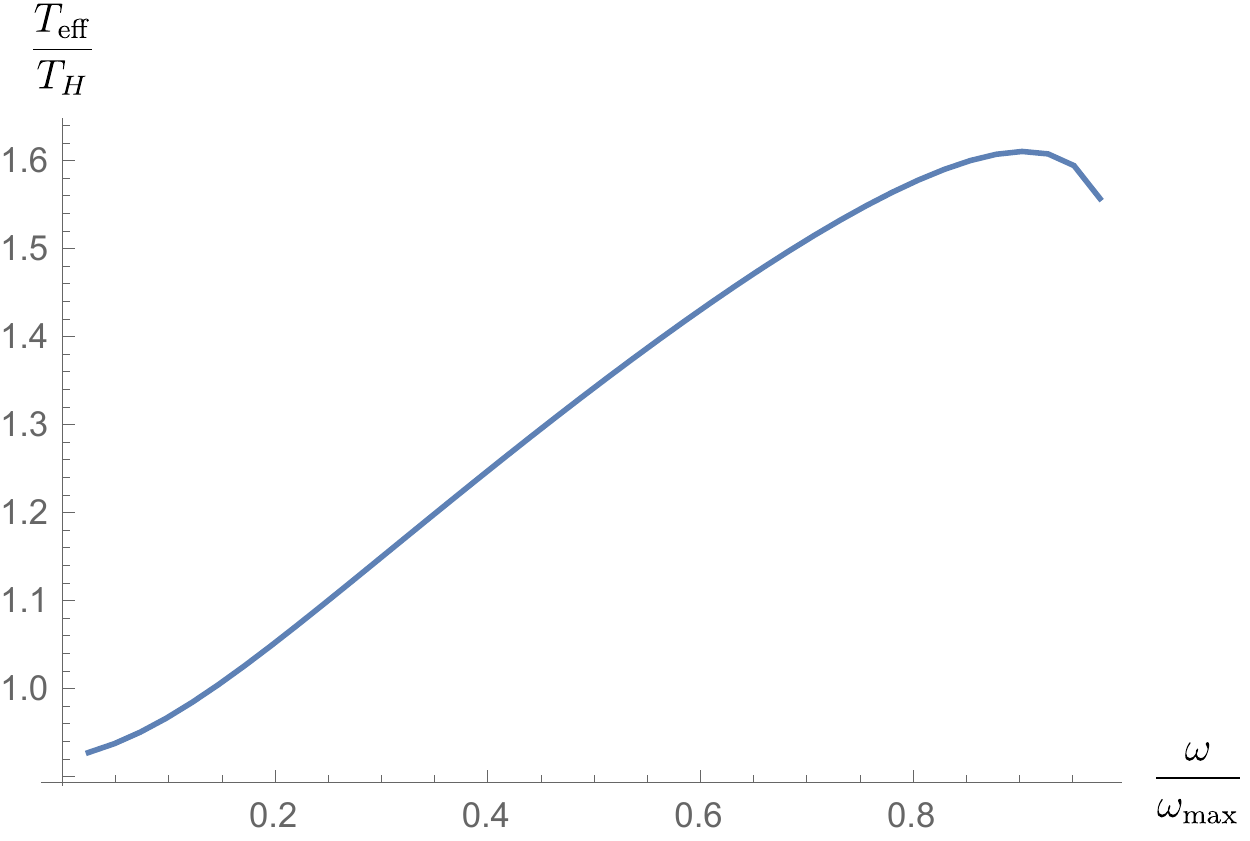}
\includegraphics[width=0.49\linewidth]{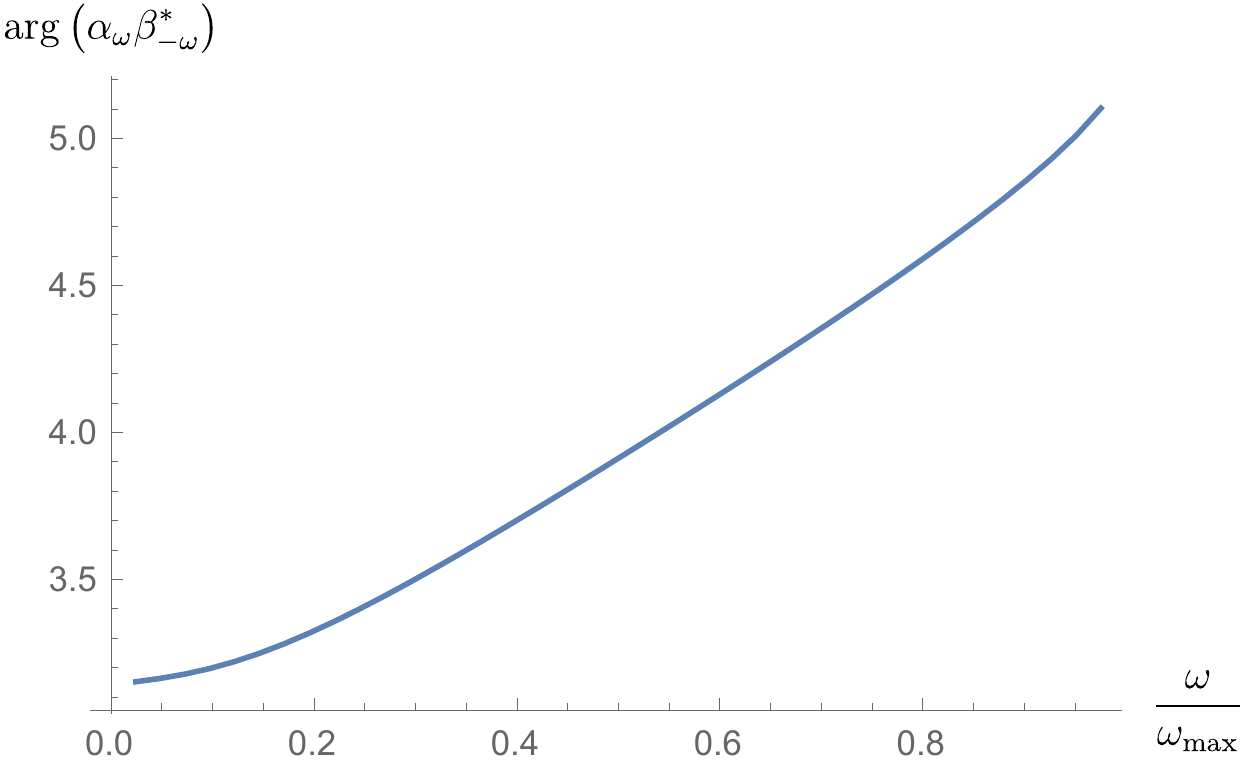}
\caption{Effective temperature (left, in units of $T_H$) and relative phase of the coefficients $\alpha_\om$ and $\beta_{-\om}$ (right) as functions of $\om / \om_{\rm max}$ for the waterfall solution shown in \fig{fig:NPGPEsol}. There are no significant differences with respect to the results obtained in the main text, see \fig{fig:Teff} right panel and \fig{fig:cartphaseandtorel}. 
}\label{fig:NPGPEspec}
\end{figure}
To characterize the deviations of the spectrum with respect to the GPE, we first compute $T_H / \om_{\rm max}$. Using the solution shown in \fig{fig:NPGPEsol}, we obtain $T_H / \om_{\rm max} \approx 0.065$. By comparison, for the flow corresponding to the blue curve in \fig{fig:bckgd}, $T_H / \om_{\rm max} \approx 0.063$, i.e., smaller by only $3 \%$. The effective temperature of \eq{eq:Teff} evaluated in the flow of \fig{fig:NPGPEsol} is shown in the left panel of \fig{fig:NPGPEspec}. We observe that its behavior closely resembles that of \fig{fig:Teff}, with maximum relative deviations of a few $\%$. The relative phase of $\beta_{-\om}$ and $\alpha_\om$, shown in the right panel, is also close to that of \fig{fig:cartphaseandtorel}, with maximum deviations close to $0.1 {\rm rad}$. 
In conclusion, although we worked here with the experimental values of the asymptotic densities~\cite{Steinhauer:2015saa}, we observed no significant deviation with respect to the results obtained when working with one-dimensional waterfall solutions. 
Even when using asymptotic values of the density twice larger than those reported in~\cite{Steinhauer:2015saa}, we still find that the deviations are smaller than $10 \%$.

\section{Stability of the waterfall solutions}
\label{App:NLstab}

In this appendix we report on numerical results confirming the stability of the waterfall solutions. We first note that the linear analysis of~\cite{Michel:2015mlr}, done for the homogeneous black hole solutions with step-like profiles for both $g$ and $V$, remains valid in our present setup as it only relies on the behavior of the scattering coefficients near $\om \to 0$ and $\om \to \om_{\rm max}$, which is the same for the homogeneous configurations considered in that reference and the waterfall ones studied here. To linear order, density perturbations thus decay polynomially in time, with an exponent equal to $3/2$. The numerical results shown below confirm that this behavior persists when considering finite perturbations.

We solved the time-dependent GPE~\eqref{eq:ndGPE} starting from a perturbed waterfall configuration at $t=0$, on a torus of radius much larger than the healing length and length scales of the initial perturbations. Explicitly, the density at $t=0$ is
\begin{equation}\label{eq:rhowat} 
\rho(x,t=0) = 
\rho_+ \left[
\lp M_+ + \frac{1-M_+}{\cosh \lp \sigma x \rp^2} + \frac{1-M_+}{\cosh \lp \sigma \lp x + x_{\rm max} \rp \rp^2} \rp \frac{1 - \tanh \lp x / \epsilon \rp}{2} + \frac{1 + \tanh \lp x / \epsilon \rp}{2} \right]
+ \delta \rho(x,0) 
\end{equation} 
where $\epsilon$ is a regulator of the order of the step of the uniform spatial grid, $x_{\rm max}$ is half the length of the integration domain (centred on $x=0$), $\delta \rho(x,t)$ is the density perturbation, and $\sigma = \sqrt{M_+ - 1} / \xi_+$. The phase $\theta \equiv \arg \psi$ is
\begin{eqnarray}\label{eq:thetawat} 
\theta(x, t=0) &=& 
\lp M_- \frac{x}{\xi_-} + \frac{M_+^{1/2}-\sqrt{M_+-1}}{\sqrt{2M_+-1-2\sqrt{M_+^2-M_+}}} \arctan \lp \frac{\tanh(\sigma x / 2)}{\sqrt{2M_+-1-2\sqrt{M_+^2-M_+}}} \rp \right. \nn
&& \left. - \frac{M_+^{1/2}+\sqrt{M_+-1}}{\sqrt{2M_+-1+2\sqrt{M_+^2-M_+}}} \arctan \lp \frac{\tanh(\sigma x / 2)}{\sqrt{2M_+-1+2\sqrt{M_+^2-M_+}}} \rp \right. \nn
&& \left. + \frac{M_+^{1/2}-\sqrt{M_+-1}}{\sqrt{2M_+-1-2\sqrt{M_+^2-M_+}}} \arctan \lp \frac{\tanh(\sigma (x+x_{\rm max}) / 2)}{\sqrt{2M_+-1-2\sqrt{M_+^2-M_+}}} \rp \right. \nn
&& \left. - \frac{M_+^{1/2}+\sqrt{M_+-1}}{\sqrt{2M_+-1+2\sqrt{M_+^2-M_+}}} \arctan \lp \frac{\tanh(\sigma (x+x_{\rm max}) / 2)}{\sqrt{2M_+-1+2\sqrt{M_+^2-M_+}}} \rp \rp \frac{1 - \tanh \lp x / \epsilon \rp}{2}  \nn
&& + M_+ \frac{x}{\xi_+} \frac{1 + \tanh \lp x / \epsilon \rp}{2} + \delta \theta(x,0). 
\end{eqnarray}
In \eqs{eq:rhowat}{eq:thetawat}, the terms in $x+x_{\rm max}$ are added to implement the periodic boundary conditions; the value of $x_{\rm max}$ is chosen so that $\theta(x_{\rm max}) - \theta(-x_{\rm max})$ is sufficiently close to an integer multiple of $2 \pi$ to avoid large perturbations originating from $x = x_{\rm max}$. The configuration is thus nearly stationary for $\delta \rho = \delta \theta = 0$, and contains a black hole horizon close to $x=0$ and a white hole horizon close to $x = x_{\rm max}$. 

\begin{figure}
\includegraphics[width=0.49 \linewidth]{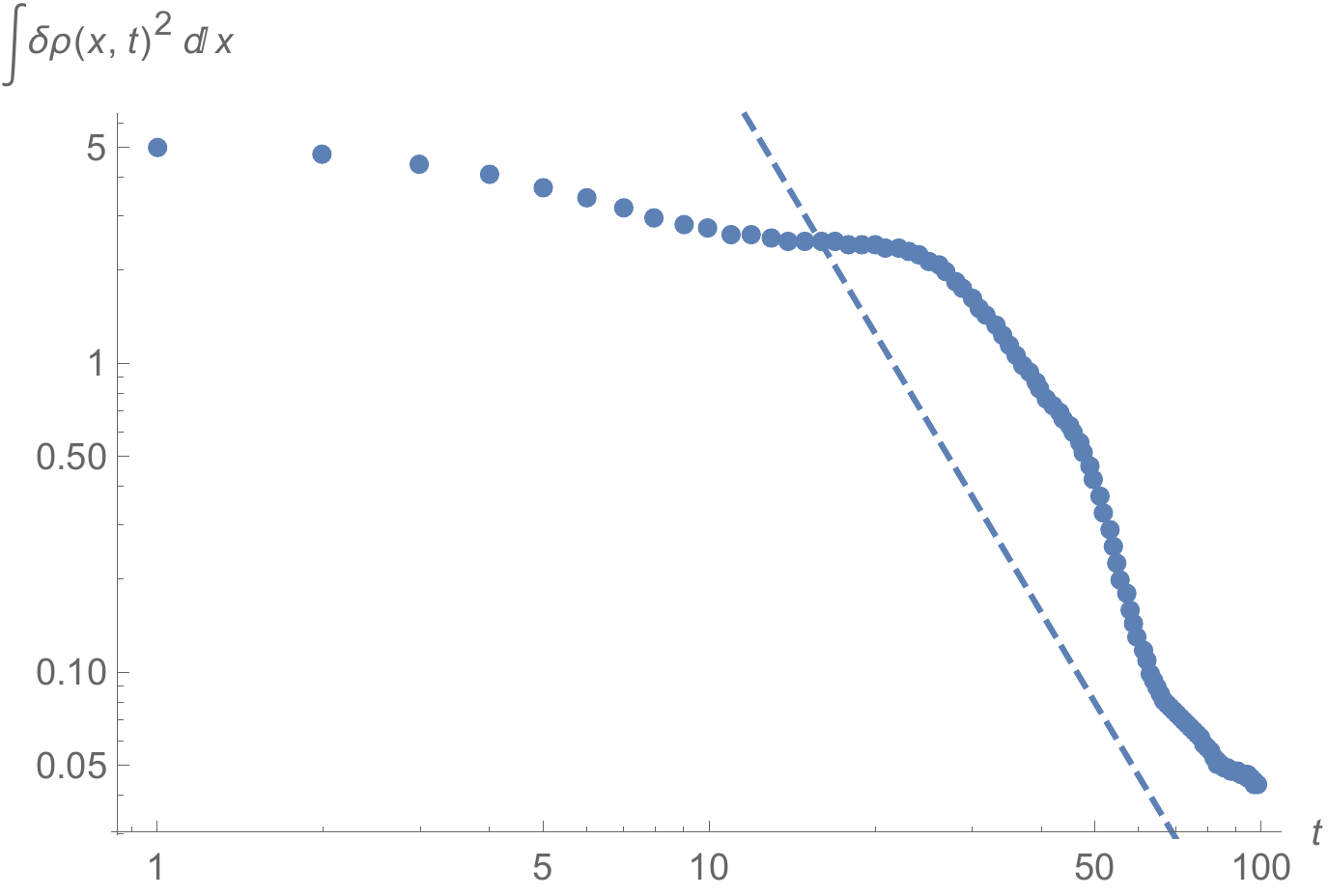}
\includegraphics[width=0.49 \linewidth]{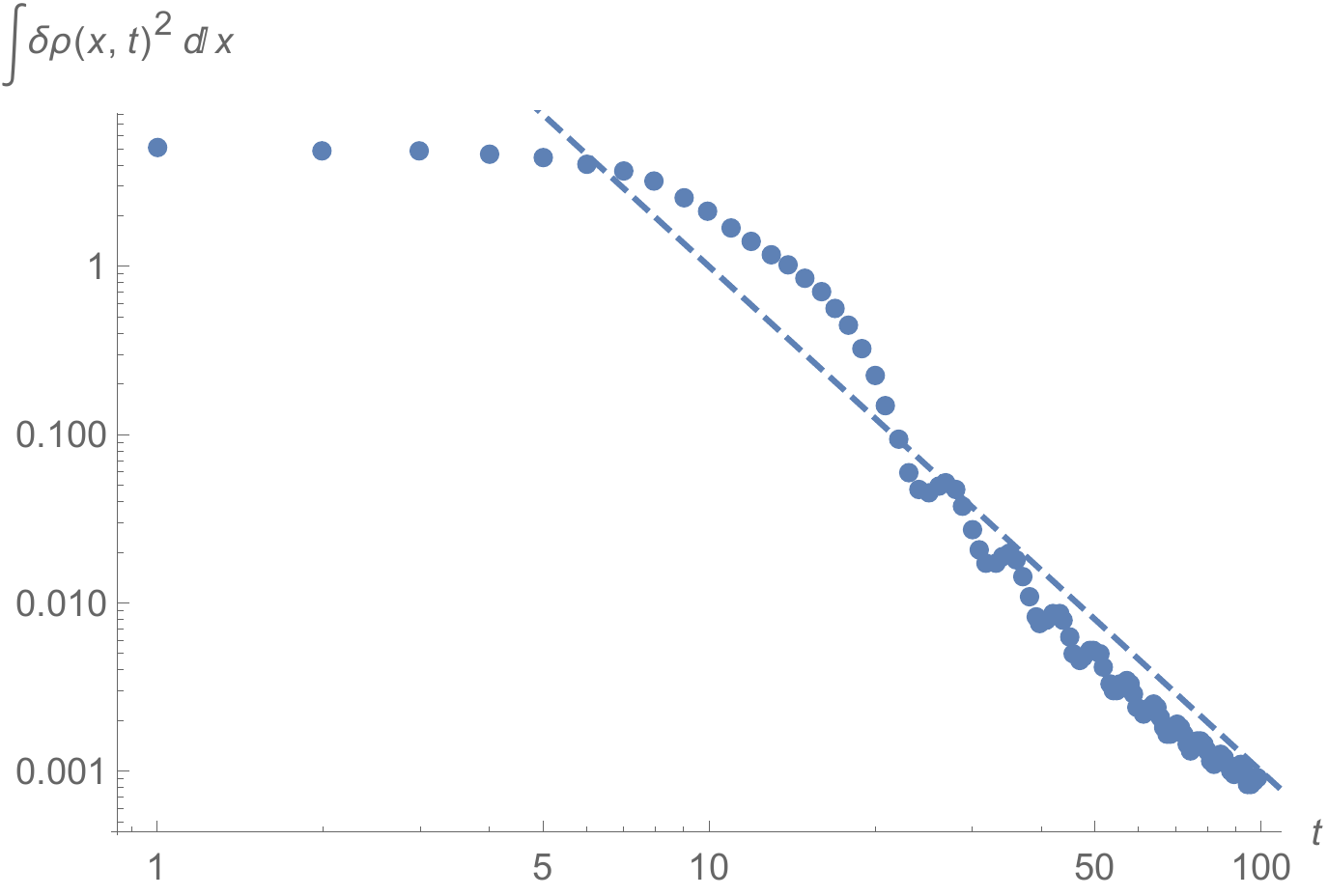}
\caption{Integrated squared density perturbation over a waterfall solution as a function of time. The background solution is the waterfall with $M_+ = 5$. The initial perturbation has the form \eq{eq:pert_rho} with $A = 1$, $\chi = 4$, and $x_p = -20$ (left) and $20$ (right). For these simulations, $x_{\rm max} \approx 249.5$ and the squared density perturbation is integrated between $-x_{\rm max}/8$ and $x_{\rm max}/8$. The spatial grid is made of $12800$ uniformly spaced points; the time step is $0.0025$. The oblique dashed lines show $10^4 t^{-3}$ (left) and $10^3 t^{-3}$ (right). (In this plot, the scale $\lambda$ of \eq{lresc} is fixed so that $\rho_+ = 1$.) 
}\label{fig:stab}
\end{figure}

The numerical integration uses a modified version of the code of~\cite{Michel:2015pra}, 
written in {\it Mathematica}~\cite{Mathematica} with a dissipative term linear in $\delta \rho(x,t) \lp \frac{\abs{x}}{x_{\rm max}} - \frac{1}{2} \rp$ added in the region $\abs{x} > x_{\rm max}/2$ to suppress the perturbations coming back to the black hole horizon after making a full turn. We verified that the residual waves going to the white hole horizon and back to the black hole one, as well as the part of the perturbation reflected around $x = \pm x_{\rm max} / 2$ because of the dissipation term, are small enough not to affect the results.

To estimate the evolution of the perturbations, we compute the integral of the squared density perturbation $\delta \rho^2$ over the interval $\left[ -x_{\rm max}/8 , x_{\rm max}/8 \right]$ as a function of time. In \fig{fig:stab} we show results for $M_+ = 5$ and an initial perturbation of the form
\begin{equation}\label{eq:pert_rho}
\delta \rho(x,0) = A \exp \lp - \lp x - x_p \rp^2/\chi^2 \rp, \; 
\delta \theta(x,0) = 0. 
\end{equation}
As can be seen in the figure, the integrated squared density perturbation decays as $t^{-3}$, in accordance with the linear theory. We verified that this behavior remains for $M_+ = 1.2, \, 2, \, \text{and} \, 3.5$, as well as for similar perturbations on the phase. Although a systematic study of the domain of stability of the waterfall solutions is beyond the scope of the present
work, our results indicate that they are stable both at a linear level and under finite initial perturbations.

\section{Dispersionless settings}

\label{App:DLS}

In this Appendix we study the propagation of a dispersionless, massless field in transcritical flows. 
Our first aim is to obtain a generalization of the asymptotic expressions of the density correlations of \eq{eq:relG2} and \eq{eq:relG3} which is valid in the near horizon region where $v$ and $c$ rapidly vary. 
This generalized expression will allow us to describe the gradual change of the correlations in this region, see \fig{F_correlations}, and the modifications of the correlations induced by the formation of a sonic horizon. 
Our second aim is to prepare the calculation of the shift of \eq{xM} which is done in the next Appendix.

\subsection{Generalized expression for \texorpdfstring{$G_2^{\rm rel}$}{G2}} 
\label{sub:gen}

Starting from \eq{eq:BdG_W}, we perform two simplifications. We first send the healing length to $0$, as was done in \eq{eq:resc_BdG} when sending $\gamma$ to $0$. We also reorder the derivatives $\pd_x$ and $x$-dependent factors to obtain the wave equation of a relativistic massless scalar $\phi(t,x)$. Then, because of conformal invariance, there is a complete decoupling of the $v$ (copropagating) sector from the $u$ sector which describes waves counterpropagating with respect to the background flow. As a result, the two-point function $G_2^{\rm rel}$ is a sum of terms $G_{2, v} + G_{2, u}$ encoding each contribution separately.

This decoupling between the $v$ and $u$ sectors can be easily understood by considering the acoustic metric associated with the background flow~\cite{Unruh:1980cg,Unruh:1994je}, and by rewriting it in terms of the light-like coordinates $V$ and $u$:
\begin{equation} \label{eq:ds}
ds^2 = -c^2(x) dt^2 + (dx - v(x) dt)^2 = - (c^2(x) - v^2(x)) dV du ,   
\end{equation}
where 
\begin{equation} \label{eq:du}
du \equiv dt + \frac{dx}{c(x) - v(x)}, \, dV \equiv dt - \frac{dx}{c(x) + v(x)}. 
\end{equation}
(The signs in the definition of our light-like coordinates $u$ and $V$ are chosen so that lines of constant $u$ give the characteristics of the counter-propagating waves for a flow from left to right. 
The differences with the notations of~\cite{Macher:2009nz} come from the fact that our flow velocity is from left to right $v(x)>0$.) 

In these settings, to obtain the Hawking radiation and its associated $uu$-correlation pattern, one should introduce the notion of ``Unruh'' vacuum~\cite{Unruh:1976db}, which is unambiguously defined as the only regular state across the horizon which is stationary with respect to time translations $\partial_t\vert_x$. In this state, at fixed $t$, regularity across the sonic horizon implies that in its near vicinity one has~\cite{Parentani:2010bn}
\begin{equation} \label{eq:Gux}
\tilde G_{2, u}(t,x; t,x') 
\sim \frac{-1}{4\pi} \ln|x - x'| .
\end{equation}
In other words the regularity of the state is expressed as a translation invariance when using the regular coordinate $x$ which is an affine coordinate at fixed $t$, since $ds^2 = dx^2$ at fixed $t$.
When considered globally, the two-point function in Unruh vacuum can be written as 
\begin{equation} \label{eq:GuU}
\tilde G_{2, u}(u; u') = \frac{-1}{4\pi} \ln|U - U'|, 
\end{equation} 
where $U$ is a light-like coordinate which is regular across the horizon at $x = x_H$, 
which means that $\pd_x U$ is continuous at $x_H$, both at fixed $V$ and at fixed $t$. 
Up to an arbitrary scale which plays no role in the physics, $U$ is uniquely defined by the regularity on the horizon and the stationarity of $G_{2, u}$. 

Before expressing $U$ in terms of the coordinate $u$ entering \eq{eq:du}, following~\cite{Balbinot:2007de}, we relate the present formalism to density correlations of phonons in a transonic flow. 
To this end, we first consider 
\begin{equation}\label{eq:Gxx1}
\tilde  G_{2, u}^{(xx)} \lp x,t ; x',t' \rp \equiv \frac{-1} 
{4\pi} \pd_x \pd_{x'} \ln \vert U(t,x) - U(t',x') \vert . 
\end{equation}
The link between this function and the relativistic limit of \eq{G2J} is given by~\footnote{It should be noticed that the ordering of the derivatives $\partial_x$ and the functions $\rho(x)$ and $\xi(x)$ adopted here to get \eq{eq:Gxx2} is different from that given in Eqs.~(5,6) of~\cite{Balbinot:2007de}. In our case, the derivatives $\partial_x$ act on the $\log$, but not on the prefactor $C(x) = 1/\sqrt{{\rho}(x)\xi(x)}$ of their Eq.~(6). We have made this choice in order to avoid the infra-red divergences which occur when acting on $C(x)$ while assuming that the two-point function can be approximated by $\ln[\Delta U \Delta V]$, see~\cite{Anderson:2014jua,Anderson:2015bza}. 
In brief, \eq{eq:Gxx2} can be viewed as a local density approximation. \label{foot:lda}} 
\begin{equation}
G^{\rm rel}_2(x,x') = \sqrt{\frac{\xi_+ \xi_-}{{\rho}_+ {\rho}_-}} \sqrt{{\rho}(x)\xi(x)\, {\rho}(x')\xi(x')} 
\left[ \tilde G_{2, u}^{(xx)} + \tilde G_{2, v}^{(xx)} \right] .
\label{eq:Gxx2} 
\end{equation}

Here $\tilde G_{2, v}^{(xx)}$ encodes the contribution of the $v$-modes. It is given by
\begin{equation}\label{eq:Gxxv2}
\tilde  G_{2, v}^{(xx)} \lp x,t ; x',t' \rp \equiv \frac{-1} 
{4\pi} \pd_x \pd_{x'} \ln \vert V(t,x) - V(t',x') \vert , 
\end{equation}
where $V(x,t)$ obeys \eq{eq:du}. The physics encoded in this choice is clear: it means that the incoming $v$-modes are in their ground state in the asymptotic left region, see \fig{fig:charact}. 

For the contribution of $u$-modes, some extra algebra is needed to relate $U$ entering \eq{eq:GuU} to the two coordinates $u_L,u_R$ which obey $du = dt + dx/(c(x) - v(x))$, and which cover respectively the left and right side of $x_H$. Indeed, due to the divergence of $1/(c-v)$ on the sonic horizon for $x \to x_H$, two $u$ coordinates must be used. When integrating $du = dt + dx/(c(x) - v(x))$, the integration constant can be chosen independently on each side of $x_H$. More precisely, using the fact that close to the horizon, $v - c \approx \kappa_H (x - x_H)$, one gets
\begin{equation}
\label{CRL}
u_{R/L} \mathop{=}_{x \to x_H} t - \frac{1}{\kappa_H} \ln \lp \abs{x - x_H} \rp + C_{R/L} + o(1), 
\end{equation}
where $C_{R/L}$ are two real constants, taking {\it a priori} different values on each side of the horizon. 
It is convenient to adopt the conventional choice $C_L =  C_R$, as it allows to express the regularity of \eq{eq:GuU} in simple terms. 
Indeed, when $C_L = C_R$, one recovers the standard relation~\cite{Brout:1995rd}
\begin{equation}\label{eq:U}
U = 
\begin{cases}
\frac{1}{\kappa_H} e^{-\kappa_H u_R} & x > x_H \\ 
-\frac{1}{\kappa_H} e^{- \kappa_H u_L} & x < x_H 
\end{cases} . 
\end{equation}
Equivalently, starting with these relations, one easily verifies that the continuity of $\pd_x U$ across the horizon is equivalent to the condition $C_L = C_R$ in \eq{CRL}. In the following, we work with $u_R$ and $u_L$ which satisfy this relation. 

We remind the reader that the above exponential relation between $U$, specifying the regular vacuum  state, and $u_L$, which is linearly related to $t$ and $x$ for asymptotic values of $x$ in the subsonic region, encodes the steady production of thermally distributed particles at the Hawking temperature $T_H = \kappa_H /2\pi$. The thermality of the $u$-phonons shows up when considering the equal-time expression of $\tilde G_{2, u}^{(xx)}$ on one side of the horizon, say on the left subsonic region. One obtains 
\begin{equation}
\tilde G_{2, u}^{(xx)} = - \frac{\kappa_H^2}{16\pi} \frac{\pd_x u_L(x) \, \pd_{x'} u_L(x')}{\sinh^2 \lp \frac{\kappa_H}{2} (u_L(x) - u_L(x')) \rp }. 
\label{eq:Gxx3}
\end{equation}
When considered sufficiently far from the horizon that $\pd_x u_L(x)$ reaches its asymptotic value, $\tilde{G}_{2,u}^{(xx)}$ becomes identical to the $u$ contribution of the two-point function in a thermal state at temperature $\kappa_H / 2 \pi$~\cite{Parentani:2010bn}.

Similarly the symmetry under the exchange $U \to - U$, $u_L \to u_R$ encodes the correlations across the horizon between phonons of opposite energy. 
This can be seen by studying wave packets of regular $in$-modes~\cite{Brout:1995wp} or, equivalently, the reduction of the state due to the detection of a localized quantum on one side of the horizon~\cite{Massar:1996tx,Brout:1995rd}.
In the present formalism, these correlations show up when considering $\tilde G_{2, u}^{(xx)}$ at equal time for $x < x_H$ and $x' >  x_H$. In the place of \eq{eq:Gxx3}, one gets
\begin{equation}
\tilde G_{2, u}^{(xx)} = - \frac{\kappa_H^2}{16\pi} \frac{ | \pd_x u_R(x) \, \pd_{x'} u_L(x')| 
}{\cosh^2 \lp \frac{\kappa_H}{2} (u_R(x) - u_L(x')) \rp }. 
\label{eq:Gxx}
\end{equation}
In the asymptotic regions where $v - c$ becomes constant, the locus of the maxima of $G_2^{(xx)}$ gives back the mirror-image relationship between two null characteristics on opposite side of the horizon, expressed here as $u_R=u_L$.  

\subsection{Time-dependent modifications of density correlations} 
\label{App:t-dep} 

We aim to study the modifications of the density correlations which result from the formation of a sonic horizon. 
These have been already studied in numerical simulations based on the Wigner truncated method~\cite{Carusotto:2008ep}. 
Here instead we shall use the dispersionless settings presented above combined with a simple analytical model
to describe the formation of the horizon~\cite{Parentani:2010bn}. The advantages of this method is that we can analytically follow both the development of the long-distance correlations and the replacement of vacuum auto-correlations of $u$-configurations by thermal ones given in \eq{eq:Gxx3}.

The model consists in working with a stationary background flow in which the phonon state at $t= 0$ is imposed to be the instantaneous local vacuum. That is, at $t= 0$, the $u$-contribution of the two-point function is given by  
\begin{equation} \label{eq:Guin}
\tilde G_{2, u}^{\rm inst.\, vac.}(t= 0,x; t= 0,x') =
\frac{-1}{4\pi} \ln|x - x'| , 
\end{equation}
see \eq{eq:Gux}. Because of the two-dimensional conformal invariance, for $t>0$, one gets
\begin{equation} \label{eq:Guit}
\tilde G_{2, u}^{\rm inst.\, vac.}(t,x; t,x') =  
\frac{-1}{4\pi} \ln|X_u(x,t) - X_u(x',t)| , 
\end{equation}
where $X_u(x,t)$ gives the value of $x$ reached at $t=0$ by the null characteristic passing through $x$ at $t$. 
Using $U(x,t)$ of \eq{eq:U}, which is well defined for any regular flow profile given by $v(x)$ and $c(x)$, 
we introduce $\mathcal{U}(x)$ defined by $\mathcal{U}(x) \equiv U(x,0)$. 
Then, by definition of $X_u$, $\mathcal{U}(X_u(x,t)) = U(x,t)$. Using \eq{eq:U}, one finds that $U(x,t) = e^{- \kappa_H t} U(x,0)$, from which 
\begin{equation}
X_u(x,t)=\mathcal{U}^{-1}\left(U(x,t)\right). 
\label{eq:X}
\end{equation}

To obtain simple equations, we work with a background flow given by 
\begin{equation} \label{eq:vcb}
v(x)-c(x)=(v_+-c_+)\frac{e^{\kappa_H x/(v_+-c_+)}-e^{-\kappa_H x/(c_--v_-)}}{e^{\kappa_H x/(v_+-c_+)}+\frac{v_+-c_+}{c_--v_-}e^{-\kappa_H x/(c_--v_-)}},
\end{equation}
where $\kappa_H$ gives the surface gravity. 
The (positive) constants ${v_+-c_+}, {c_--v_-}$ and $\kappa_H$ can be adjusted so as to match the properties of some transcritical flow one wishes to consider. For instance, for the waterfall solution with $M_+ = 5$, one finds $\frac{v_+-c_+}{c_+} \approx 4, \frac{c_--v_-}{c_+} \approx 1.2$, and $\frac{\kappa_H\xi_+}{c_+} \approx 5.1$. 
To be able to compute the contribution of the $v$-modes, the profile needs to be completely fixed, namely $v(x)+ c(x)$ should also be given. Here we chose to work with  
\begin{equation} \label{eq:vcq}
\begin{array}{l}
v(x) + c(x) = 2 c_1+(1- 2q)\left(v(x)-c(x)\right),
\end{array}
\end{equation}
where the constants $c_1$ and $q$ are fixed by the asymptotic values in the background flow. For the waterfall solution with $M_+ = 5$, we get $\frac{c_1}{c_+} \approx  1.9$ and $q \approx 0.24$. 
\begin{figure}
	\centering
	\includegraphics[width=0.8\linewidth]{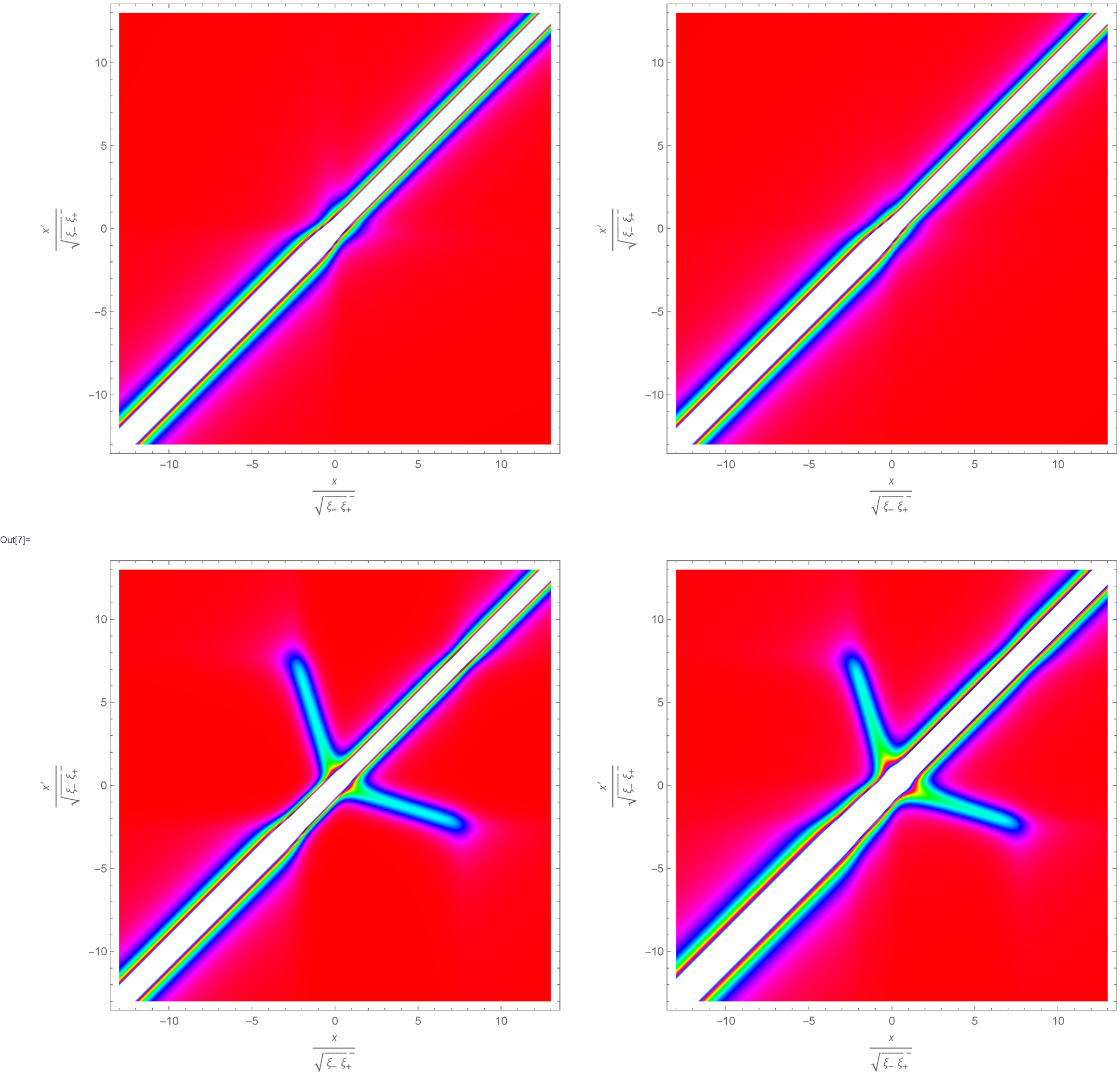} 
	\includegraphics[width=0.15\linewidth]{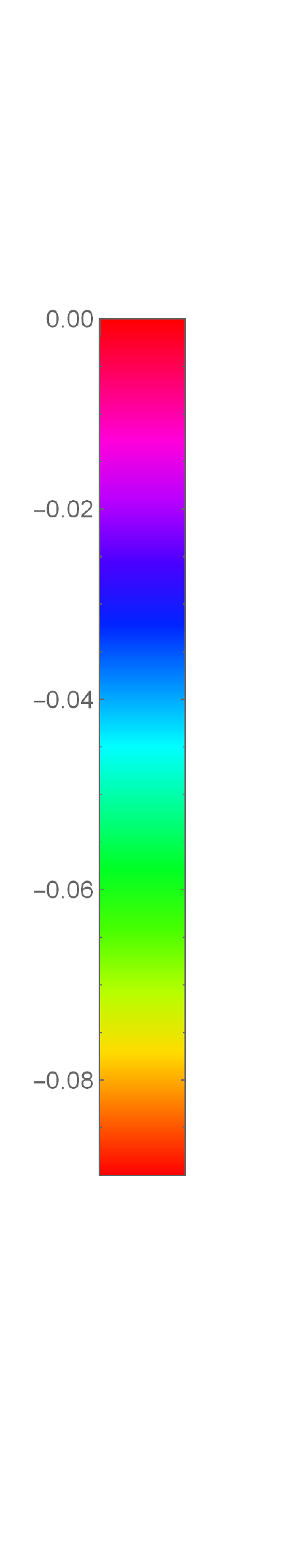} 
	\caption{Plots of equal time correlation functions evaluated in a background flow given by \eq{eq:vcb} and \eq{eq:vcq}. 
In the upper left panel, we show the stationary contribution of $v$-modes given by \eq{eq:Gxxv2}. 
The contribution of $u$-modes given by \eq{eq:Guit} evaluated at $t=0$ is shown on the upper right panel, whereas that evaluated at $\kappa_H t\approx 5.9$ is shown on the lower left panel. The growth of the correlations on opposite sides is the main signal. One should also notice that on both sides, there is a narrowing of the auto-correlations which encodes the thermal radiation emitted from the horizon. On the lower right panel, we show the sum of the $u$ and $v$ contributions evaluated again at $\kappa_Ht\approx 5.9$. \label{F_correlations}
}
\end{figure}

Knowing the profile, we can integrate \eq{eq:du} to find $X_u(t,x)$ and $V(t,x)$ governing the $u$- and $v$-contribution of the two-point function of \eq{eq:Gxx2} in the instantaneous vacuum. 
On the left upper plot of Fig.~\ref{F_correlations}, we represent $G_{2,v}^{\rm rel}$, the equal-time correlations of $v$-modes. 
When using $V(t,x)$ of \eq{eq:du} to fix the initial state of $v$-modes, this term is independent of the time lapse after the formation of the horizon. On the right upper plot, using \eq{eq:X}, we represent the equal-time correlations of $u$-modes just after the formation of the horizon. As expected $G_{2,v}^{\rm rel}$ and $G_{2,u}^{\rm inst.\, vac.}$ evaluated at $t= 0$ are very similar, and contain only a vacuum-like diverging contribution $\sim -\xi^2(x)/(x-x')^2$. 
On the lower left plot of \fig{F_correlations}, we show the equal-time correlations $G_{2,u}^{\rm inst.\, vac.}$ evaluated after a finite time ($\kappa_H t \approx 5.9$), chosen so as to match the observations reported in~\cite{Steinhauer:2015saa}. 
On opposite sides of the horizon, we see the propagation of correlated pairs of $u$-phonons. 
In addition, on both the subsonic and supersonic sides, we see a narrowing of the auto-correlations associated with the replacement of vacuum-like correlations $\sim -\xi^2/(x-x')^2$ by thermal correlations given in \eq{eq:Gxx3}. 
On the lower right plot, we represent the sum of $u$ and $v$ contributions. 
Whereas the non-local correlations are hardly affected, we see that the narrowing of the $u$-correlations is now less visible. 
For $\kappa_H t \to \infty$, one obtains a stationary situation very similar to that represented in the right plot of \fig{fig:G2rel}.

\section{Shift of correlations induced by the flow asymmetry} 
\label{App:shift}

In subsection~\ref{sub:relshift}, we present the general expressions for the shift of \eq{xM} using the settings of the former Appendix. 
In subsection~\ref{sub:tanh_shift} we consider an asymmetric $\tanh$ profile and show that, surprisingly, the shift vanishes. We then turn to two asymmetrical profiles which display a non-trivial shift: a perturbed $\tanh$ profile in~\ref{sub:pert_tanh_shift} and a linear profile in~\ref{sub:lin_shift}. 

\subsection{Generalized expression for the shift} 
\label{sub:relshift}

To compute the shift $x_M$, we first assume that the coordinate $u$, solution of \eq{eq:du}, has the following asymptotic expansion:
\begin{equation} \label{eq:expu}
u(x,t) \mathop{=}_{x \to \pm \infty} t + \frac{x - x_H + b_\pm}{v^{\rm gr}_\pm } + o(1),
\end{equation}
where $v^{\rm gr}_\pm \equiv - c_\pm + v_\pm$ gives the group velocity in the laboratory frame in each asymptotic region, which satisfies $v^{\rm gr}_- < 0$ and  $v^{\rm gr}_+ > 0$. 
In the limit $x \to -\infty$ and $x' \to +\infty$, the maximum of $G_2^{(xx)}$ is located where
\begin{equation}\label{c10}
\frac{x - x_H + b_-} {v^{\rm gr}_-} =  \frac{x' - x_H + b_+}{v^{\rm gr}_+}. 
\end{equation}
To get the point where these asymptotes cross each other, we set $x = x'$, giving the shift
\begin{equation} \label{eq:relshift} 
\Delta x \equiv x_M - x_H = \frac{v_-^{\rm gr} b_+ - v_+^{\rm gr} b_-}{v_+^{\rm gr} - v_-^{\rm gr}}.
\end{equation}
As expected, the correction terms $b_\pm$ in \eq{eq:expu} fix the shift $\Delta x$. 

To complete the calculation, we now relate $\Delta x$ to the profile of $c(x) - v(x)$.
To this end, we use that, for $\epsilon > 0$ and $x > x_H$, and up to a global constant,
\begin{equation}
u_R = t + \int_{x_H + \epsilon}^x \frac{dy}{c(y) - v(y)} + \frac{1}{\kappa_H} \ln \lp \abs{\epsilon} \rp + O(\epsilon),
\end{equation}
while for $x < x_H$,
\begin{equation}
u_L = t + \int_{x_H - \epsilon}^x \frac{dy}{c(y) - v(y)} + \frac{1}{\kappa_H} \ln \lp \abs{\epsilon} \rp + O(\epsilon).
\end{equation}
Using these expressions and the definition of $b_\pm$ \eq{eq:expu}, we obtain
\begin{equation} \label{eq:diffB}
\frac{b_+}{v_+^{\rm gr}} - \frac{b_-}{v_-^{\rm gr}}
= \mathop{\rm lim}_{\epsilon \to 0} \mathop{\rm lim}_{X \to \infty} \lp \int_{\left[x_H - X, x_H + X \right] - \left[ x_H - \epsilon, x_H + \epsilon \right]} \frac{dx}{c(x) - v(x)} - \frac{X}{ v^{\rm gr}_+ } - \frac{X}{v^{\rm gr}_-} \rp.
\end{equation}
The near-horizon contribution is given by a principal value (and vanishes in the case one uses the approximation $c - v = - \kappa_H x$, 
which globally describes de Sitter space). The overall shift is related to the asymmetry in the ways $c - v$ approaches its asymptotic values on the two sides of the horizon. 

\begin{figure}[h]
\begin{center}
\includegraphics[width=0.49 \linewidth]{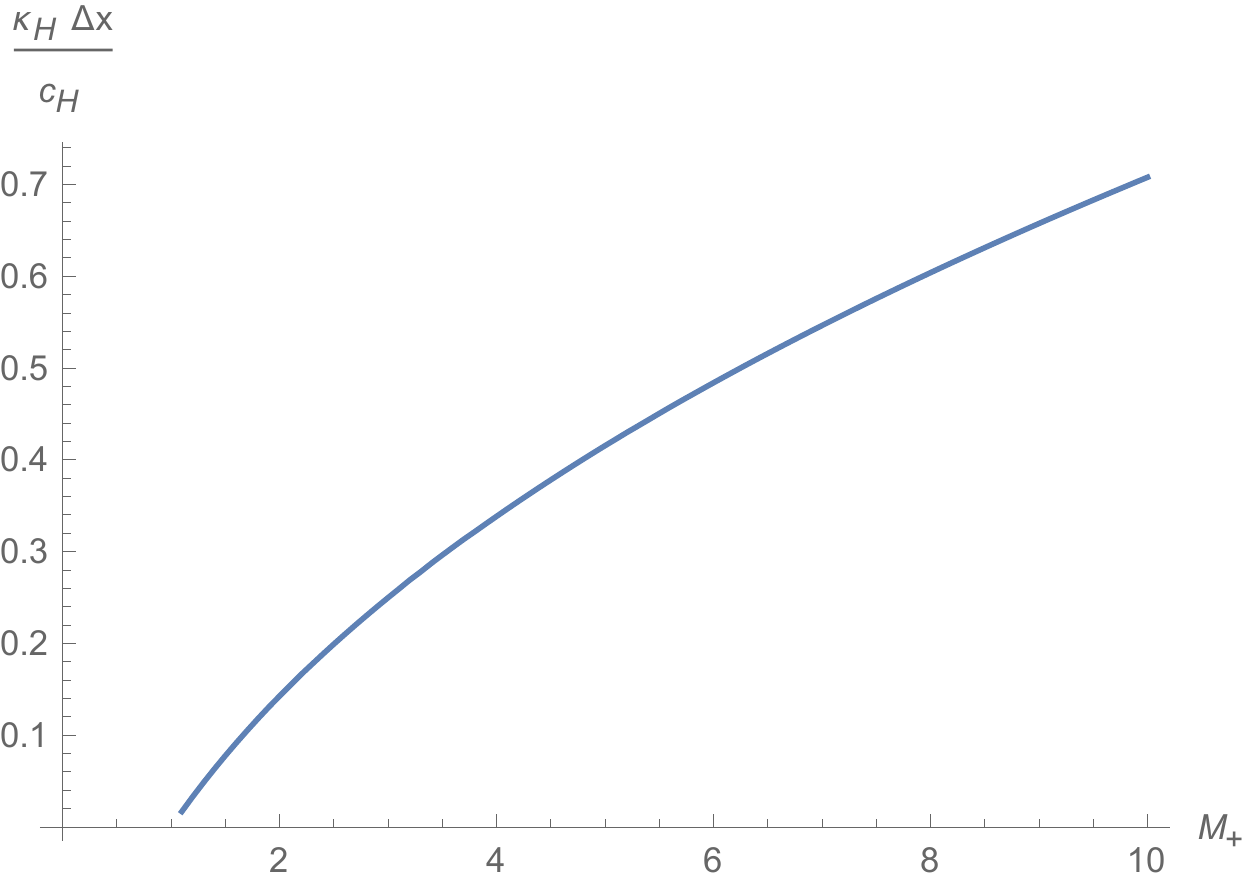}
\end{center}
\caption{As a function of the downstream Mach number $M_+$, we show the relativistic shift $\Delta x$ of \eq{eq:relshift} adimensionalized by the length associated with the surface gravity $c_H/\kappa_H$. 
}\label{fig:shift_waterfall}
\end{figure}

In \fig{fig:shift_waterfall}, we show $\Delta x$ as a function of the downstream Mach number $M_+$ for the waterfall solution \eq{eq:waterfall}, in units of $c_H / \kappa_H$. We notice that it is positive for all values of $M_+$. Its behaviors at $M_+ \to 1$ and $M_+ \to \infty$ are 
\begin{equation}
\frac{\kappa_H}{c_H} \Delta x  \mathop{\sim}_{M_+ \to 1^+} \sqrt{\frac{2}{3}} \lp 1 - \sqrt{\frac{2}{3}} \rp \text{argtanh} \lp \sqrt{\frac{2}{3}} \rp \lp M_+ - 1 \rp
\end{equation}
and 
\begin{equation} 
\frac{\kappa_H}{c_H} \Delta x \mathop{\sim}_{M_+ \to \infty} \frac{1}{2} M_+^{1/6} \ln \lp M_+ \rp.
\end{equation}
In the following we compute the same shift for $\tanh$, perturbed $\tanh$, and linear profiles. 

\subsection{The \texorpdfstring{$\tanh$}{tanh} case} 
\label{sub:tanh_shift}

We consider a profile of the form
\begin{equation} \label{eq:tanhprofile}
c(x) - v(x) = a - b \tanh(x  \sigma), 
\end{equation}
where $\lp a,b, \sigma \rp \in \mathbb{R}^3$ and $b^2 > a^2$. \eq{eq:du} can be integrated explicitly, giving
\begin{align} \label{eq:u_tanh}
\begin{cases}
u_R = t - \dfrac{1}{\kappa} \lp \log \lp \dfrac{b \sinh(x / \sigma) - a \cosh(x / \sigma)}{b \sinh(x_R / \sigma) - a \cosh(x_R / \sigma)} \rp + \dfrac{a}{b} \dfrac{x-x_R}{\sigma} \rp \\[2.ex]
u_L = t - \dfrac{1}{\kappa} \lp \log \lp \dfrac{b \sinh(x / \sigma) - a \cosh(x / \sigma)}{b \sinh(x_L / \sigma) - a \cosh(x_L / \sigma)} \rp + \dfrac{a}{b} \dfrac{x-x_L}{\sigma} \rp
\end{cases},
\end{align}
where $x_R$ and $x_L$ are two integration constants. $\kappa$ is the surface gravity, equal to
\begin{equation}
\kappa = \frac{b}{\sigma} \lp 1 - \lp \frac{a}{b} \rp^2 \rp.
\end{equation}
Continuity of $\pd_x U$ across the horizon imposes
\begin{equation} 
\frac{b \sinh \lp x_L / \sigma \rp - a \cosh \lp x_L / \sigma \rp}{b \sinh \lp x_R / \sigma \rp - a \cosh \lp x_R / \sigma \rp} = - \exp \lp \frac{a}{b} \frac{x_R - x_L}{\sigma} \rp.
\end{equation}
Using this relation and the position of the horizon: $x_H = (\sigma / 2) \log \lp (a+b)/(b-a) \rp$, $u_R$ and $u_L$ can be written in the limits $x / \sigma \to \pm \infty$
\begin{equation} 
\begin{cases}
u_R \mathop{=}\limits_{x \to + \infty} t - \dfrac{1}{\kappa} \lp \log \lp \dfrac{-\displaystyle{\sqrt{b^2-a^2}}}{b \sinh(x_L / \sigma) - a \cosh(x_L / \sigma)} \rp -\dfrac{a}{2b} \log \lp \dfrac{b-a}{b+a} \rp - \dfrac{a}{b} \dfrac{x_L}{\sigma} + \lp 1+\dfrac{a}{b} \rp \dfrac{x-x_H}{\sigma} \rp + O \lp e^{-2x/\sigma} \rp \\[2.2ex]
u_L \mathop{=}\limits_{x \to - \infty} t - \dfrac{1}{\kappa} \lp \log \lp \dfrac{-\displaystyle{\sqrt{b^2-a^2}}}{b \sinh(x_L / \sigma) - a \cosh(x_L / \sigma)} \rp -\dfrac{a}{2b} \log \lp \dfrac{b-a}{b+a} \rp - \dfrac{a}{b} \dfrac{x_L}{\sigma} - \lp 1-\dfrac{a}{b} \rp \dfrac{x-x_H}{\sigma} \rp + O \lp e^{2x/\sigma} \rp
\end{cases}
\end{equation}
Notice that the constant terms in $u_R$ and $u_L$ are the same. So, with the notations of \eq{eq:expu}, $b_+ / v_+^{\rm gr} = b_- / v_-^{\rm gr}$, hence a vanishing shift $\Delta x = 0$.

\subsection{Perturbed \texorpdfstring{$\tanh$}{tanh} profile} 
\label{sub:pert_tanh_shift}

\begin{figure}
\includegraphics[width=0.49 \linewidth]{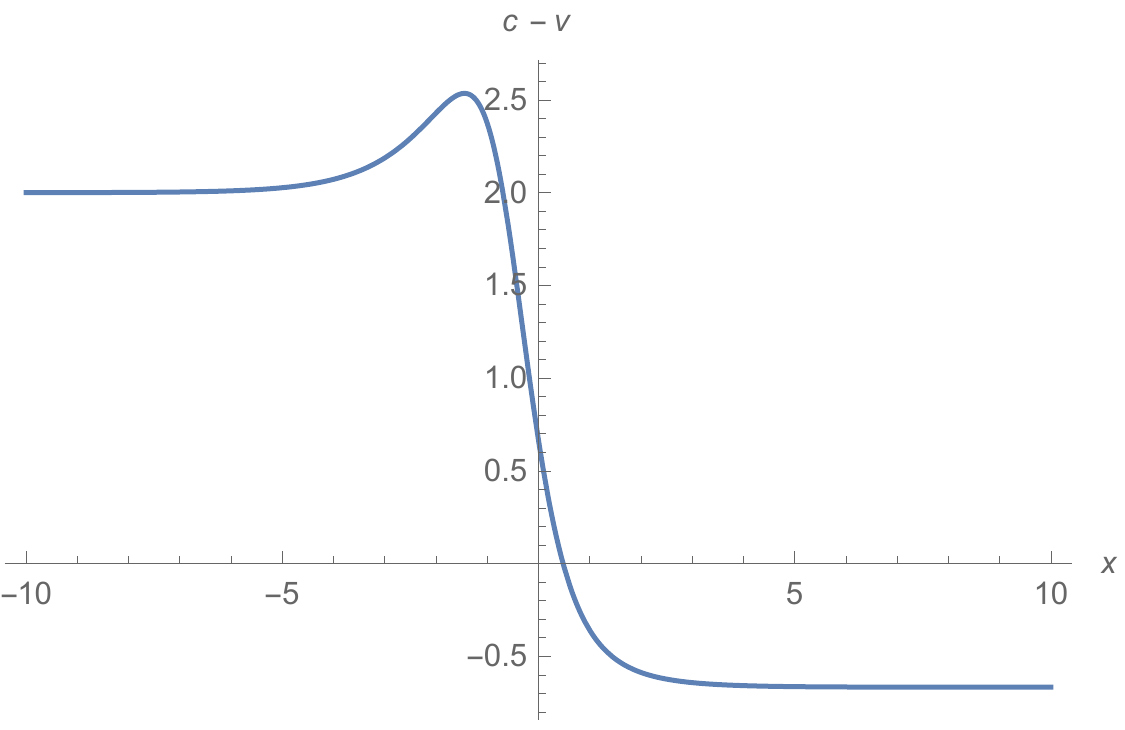}
\caption{Profile of $c-v$ for the perturbed $\tanh$ flow of \eq{eq:u_pert_tanh} with $\gamma = \eta = 0.5$. The horizon is located at $x \approx 0.48$. 
} \label{fig:vpc_pert_tanh}
\end{figure}

The result of the subsection~\ref{sub:tanh_shift} can be misleading as they seem to suggest that the vanishing of the shift $\Delta x$ is a generic property of flows with smooth $c-v$. To show that this is not the case, let us consider the following generalization of \eq{eq:u_tanh}:
\begin{equation}\label{eq:u_pert_tanh_dim}
u(t,x) = t + \frac{1}{\kappa} \lp \log \lp \frac{\abs{b \sinh \lp x / \sigma \rp - a \cosh \lp x / \sigma \rp - c}}{A_0} \rp - \eta \frac{x}{\sigma} \rp,
\end{equation}
where $\lp a,b,c,\kappa, A_0, \eta, \sigma \rp \in \mathbb{R}^7$. One can check that the function $U$ thus defined is smooth.\footnote{Because of the absolute value, $u$ can be continued in the complex plane to a function which is smooth just above the real axis, up to a term $i \pi / \kappa$ which arises when passing above $x_H$. This additional term exactly compensates the relative sign in the definition of $U$ on both sides of the horizon, see \eq{eq:U}.} As in subsection~\ref{sub:tanh_shift}, we assume $b^2 > a^2$. One can reduce the number of free parameters by defining the non-dimensional variables
\begin{equation}
\bar{t} \equiv \kappa t, \; \bar{x} \equiv \frac{x}{\sigma} + \frac{1}{2} \log \lp \frac{b-a}{b+a} \rp \; \text{and} \; \bar{u} \equiv \kappa u.
\end{equation}
In the remainder of this subsection we will work only with these non-dimensional variables and remove the bars to simplify the notations. \eq{eq:u_pert_tanh_dim} then becomes
\begin{equation}\label{eq:u_pert_tanh}
u(x,t) = t + \log \lp \abs{\sinh(x)-\gamma} \rp + \eta x,
\end{equation}
where $\gamma \equiv c/b$. Differentiating \eq{eq:u_pert_tanh} gives
\begin{equation}\label{eq:vpc_pert_tanh}
\frac{1}{c(x)-v(x)} = \frac{\cosh(x)}{\sinh(x) - \gamma} + \eta.
\end{equation}
The horizon is located at $x_H = {\rm argsinh}(\gamma)$. A straightforward calculation shows that $c(x) - v(x)$ 
has no divergence, and thus can correspond to a physical flow, if and only if 
\begin{equation}
\eta^2 < 1 \wedge \eta \gamma < \sqrt{1-\eta^2}.
\end{equation} 
Notice however that $c - v$ is not a monotonic function of $x$ when $\gamma \neq 0$: $c' - v'$ changes sign at $x = \text{argsinh} \lp 1 / \gamma \rp$, see \fig{fig:vpc_pert_tanh}. 

\begin{figure}
\begin{center}
\includegraphics[width=0.49 \linewidth]{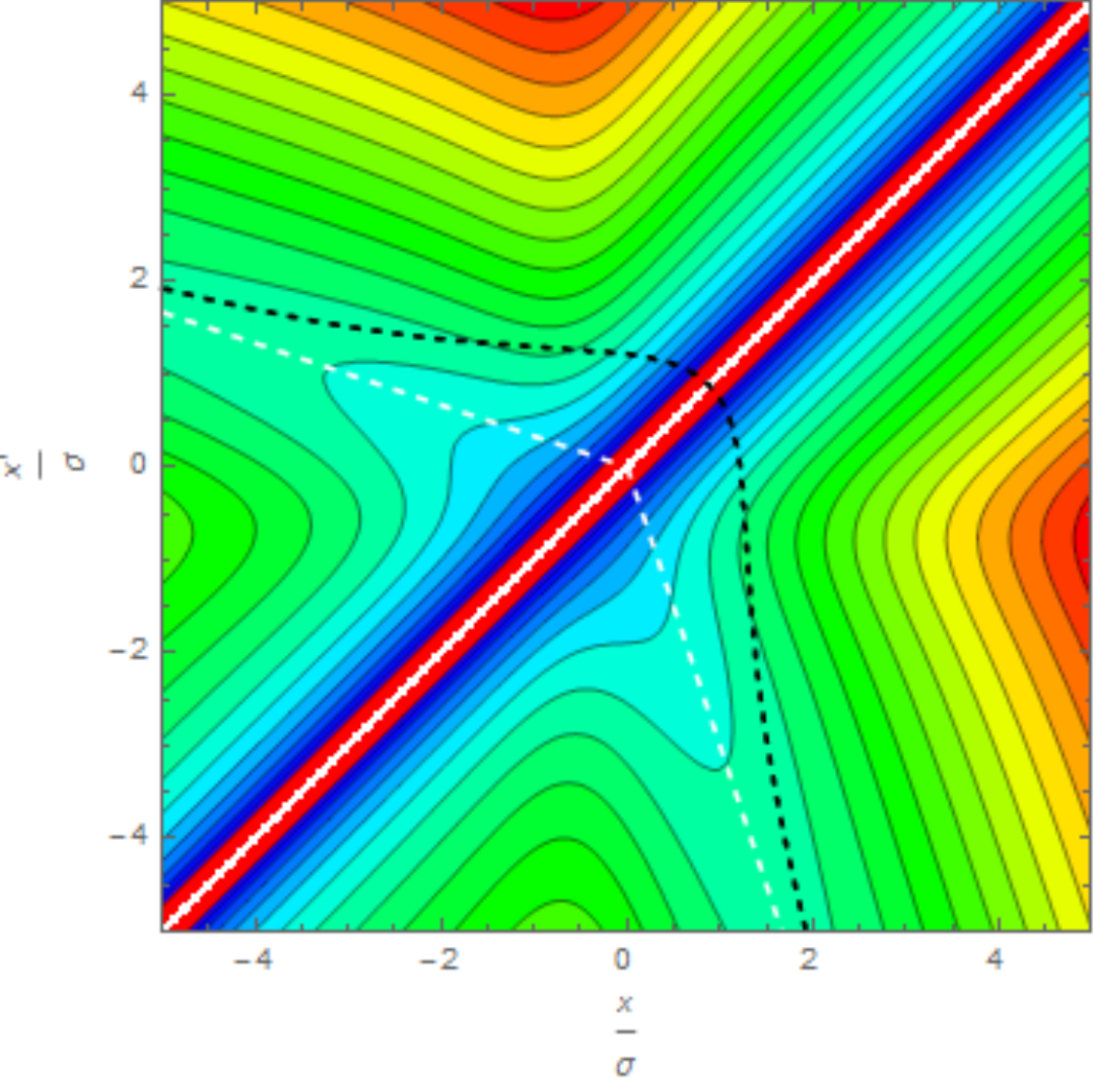}
\includegraphics[width=0.49 \linewidth]{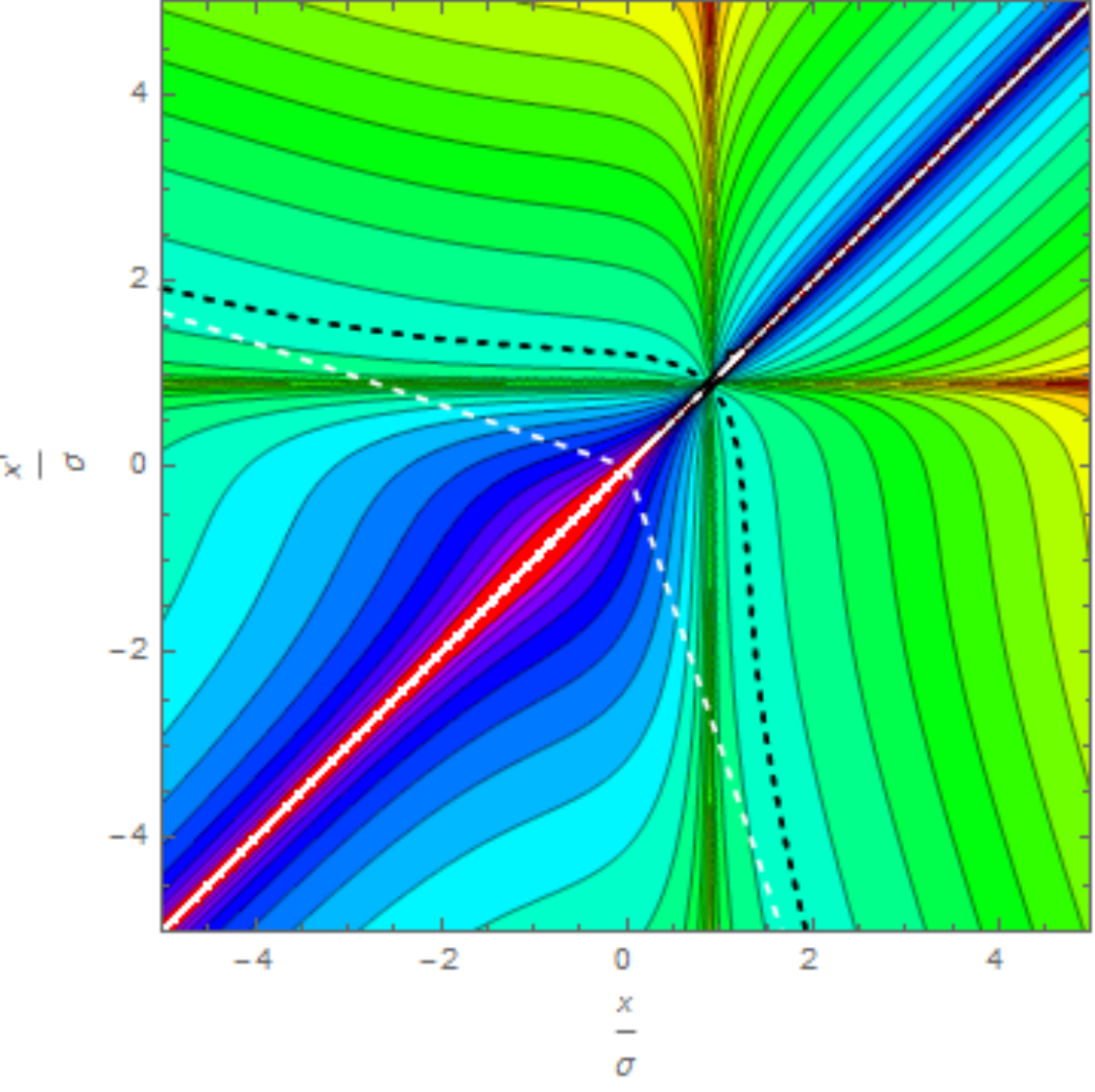} 
\includegraphics[width=0.4 \linewidth]{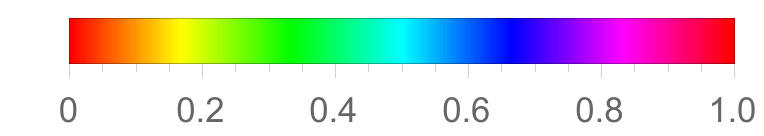} 
\end{center}
\caption{Two-point correlation function \eq{eq:Gxx1} (left) and \eq{eq:Gtt} (right) (rescaled so that the maximum represented value is $1$) for a perturbed $\tanh$ profile given by \eq{eq:vpc_pert_tanh}, with $\gamma = 2 \eta = 1$. The black, dashed line shows the locus $u_R(x) = u_L(x')$ and $u_L(x) = u_R(x')$. 
The white, straight, dashed half-lines are their asymptotes. In these plots, the origin of $x$ has been put where they meet, i.e., $x_M = 0$. Hence the sonic horizon is located at $x_H = - \Delta x$. 
On the right plot, we clearly see that the maxima of $G_2^{(tt)}$, which give $x(x')$, start from $x= x_H$. On the left plot, we see that the maxima of $G_2^{(xx)}$ emerge from a broad region which is approximately centered on $x_M = 0$, and not on $x_H$. 
}\label{fig:corr_xx_pert_tanh}
\end{figure}

Setting $u(t,x) = u(t,x')$ with $x > x_H$ and $x' < x_H$ and taking the limit $x \to \infty$ gives 
\begin{equation}
\frac{x}{ v_+^{\rm gr}} - \frac{x'}{ v_-^{\rm gr} } = 0. 
\end{equation}
Using \eq{c10}, we get $\Delta x = - x_H$. In other words, the asymptotes of the lines $u_R = u_L$ intersect at $x_M = 0$, which is not the position of the horizon for $\gamma \neq 0$. The corresponding two-point correlation function of \eq{eq:Gxx} is shown in \fig{fig:corr_xx_pert_tanh} for a positive $\gamma$. We also represent the equal-time 
correlation $G_2^{(tt)}$ given by 
\begin{eqnarray}
\label{eq:Gtt}
G_2^{(tt)} \lp x,t ; x',t  \rp &\equiv& \frac{-1}{4\pi} \pd_t \pd_{t'} \ln \lp U(t,x) - U(t',x') \rp_{t = t'} \\ \nonumber
&=& \frac{\kappa_H^2}{ 16 \pi\lp\cosh \lp 
\frac{\kappa_H}{2} (u_R(x) - u_L(x')) \rp \rp^2}, 
\end{eqnarray}
which behaves differently in the near horizon region, as it only depends on $u_R(x)$ and $u_L(x)$, and not on their space derivatives. Hence the maximal value of $G_2^{(tt)} \lp x,t ; x',t \rp$ exactly follows the $x(x')$ giving 
the image of the pair of null geodesics in the $x,x'$ plane.

\subsection{Linear profile}
\label{sub:lin_shift}
As a third example, we consider a profile of $v + c$ which is linear in a ``near-horizon'' region and uniform outside it. To be specific, we define three real numbers $\kappa$, $x_+ > 0$, and $x_- < 0$, and choose the profile
\begin{equation}\label{eq:vpc_lin}
c(x) - v(x) = \left\lbrace
\begin{array}{ll}
-\kappa x_- & x < x_- \\
-\kappa x & x_- \leq x \leq x_+ \\
-\kappa x_+ & x_+ < x
\end{array}
\right. .
\end{equation}
$\kappa$ is then the surface gravity. Integrating $1/ (c - v)$ over $x$, imposing that $U$ be a smooth function of $x$ across the horizon and at $x \in \left\lbrace x_-, x_+ \right\rbrace$ gives 
\begin{equation}
u(x) = \left\lbrace
\begin{array}{ll}
t - \dfrac{x-x_-}{\kappa x_-} - \dfrac{1}{\kappa} \log \lp \dfrac{-x_-}{\abs{x_0}} \rp & x < x_- \\
t - \dfrac{1}{\kappa} \log \lp \abs{\dfrac{x}{x_0}} \rp & x_- \leq x \leq x_+ \\
t - \dfrac{x - x_+}{\kappa x_+} - \dfrac{1}{\kappa} \log \lp \dfrac{x_+}{\abs{x_0}} \rp & x_+ < x
\end{array}
\right. ,
\end{equation}
where $x_0$ is an arbitrary integration constant. Considering two points $x > x_+$ and $x' < x_-$, setting $u(t,x) = u(t,x')$ gives
\begin{equation}
\frac{x}{\kappa x_+} - \frac{x'}{\kappa x_-} = \frac{1}{\kappa} \log \lp \frac{-x_-}{x_+} \rp.
\end{equation}
Since the horizon is at $x_H=0$, we obtain
\begin{equation}
\Delta x = \frac{\log \lp - x_- / x_+ \rp}{(1/x_+) - (1/x_-)}.
\end{equation}

\section{Phase of the coefficients \texorpdfstring{$\alpha_\om$}{alpha} and \texorpdfstring{$\beta_\om$}{beta}} 
\label{App:phase} 

In this appendix we consider the phases of the individual coefficients $\alpha_\om$ and $\beta_{-\om}$ 
appearing in the first line of \eq{eq:scatt_coeff}. 
When working with the vacuum, it is unclear whether these phases can be measured experimentally. However they can be measured when working in the stimulated regime by sending a classical wave~\cite{Rousseaux:2007is,Weinfurtner:2010nu,Euve:2015vml}, or a coherent state~\cite{Macher:2009nz}, towards the horizon. We hope they will be measured in forthcoming experiments based on water waves. 
We wish to emphasize here that the comparison of their measured values with those theoretically computed could provide an important additional check that the observed phenomena are indeed due to the particular mode conversion one wishes to probe. 
So far, only the norms of the scattering coefficients have been used, and there is a need for alternative ways to discriminate between different possible explanations.

\begin{figure} 
\begin{center}
\includegraphics[width=0.49 \linewidth]{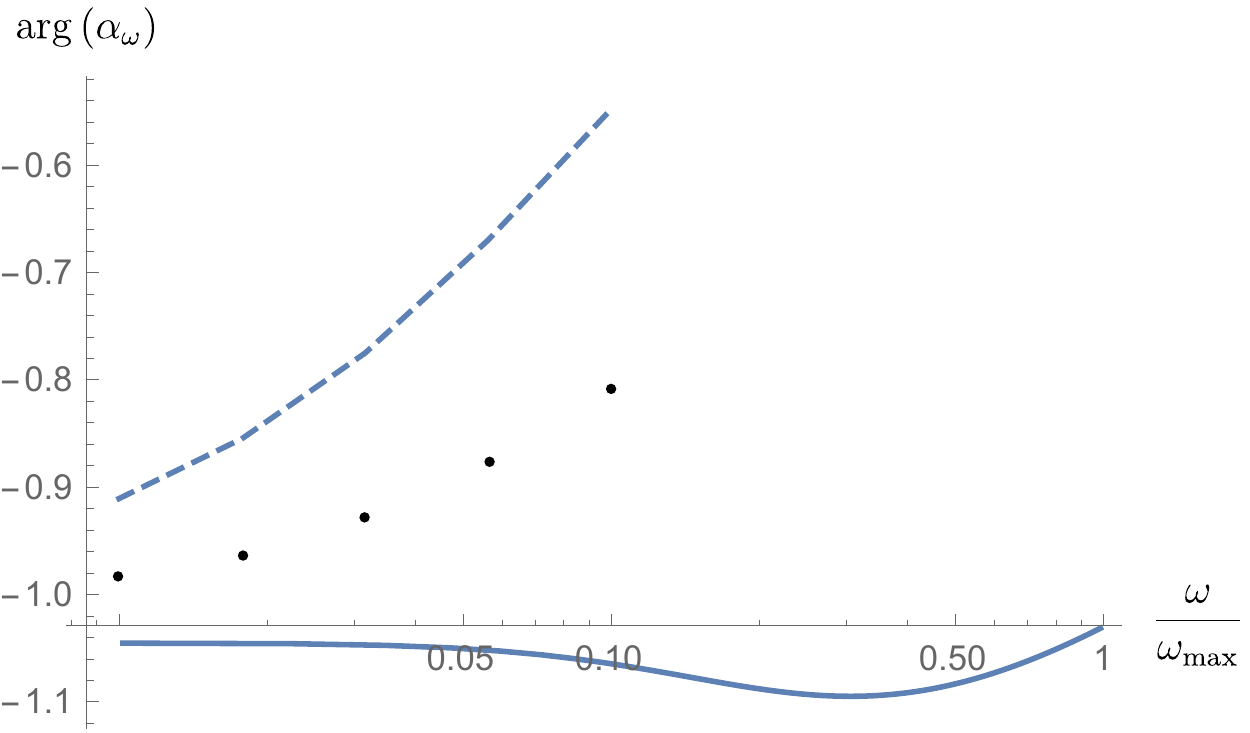} Fphia
\includegraphics[width=0.49 \linewidth]{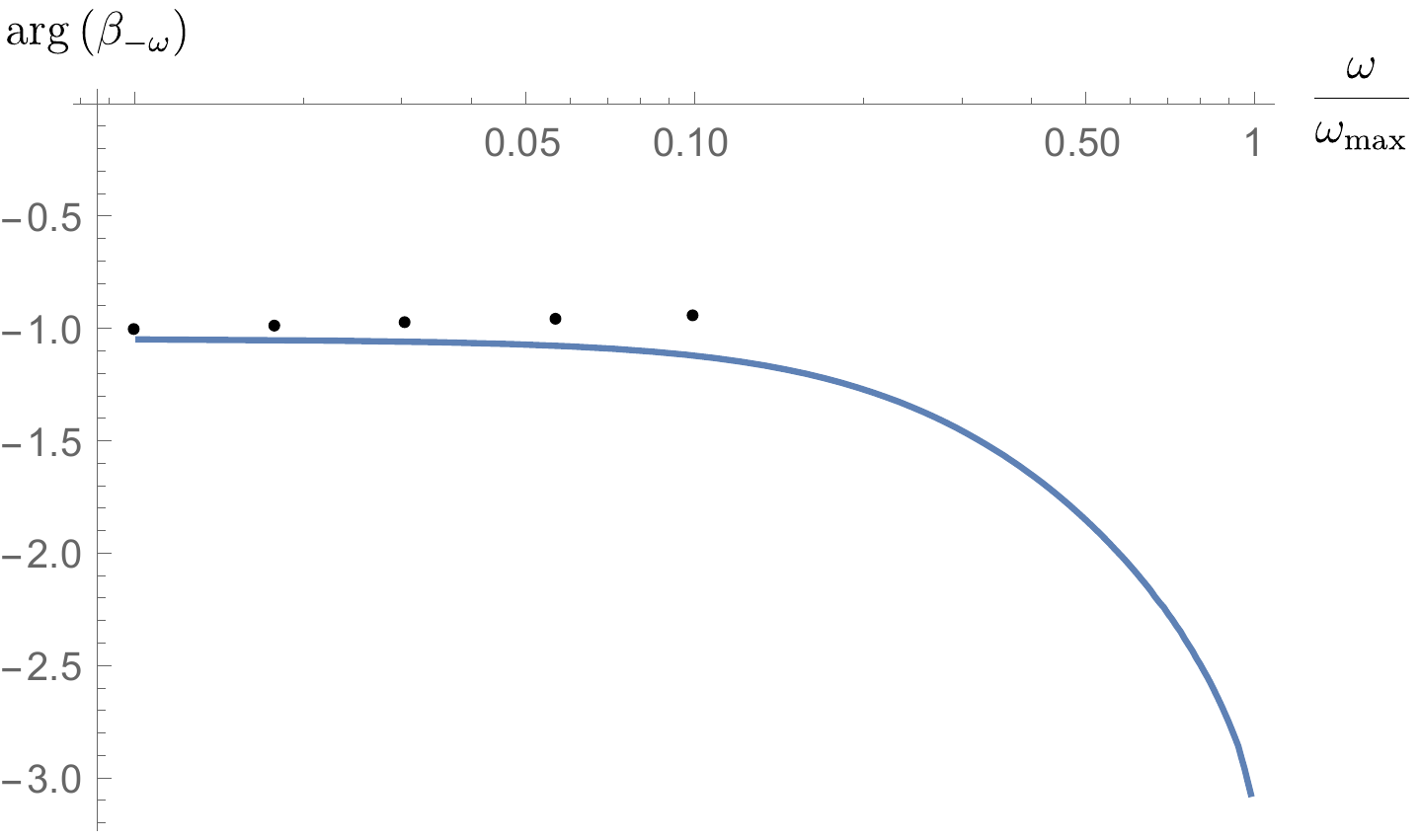} 
\caption{Phase of the scattering coefficients $\alpha_\om$ (left) and $\beta_{-\om}$ (right) 
for the waterfall solution with $M_+ = 5$. 
Continuous curves show the phases of $\alpha_\om$ and $\beta_{-\om}$ obtained by numerical integration and working with the definition of the asymptotic modes used in the main text. 
Dots show the analytical prediction of~\cite{Coutant:2011in} re-expressed with the same modes. The dashed line on the left plot shows the result of the naive WKB approximation, see text for explanation. 
}\label{fig:phasealphabeta} 
\end{center}
\end{figure} 
To compute the phases of individual coefficients, one needs to choose the phases of the asymptotic modes that are involved. 
For the coefficients $\alpha_\om$ and $\beta_{-\om}$, the latter are the incoming mode $\phi_\om^{d, \rm in}$, governed by the dispersive root $k^d_\om$, and the two out modes $\phi_\om^{u}$ and $(\varphi_{-\om}^{u})^*$, see \fig{fig:charact}. 
As done in \eq{3chi}, we work with modes whose phase is (on the side where the mode is defined) 
asymptotically equal to $k^a_\om x + o(1)$ (we remind the reader that $o(1)$ means ``up to terms going asymptotically to zero''). The phases of the scattering coefficients $\alpha_\om$ and $\beta_{-\om}$ as functions of $\om/\om_{\rm max}$ and obtained by solving numerically \eq{eq:BdG_W} are shown by continuous lines in \fig{fig:phasealphabeta} for the waterfall solution with $M_+ = 5$. We see that both $\arg \alpha_\om$ and $\arg \beta_{-\om}$ display a regular smooth behavior which could be compared with experimental data. 

To complete this study, we now compare these curves with the predictions of~\cite{Coutant:2011in}. This theoretical analysis involves modes evaluated in an intermediate region satisfying the two following properties, see Eqs.~(42-45) and Fig.~2 of this reference. 
First, it must be sufficiently far away from the turning point $x_{\rm t.p.}(-\om)$ of the negative-frequency mode, see \fig{fig:charact}, for the WKB approximation to be valid. Second, it must be close enough to the sonic horizon for $c(x)-v(x) \approx \kappa_H x$ to provide a good approximation. 
To perform the comparison with the above curves, the modes defined in this intermediate region need to be propagated further away from the horizon towards $|x| \to \infty$. 
Assuming that this propagation can be described under the WKB approximation, the phases accumulated from the near horizon region to the asymptotic ones are given by
\begin{eqnarray}\label{eq:WKBphase} 
&& \arg \lp \phi_{\om}^{d, \rm WKB} (x) \rp \mathop{=}_{x \to + \infty} 
\int_{x_d(\om)}^x k_{\om}^d(y) dy + C_d(\om) + o(1), \nonumber \\ 
&&\arg \left( \varphi_{-\om}^{u, \rm WKB} (x) \right) \mathop{=}_{x \to +\infty} \int_{x_{u}(-\om)}^x k^u_{-\om}(y) dy + C_u(-\om) + o(1). 
\label{2C}
\end{eqnarray}
The phases $C_d(\om)$ and $C_u(-\om)$ are then chosen to match Eqs.~(42,44) of~\cite{Coutant:2011in} at the corresponding locations $x_d(\om)$ and $x_u(-\om)$ situated in the intermediate region, on the supersonic side. 
(The outgoing mode $\phi_\om^u$ can be matched at the same point as the incoming mode $\phi_{\om}^{d,\rm in}$ because their wave vectors belong to the same branch of the dispersion relation.) When the intermediate region is sufficiently extended, the results should hardly depend on the exact location of $x_d(\om)$ and $x_u(-\om)$. However, in the waterfall solution with $M_+ = 5$, the scale of variation of $\kappa(x) = \partial_x(c-v)$ is of the same order as the dispersive length scale, see the discussion in the paragraph after \eq{eq:surgra}. Considering for instance the value $\om = 0.1\om_{\rm max} \approx 0.8 T_H$, we find that $\kappa_H$ and $\kappa(x_{\rm t.p.})$ differ by more than $60 \%$. As a result, the treatment used in~\cite{Coutant:2011in} is not expected to give accurate results in the present flow, and can only be applied to frequencies lower than $\om_{\rm max} / 10$. 
The phases given in Eq.~(76) of~\cite{Coutant:2011in} are shown by dots in \fig{fig:phasealphabeta} for $\om \leq 0.1 \om_{\rm max}$. 
They are re-expressed in terms of the asymptotic modes obeying \eq{2C}. (To get the values represented, we adjusted the location of $x_d(\om)$ and $x_u(-\om)$ to minimize the residual phase dependence in the mode matching.) The dashed line on the left plot shows the result of the naive WKB approximation, assuming the mode $\phi_{\om}$ is given by its WKB expression everywhere up to a phase jump of $\pi / 4$ (which is the $\om \to 0$ limit of the analytical result, see Eqs.~(71) and~(76) in~\cite{Coutant:2011in}) around the horizon. This naive treatment is unable to predict the phase of the $\beta_\om$ coefficient. 

As can be seen in the figure, in spite of the fact that the waterfall solution with $M_+ = 5$ does not meet the regularity conditions used in the analytical treatment, its predictions are in rather good agreement with numerical values. 
We also see that the error significantly increases when $\om$ approaches $0.1 \om_{\rm max}$, in agreement with the above remark. 
These results indicate that the phases of the scattering coefficients seem rather robust quantities, which could ease the comparison between observational and numerical values. 

\bibliography{../bib/biblio}
\end{document}